\documentclass{elsart}
\usepackage{amsfonts}
\usepackage{epsfig}
\newlength{\wordlength}

\newlength{\onewordlength}

    \newcommand{\ba}{\begin{eqnarray}}
    \newcommand{\ea}{\end{eqnarray}}
    \newcommand{\be}{\begin{equation}}
    \newcommand{\ee}{\end{equation}}

    \newcommand{\AmS}{{\protect\the\textfont2%
  A\kern-.1667em\lower.5ex\hbox{M}\kern-.125emS}}

\newcommand{\bn}{{\bf n}}

\newcommand {\bk} {{\mathbf k}}
\newcommand {\bq} {{\mathbf q}}

\newcommand {\bp} {{\mathbf p}}

\newcommand{\bx}{{\bf x}}
\newcommand{\by}{{\bf y}}

\newcommand{\calO}{{\mathcal O}}
\newcommand{\calR}{{\mathcal R}}
\newcommand{\calM}{{\mathcal M}}
\newcommand{\calZ}{{\mathcal Z}}

\begin{document}
\runauthor{PKU}
\begin{frontmatter}

\title{Low-energy $D^{*+}D^0_1$ Scattering and the Resonance-like Structure $Z^+(4430)$}
 CLQCD Collaboration\\
 \author[PKU]{Guo-Zhan Meng},
 \author[PKU]{Ming Gong},
 \author[IHEP]{Ying Chen},
 \author[PKU]{Song He},
 \author[IHEP]{Gang Li},
 \author[PKUC]{Chuan Liu},
 \author[NANKAI]{Yu-Bin Liu},
 \author[ITP]{Jian-Ping Ma},
 \author[NANKAI]{Xiang-Fei Meng},
 \author[PKU]{Zhi-Yuan Niu},
 \author[PKU]{Yan Shen},
 \author[ZHEJIANG]{Jian-Bo Zhang},
 \author[IHEP]{Yuan-Jiang Zhang}

 \address[PKU]{School of Physics, Peking University\\
               Beijing, 100871, P.~R.~China}
 \address[IHEP]{Institute of High Energy Physics\\
                Academia Sinica, P.~O.~Box 918\\
                Beijing, 100039, P.~R.~China}
 \address[PKUC]{School of Physics and Center for High Energy Physics\\
               Peking University, Beijing, 100871, P.~R.~China}
 \address[NANKAI]{Department of Physics, Nankai University\\
                  Tianjin, 300071, P.~R.~China}
 \address[ITP]{Institute of Theoretical Physics\\
                Academia Sinica, Beijing, 100080, P.~R.~China}
 \address[ZHEJIANG]{Department of Physics, Zhejiang University\\
                  Hangzhou, 310027, P.~R.~China}
                \thanks{Work supported in part by NSFC under grant No.10835002, No.10675005 and No.10721063.}

 \begin{abstract}
 Low-energy scattering of $D^*$ and $D_1$ meson are studied using quenched
 lattice QCD with improved lattice actions on anisotropic lattices.
 The calculation is performed within L\"uscher's finite-size
 formalism which establishes the relation between the scattering phase
 in the infinite volume and the exact energy level in the finite
 volume. The threshold scattering parameters, namely the scattering
 length $a_0$ and the effective range $r_0$, for the $s$-wave scattering in
 $J^P=0^-$ channel are extracted.
 After the chiral and  continuum extrapolations, we obtain:
 $a_0=2.52(47)$fm and $r_0=0.7(1)$fm where the errors are purely
 statistical. Based on these results, we discuss the possibility of a shallow bound state
 for the two charmed mesons within the non-relativistic potential scattering model.
 It is argued that, albeit the interaction between the two
 charmed mesons being attractive, it is unlikely that
 they can form a shallow bound state in this channel. This calculation
 provides some useful information on the nature of the newly
 discovered resonance-like structure $Z^+(4430)$ by the Belle
 Collaboration.
 \end{abstract}
 \end{frontmatter}


 \section{Introduction}

 Recently, a charged resonance-like structure $Z^+(4430)$ has been observed at
 Belle in the $\pi\psi'$ invariant mass spectrum of $B\rightarrow
 K\pi^+\psi'$ decays~\cite{:2007wga}. This discovery has triggered
 many theoretical investigations on the nature of this
 structure~\cite{Rosner:2007mu,Liu:2007bf,Liu:2008xz,Bugg:2008wu,Qiao:2007ce,Lee:2007gs,Braaten:2007xw,Liu:2008qx,Godfrey:2008nc}.
 Since the invariant mass of the resonance is very close to the $D^*D_1$ threshold,
 one possible interpretation is a molecular bound state formed by the $D^*$ and $D_1$
 mesons~\cite{Liu:2007bf,Liu:2008xz}.
 To further investigate this possibility, the interaction between
 $D^*$ and $D_1$ mesons becomes crucial.
 As is known, the interaction of two hadrons can be studied via
 the scattering process of the hadrons. Since the energy being
 considered here is very close to the threshold of the $D^*D_1$ system,
 only threshold scattering parameters, i.e. scattering length $a_0$ and
 effective range $r_0$, are relevant for this study.
 In phenomenological studies, the interaction between the mesons can be computed
 by assuming meson exchanges models.
 However, since the interaction between the mesons at low-energies
 is non-perturbative in nature, it is tempting
 to study the problem using a genuine non-perturbative method like
 lattice QCD.

 In this paper, we study the scattering threshold parameters of
 $D^*-D_1$ system using quenched lattice QCD within
 the so-called L\"uscher's formalism, a finite-size technique
 developed to study scattering processes in a finite
 volume~\cite{luscher86:finitea,luscher86:finiteb,luscher90:finite,luscher91:finitea,luscher91:finiteb}.
 Within this approach, it is also feasible to investigate
 the possible bound state of the two mesons~\cite{luscher91:finitea,Sasaki:2007kr}.
 We have used improved gauge and fermion lattice actions on
 anisotropic lattices. The usage of anisotropic lattices with asymmetric volumes
 has enhanced our resolution in energy and the momentum. The computation
 is carried out in all possible angular momentum channels, although
 only the $J^P=0^-$ channel yields definite results. We find that,
 in this particular channel, the interaction between a $D^*$ and
 a $D_1$ meson is attractive in nature. The scattering length after
 continuum and chiral extrapolation is $a_0=2.52(47)$fm while the effective range
 is $r_0=0.7(1)$fm. Possibility of a bound state can also
 be addressed within L\"uscher's formalism. Our simulation results
 indicate that the two-particle system of the two charmed mesons
 resembles more like an ordinary scattering state rather
 than a shallow bound state.

 This paper is organized as follows. In Section~\ref{sec:method}, we briefly
 introduce L\"uscher's formalism and its extensions to the asymmetric volumes.
 In Section~\ref{sec:operators},  we discuss possible one-particle
 and two-particle interpolating operators and their correlation matrices
 are defined. In section~\ref{sec:simulation_details}, simulation
 details are given and the results for the single- and two-meson systems
 are analyzed. After verifying the single-particle states, we extract
 the exact energy of the two-particle system. By applying L\"uscher's
 formula the scattering phases are extracted for various lattice momenta.
 When fitted to the known low-energy behavior, the threshold parameters
 of the scattering system, i.e. the scattering length $a_0$ and the effective range $r_0$
 are obtained in the $s$-channel. We also discuss various interpolation and
 extrapolations which bring our results to the chiral and continuum limit.
 Based on our simulation results, the possibility of
 a bound state in this channel is discussed.
 In Section~\ref{sec:conclude}, we will conclude with some general remarks.

 \section{Strategies for the computation}
 \label{sec:method}
 \subsection{L\"uscher's finite volume technique and its generalization}
 \label{subsec:formulae}
 Within L\"uscher's formalism, the exact energy eigenvalue
 of a two-particle system in a finite box of size $L$ is
 related to the elastic scattering phase of the two particles
 in the infinite volume. Consider two interacting particles
 with mass $m_1$ and $m_2$ enclosed in a cubic box of size $L$,
 with periodic boundary conditions applied in all three directions.
 The spatial momentum $\bk$ is quantized according to:
 \be
 \label{eq:free_k}
 \bk=\left({2\pi\over L}\right)\bn\;,
 \ee
 with $\bn$ being a three-dimensional integer.
 Now consider the two-particle system in this finite box
 and let us take the center-of-mass frame of the system
 so that the two-particles having three-momentum
 $\bk$ and $-\bk$ respectively.
 The exact energy of the two-particle
 system in this finite volume is denoted as: $E_{1\cdot2}(\bk)$.
 We now define a variable $\bar{\bk}^2$ via:
 \be
 E_{1\cdot 2}(\bk)=\sqrt{m^2_1+\bar{\bk}^2}
 +\sqrt{m^2_2+\bar{\bk}^2}\;.
 \ee
 Note that due to interaction between the two particles,
 the value of $\bar{\bk}^2$ differs from its free counter-part
 $\bk^2$ with $\bk$ being quantizes according to
 Eq.~(\ref{eq:free_k}). It is also convenient to further
 define a variable $q^2$ as:
 \be
 q^2=\bar{\bk}^2L^2/(2\pi)^2\;.
 \ee
 which differs from $\bn^2$ due to interaction.
 What L\"uscher's formula tells  us  is a direct relation
 of $q^2$ and the elastic scattering phase shift $\tan\delta(q)$
 in the infinite volume and it reads:~\cite{luscher91:finitea}
 \be
 \label{eq:luscher_cube}
 \tan\delta(q)={\pi^{3/2}q\over\calZ_{00}(1;q^2)}\;,
 \ee
 where $\calZ_{00}(1;q^2)$ is the zeta-function which
 can be evaluated numerically once its argument $q^2$ is given.
 Therefore, if we could obtain the exact two-particle energy
 $E_{1\cdot 2}(\bk)$ from numerical simulations, we could
 infer the elastic scattering phase shift by applying L\"uscher's
 formula given above.

 Here we would like to point out that,
 the above relation is in fact only valid under certain assumptions.
 For example, the size of the box cannot be too small. In particular, it
 has to be large enough to accommodate free single-particle states.
 Therefore, in a practical simulation, one should check whether this
 is indeed realized in the simulation. Polarization effects are also
 neglected which are suppressed exponentially by $\calO(e^{-mL})$ where
 $m$ being the single-particle mass gap. Also neglected are mixtures
 from higher angular momenta.

 In the case of attractive interaction, the lowest two-particle
 energy level might be lower than the threshold which then
 renders the quantity $q^2<0$. The phase shift in the continuum, $\delta(q)$,
 is only defined for positive $q^2$, i.e. energies above the
 threshold. When $q^2<0$, it is related to the phase $\sigma(q)$ via:
 \be
 \tan\sigma(q)={\pi^{3/2}(-iq)\over\calZ_{00}(1;q^2)}\;,
 \ee
 where $(-iq) >0$ and the phase $\sigma(q)$ for pure imaginary $q$
 is obtained from $\delta(q)$ by analytic continuation:
 $\tan\sigma(q)=-i\tan\delta(q)$~\cite{luscher91:finitea,Sasaki:2007kr}.
 The phase $\sigma(q)$ for pure imaginary $q$ is of
 physical significance since if there exists a true
 bound state at that particular energy, we have $\tan\sigma(q)=-1$
 in the infinite volume and continuum limit.
 In the finite volume, the relation above is modified as:~\cite{Sasaki:2007kr}
 \be
 \cot\sigma(q)=-1+{6\over
 2\pi\sqrt{-q^2}}e^{-2\pi\sqrt{-q^2}}+\cdots\;,
 \ee
 where the finite-volume corrections are assumed to be small.
 Therefore, for $q^2<0$, we could compute $\tan\sigma(q)$ from
 Monte Carlo simulations and check the possibility of a
 bound state at that energy.

 The above formulae apply to the case of a box with cubic
 symmetry. In real calculations, in order to have more
 accessible low-momentum modes, it is advantageous to use
 asymmetric volumes in the study of hadron
 scattering~\cite{chuan04:asymmetric,chuan04:asymmetric_long,Li:2007ey}.
 If the rectangular box is of size
 $L\times (\eta_2L)\times (\eta_3L)$, then
 Eq.~(\ref{eq:luscher_cube}) is modified to:
 \be
 \label{eq:phase_rectangular}
 \tan\delta(q)={\pi^{3/2}q\eta_2\eta_3\over\calZ_{00}(1;q^2;\eta_2,\eta_3)}\;,
 \ee
 where the modified zeta-function $\calZ_{00}(1,q^2;\eta_2,\eta_3)$
 is the analogue of $\calZ(1;q^2)$ and its explicit definition
 can be found in Refs.~\cite{chuan04:asymmetric,chuan04:asymmetric_long}.
 Similarly, for negative $q^2$, the formula is modified to:
 \be
 \label{eq:sigma_rectangular}
 \tan\sigma(q)={\pi^{3/2}(-iq)\eta_2\eta_3\over\calZ_{00}(1;q^2;\eta_2,\eta_3)}\;.
 \ee

 \section{One- and two-particle operators and correlators}
 \label{sec:operators}

 Single-particle and two-particle energies are measured
 in Monte Carlo simulations using appropriate correlation functions.
 These correlation functions are constructed from
 corresponding interpolating operators with definite symmetries.
 Since we are interested in the interaction between a $D^*$ and a $D_1$
 meson, we need one-particle operators which would create a single $D^*$ and a single $D_1$
 meson and two-particle operators which create both $D^*$ and $D_1$
 from the QCD vacuum.
 Below we will first list these one-particle and two-particle operators
 and then proceed to discuss their correlation functions.

 \subsection{One- and two-particle operators with definite symmetries}

 Let us first construct the single meson operators for
 $D^\ast(2010)^+$ and $\bar{D}_1(2420)^0$ whose quantum numbers $J^P$
 are $1^-$ and $1^+$, respectively. Just to simplify the notation, we
 will use $Q_i$ and $P_i$ for these meson operators respectively,
 where $i = 1, 2, 3$ being the index to specify different spatial components.
 We use local interpolating fields as follows:
 \begin{eqnarray}
  Q_i(x)=[\bar{d} \gamma^i c](x), P_i(x)=[\bar{c} \gamma^i\gamma^5 u](x)
 \end{eqnarray}
 where $Q_i(x)$ stands for $D^\ast(2010)^+$
 while $P_i(x)$ stands for $\bar{D}_1(2420)^0$.
 A single-particle state with definite three-momentum $\bk$ is
 represented by the Fourier transform of the above operators:
 \begin{eqnarray}
  Q_i(t,\textbf{k})=\sum_\textbf{\scriptsize{x}} Q_i(t,\textbf{x})e^{-i \textbf{\scriptsize{k}} \cdot \textbf{\scriptsize{x}}},
  P_i(t,\textbf{k})=\sum_\textbf{\scriptsize{x}} P_i(t,\textbf{x})e^{-i \textbf{\scriptsize{k}} \cdot \textbf{\scriptsize{x}}}.
 \end{eqnarray}
 Obviously, the operators $Q_i(t,k)$ and $P_i(t, k)$ fall
 into the vector representation of the rotational group $SO(3)$ (i.e.
 their angular momentum quantum number is $1$) in the continuum.

 On the lattice, the rotational symmetry group $SO(3)$ is broken down
 to the corresponding point group.
 Usually, one utilizes an symmetric cubic box. In this case, the corresponding
 point group is the cubic group $O(\mathbb{Z})$.
 However, in order to access more non-degenerate low-momentum modes,
 it would be advantageous to use asymmetric box (although
 the lattice spacings in spatial directions are still symmetric).
 This is particularly useful for scattering processes,
 as advocated in Ref.~\cite{Li:2007ey}.
 Following this strategy, we have adopted a rectangular box
 of size $L\times (\eta_2 L)\times (\eta_3L)$ with $\eta_2=1$ and
 $\eta_3\neq 1$. In this  case, the rotational group in the continuum is broken
 down to the basic point group $D_4$. In what follows, we will construct operators that
 transform according to different irreducible representations (irreps)
 of the $D_4$ group.

 The basic point group $D_4$ has four one-dimensional irreducible
 representations: $A_1$, $A_2$, $B_1$, $B_2$ and one two-dimensional
 irreducible representation: $E$. With these notations, it is easy to verify that
 three components of an ordinary vector in the continuum, like $Q_i$'s and $P_i$'s given above,
 now falls into two irreps: $A_2$ and $E$. In particular, we have
 the following decomposition rules:
 \be
 \label{eq:decomposition}
  \textbf{0}=A_1,~\textbf{1}=E\oplus A_2,~\textbf{2}=A_1\oplus B_1\oplus B_2\oplus E.
 \ee

 For the two-particle system formed by a $D^*$ and a $D_1$ meson,
 the quantum number $J^P$ of the two-particle system can be:
 $J^P= 0^-,1^-,2^-$.
 Now, we consider the vector space
 $\{Q_1,Q_2,Q_3\}\otimes\{P_1,P_2,P_3\}$, which is 9-dimensional. Using
 standard group-theoretical methods, it is easy to find out that
 this 9-dimensional vector space is made up of two copies of $A_1$,
 one copy of $A_2$, $B_1$ and $B_2$ each and two copies of $E$.
 The basis operators of each irrep mentioned above are listed as follows:
 \begin{eqnarray}
 \label{eq:operators_def}
 O^{(A_1)(1)}(t)=&\sum_{R \in G}&[Q_1(t+1,-R \circ \textbf{k})P_1(t,R \circ \textbf{k})\nonumber\\
                             &&+Q_2(t+1,-R \circ \textbf{k})P_2(t,R \circ \textbf{k})\nonumber \\
                             &&+Q_3(t+1,-R \circ \textbf{k})P_3(t,R \circ \textbf{k})],\nonumber\\
 O^{(A_1)(2)}(t)=&\sum_{R \in G}&[Q_1(t+1,-R \circ \textbf{k})P_1(t,R \circ \textbf{k})\nonumber\\
                             &&+Q_2(t+1,-R \circ \textbf{k})P_2(t,R \circ \textbf{k})\nonumber \\
                             &&-2Q_3(t+1,-R \circ \textbf{k})P_3(t,R \circ \textbf{k})],\nonumber\\
 O^{(A_2)}(t)=&\sum_{R \in G}&[Q_1(t+1,-R \circ \textbf{k})P_2(t,R \circ \textbf{k})\nonumber\\
                           &&-Q_2(t+1,-R \circ \textbf{k})P_1(t,R \circ \textbf{k})],\nonumber\\
 O^{(B_1)}(t)=&\sum_{R \in G}&[Q_1(t+1,-R \circ \textbf{k})P_1(t,R \circ \textbf{k})\nonumber\\
                           &&-Q_2(t+1,-R \circ \textbf{k})P_2(t,R \circ \textbf{k})],\nonumber\\
 O^{(B_2)}(t)=&\sum_{R \in G}&[Q_1(t+1,-R \circ \textbf{k})P_2(t,R \circ \textbf{k})\nonumber\\
                           &&+Q_2(t+1,-R \circ \textbf{k})P_1(t,R \circ \textbf{k})],\nonumber\\
 O^{(E)(1)}_1(t)=&\sum_{R \in G}&[Q_1(t+1,-R \circ \textbf{k})P_3(t,R \circ \textbf{k})\nonumber\\
                           &&-Q_3(t+1,-R \circ \textbf{k})P_1(t,R \circ \textbf{k})],\nonumber\\
 O^{(E)(1)}_2(t)=&\sum_{R \in G}&[Q_2(t+1,-R \circ \textbf{k})P_3(t,R \circ \textbf{k})\nonumber\\
                           &&-Q_3(t+1,-R \circ \textbf{k})P_2(t,R \circ \textbf{k})],\nonumber\\
 O^{(E)(2)}_1(t)=&\sum_{R \in G}&[Q_1(t+1,-R \circ \textbf{k})P_3(t,R \circ \textbf{k})\nonumber\\
                           &&+Q_3(t+1,-R \circ \textbf{k})P_1(t,R \circ \textbf{k})],\nonumber\\
 O^{(E)(2)}_2(t)=&\sum_{R \in G}&[Q_2(t+1,-R \circ \textbf{k})P_3(t,R \circ \textbf{k})\nonumber\\
                           &&+Q_3(t+1,-R \circ \textbf{k})P_2(t,R \circ \textbf{k})],
 \label{operators}
 \end{eqnarray}
 where $\bk$ is a chosen three-momentum mode and
 $G$ is the group $D_4$ and $R\in G$ is an element of the group.
 $O^{(A_1)(i)}(t)$ with $i=1,2$ in this case designates different copies
 of the $A_1$ representations occurring in the decomposition.
 Note that in the above definitions we have not included orbital
 angular momentum of the two-particles. Therefore we are only studying
 the $s$-wave scattering of the two mesons. This is sufficient for this
 particular case since near the threshold, the scattering is always
 dominated by $s$-wave contributions.
 Using the correspondence in Eq.~(\ref{eq:decomposition}), it
 is easy to figure out the continuum quantum numbers for
 these operators which are tabulated in table \ref{tab:operator}.
 \begin{table}
 \caption{The two-particle operators defined
 in Eq.~(\ref{eq:operators_def}) and their corresponding angular momentum quantum
 number $J$ in the continuum.}
  \centering
  \begin{tabular}{|c|l|}
    \hline
    $J^P$   &   Two-particle operators\\
    \hline\hline
    ${\bf 0}^-$       &$O^{(A_1)(1)}(t)$\\
    \hline
    ${\bf 1}^-$       &$O^{(A_2)}(t)$,$O_1^{(E)(1)}(t)$,$O_2^{(E)(1)}(t)$\\
    \hline
    ${\bf 2}^-$       &$O^{(A_1)(2)}(t)$,$O^{(B1)}(t)$,$O^{(B_2)}(t)$,$O_1^{(E)(2)}(t)$,$O_2^{(E)(2)}(t)$\\
    \hline
  \end{tabular}
  \label{tab:operator}
 \end{table}

 \subsection{Correlation functions}

 We then proceed to discuss one-particle and two-particle correlation functions, respectively.
 As already mentioned, in the lattice study of hadron-hadron
 scattering, one first have to make sure that asymptotically free
 one-particle states are realized in the volume being considered. We
 therefore construct the one-particle correlation function
 $C^Q(t,{\bf k})$ and $C^P(t,{\bf k})$ for the $D^{\ast+}$ and $D_1^0$ meson as:
 \begin{eqnarray}
  C^Q(t,{\bf k})&=&\langle Q_i(t,{\bf k})Q_i(0,{\bf k})^\dagger\rangle \nonumber\\
  &=&-\sum_{\bf x}e^{-i{\bf k}\cdot{\bf x}}(\gamma_i\gamma_5)_{\alpha\beta}(\gamma_i\gamma_5)_{\gamma\delta}
  \left(X^{(c)(\gamma b 0)}_{\beta a {\bf x} t}({\bf -k})\right) \left(X^{(d)(\delta b 0)}_{\alpha a{\bf x}t}\right)^\ast,\label{cqita}\nonumber\\
  C^P(t,{\bf k})&=&\langle P_i(t,{\bf k})P_i(0,{\bf k})^\dagger\rangle\nonumber\\
  &=&-\sum_{\bf x}e^{-i{\bf k}\cdot{\bf x}}(\gamma_i)_{\alpha\beta}(\gamma_i)_{\gamma\delta}
  \left(X^{(c)(\gamma b 0)}_{\beta a {\bf x} t}({\bf -k})\right) \left(X^{(u)(\delta b 0)}_{\alpha a{\bf x}t}\right)^\ast,\label{cpita}
 \end{eqnarray}
 where $\bk$ is the three-momentum of a single meson. The
 quantities like $X^{(f)(\gamma b 0 )}_{\beta a \bx t}$ stands
 for the quark propagator for a particular flavor $f$. For example:
 \ba
 \label{eq:contractions_def}
 X^{(c)(\gamma b 0 )}_{\beta a \bx t}(-\bk) &=&
 \sum_\by e^{i\bk\cdot\by}
 \left[\calM^{(c)}\right]^{-1}_{\beta,a,\bx,t;\gamma,b,\by,0}\;,
 \nonumber \\
 X^{(d)(\delta b 0 )}_{\alpha a \bx t} &=&
  \sum_\by \left[\calM^{(d)}\right]^{-1}_{\alpha,a,\bx,t;\delta,b,\by,0}\;.
 \ea
 where we have assumed that the fermion matrix $\calM^{(f)}$ satisfying:
 $\calM^{(f)\dagger} =\gamma_5\calM^{(f)}\gamma_5$ for any flavor $f$.
 In the large temporal separation limit,
 the energy $E(\bk)$ of a single meson with definite three-momentum $\bk$
 can be extracted from the effective mass plateau of the
 corresponding correlation functions as usual.

 Next, we will discuss the more complicated two-particle correlation
 functions. Generally speaking, we need to evaluate a correlation
 matrix of the form:
 \be
   \label{eq:correlation}
  \langle O^{(\Gamma)\dagger}_{\alpha}(t) O^{(\Gamma^\prime)}_{\beta}(0) \rangle,
 \ee
 where $\Gamma$ and $\Gamma^\prime$ labels the irreducible
 representation of the group (i.e. $\Gamma=A_1$, $A_2$, $B_1$, $B_2$ and $E$ for
 group $D_4$ ). However, as we show below, we do not need to calculate the
 whole matrix in Eq.~(\ref{eq:correlation}). Since
 these point group representations are all real, the hermitian conjugate of
 an operator transforms in the same manner as the original operator.
 Furthermore, since the $QCD$ vacuum is invariant under any group
 transformations, it is therefore seen that, only the invariant
 sector(i.e. $A_1$ sector), decomposed from the product of two
 irreducible representations: $\Gamma\otimes\Gamma^\prime$, can make
 a non-vanishing contribution to the correlation matrix defined above.
 For the group
 $D_4$, all irreducible representations are one-dimensional except
 $E$ which is two-dimensional. Therefore, the direct products of two
 irreducible representations are particularly simple. For example, we
 easily verify that, the direct product of any two {\em different}
 one-dimensional irreducible representation cannot contain the $A_1$
 representation while the direct product of any one-dimensional
 irreducible representation with itself is exactly the $A_1$ representation.
 It is also seen that, the direct product of any one-dimensional
 irreducible representation with $E$ also contain no $A_1$ components.
 We therefore only have to consider the combination $E \otimes E$
 which reads:
 \be
  E \otimes E = A_1\oplus A_2\oplus B_1\oplus B_2
 \ee
 So two $E$ operators can yield an invariant representation $A_1$.
 But the other three ingredients cannot contribute.
 From this discussion we conclude that, we only have to consider the
 case $\Gamma = \Gamma^\prime$ and in particular, if $\Gamma = E$, we
 only have to consider something like:
 $\sum_{\alpha=1}^{2} \langle O^{(E)\dagger}_{\alpha}(t)O^{(E)}_\alpha(0) \rangle$.
 However, one should keep in mind that, different momentum modes $\bk$ do
 mix. This is what causes the scattering. The argument given above
 implies that, we only have to compute one correlation matrix in each
 channel. The size of this matrix is $n \times n$ where $n$ is the
 number of momentum modes being considered.

 For the operator $O^{(A_1)(1)}(t)$, the correlation function is as follows:
 \be
  C_{mn}^{(A_1)(1)}(t)=\langle O_m^{(A_1)(1)\dagger}(t)O_n^{(A_1)(1)}(0)\rangle,
 \ee
 where $m$ and $n$ are indices for different momentum modes. We
 notice that this correlation function has a disadvantage in practical calculations.
 The summation over the group element $R$ in the definition of the operator
 $O^{(A_1)(1)}(t)$ cannot be absorbed into the source-setting when
 solving the propagators. This drawback can be cured by using a
 slightly modified operator:
 \begin{eqnarray}
  \tilde{O}^{(A_1)(1)}(t)=
  \sum_{i=1}^3 \sum_{R',R''\in G} Q_i(t+1,-R'\circ {\bf k})
  P_i(t,R''\circ{\bf k})
 \end{eqnarray}
 The difference of this operator as compared with the original
 operator is that, this operator contains also non-zero total
 three-momentum components. To be specific, those terms with
 $R'\neq R''$, will create states with non-zero
 total three-momentum. However, if we form the correlation function:
 \begin{eqnarray}
  C_{mn}^{(A_1)(1)}(t)=\langle O_m^{(A_1)(1)\dagger}(t)\tilde{O}_n^{(A_1)(1)}(0)\rangle,
 \end{eqnarray}
 then since the sink operator has total three-momentum zero, and the
 vacuum also has total three-momentum zero, only the zero momentum
 terms in $\tilde{O}^{(A_1)(1)}(0)$ will contribute to the
 correlation function. That is to say, this will yield same
 correlation function as the original operator. However, using the
 operator $\tilde{O}^{(A_1)(1)}(0)$ at the source has a big
 advantage. It will allow us to complete the summation over
 $R'$ and $R''$ in one step. As a result, instead of solving
 for the quark propagators for each $R$, we only have to solve the
 quark propagator once, with $R$ being summed over and absorbed
 into the source definition.

 Implementing the trick mentioned above,
 the final result of correlation function according for the
 operator $O^{(A_1)(1)}(t)$ is as follows:
 \begin{eqnarray}
 \label{eq:C_A1}
  C_{mn}^{(A_1)(1)}(t)
  &=&\sum_{R\in G}\sum_{i,j=1}^3 \Big[\sum_{\bf x}e^{-i(R\circ {\bf p})\cdot {\bf x}}
  \cdot (\gamma_{i}\gamma_{5})_{\sigma\delta}\cdot(\gamma_{j}\gamma_{5})_{\alpha^\prime \rho^\prime}
  \cdot X_{\delta b {\bf x} t+1}^{(d) (\alpha^\prime a^\prime 1)}\nonumber\\
  &&\cdot  (\sum_{R^\prime\in G}X^{(c) (\rho^\prime a^\prime 1)}_{\sigma b {\bf x} t+1}(R^\prime\circ {\bf q}))^\ast\Big]
  \cdot\Big[\sum_{\bf y}e^{i(R\circ {\bf p})\cdot {\bf y}}
  \cdot (\gamma_{i})_{\rho\beta}\cdot(\gamma_{j})_{\gamma^\prime \sigma^\prime}\nonumber\\
  &&\cdot (\sum_{R^{\prime\prime}\in G}X_{\beta a {\bf y} t}^{(c) (\gamma^\prime b^\prime 0)}(R^{\prime\prime}\circ {\bf q}))
  \cdot (X^{(u)(\sigma^\prime b^\prime 0)}_{\rho a {\bf y} t})^\ast\Big],
 \end{eqnarray}
 with $m$ and $n$ being momentum mode indices with corresponding
 three-momenta $\bp$ and $\bq$, respectively; $R\in G$ being a group element of
 $D_4$; $X$ being the quark propagators with appropriate sources as defined
 in Eq.~(\ref{eq:contractions_def}).

 Another important feature that has became clear from the above
 expression is that, the light quark propagators are needed for the
 zero momentum mode {\itshape only}. Different momentum modes enters
 the heavy quark propagators. Since in the quark propagator
 inversions, light quarks cost most of the computer time, this
 separation means that we only have to solve the most time-consuming
 part of the propagator, which is the light quark propagator, for
 vanishing three-momentum. Heavy quark propagators are needed for
 each momentum mode, however, it is not costly since the quark mass is heavy.

 Two tricks mentioned above, one being the reduction to
 non-degenerate momentum modes, i.e. different three-momentum modes that
 are related by $D_4$ group transformations requires only one quark propagator
 inversion; the other being solving light quark
 propagators for zero momentum only, have offered us enormous amount of
 acceleration in the calculation. For example, taking highest
 three-momentum up to (1,1,0), the number of different three-momenta is
 21. But if we only count the non-degenerate momentum modes, it is
 only 6, gaining more than a factor of 3. Now that we only have to
 compute the zero momentum mode for the light quark, this gives again
 a factor of almost 6 (neglecting computer time for heavy quark inversions).
 Altogether, we expect a factor of about 15-20 gaining in the speed
 of the simulation. Note also that, these tricks are generally
 applicable for any type of calculations involving mesons with one
 heavy and one light quark.

 \section{Simulation details}
 \label{sec:simulation_details}
 \subsection{Lattice actions and simulation parameters}
 The gauge action use in this study is the tadpole improved gauge
 action on anisotropic lattices:~\cite{colin99,chuan01:gluea,chen_liu06:glueball}
 \begin{eqnarray}
  S&=&-\beta\sum_{i>j}\left[{5\over9}{ {\rm Tr}P_{ij}\over \xi u_s^4}-{1\over 36}{ {\rm Tr}R_{ij}\over\xi u_s^6}
  -{1\over 36}{ {\rm Tr}R_{ji}\over \xi u_s^6}\right]\nonumber\\
  &&-\beta\sum_i\left[{4\over 9}{\xi {\rm Tr}P_{01}\over u_s^2}-{1\over 36}{\xi{\rm Tr}R_{i0}\over u_s^4}\right],
  \label{action_u_final}
 \end{eqnarray}
 where $P_{ij}$ is the usual spatial plaquette variables and $R_{ij}$
 is the $2\times 1$ spatial Wilson loop on the lattice. The parameter
 $u_s$, which we take to be the 4-th root of the average spatial
 plaquette value, incorporates the so-called tadpole improvement and
 $\xi$ designates the aspect ratio of the anisotropic lattice.
 The parameter $\beta$ is related to the bare gauge coupling which
 controls the spatial lattice spacing $a_s$ in physical units.

 The fermion action used in this study is the tadpole improved clover
 Wilson action on anisotropic lattice whose fermion matrix
 is~\cite{chuan01:tune,chuan06:tune_v}:
 $\calM_{xy}=\delta_{xy}\sigma+A_{xy}$ with:
 \begin{eqnarray}
  {A}_{xy}=\delta_{xy}[{1\over2\kappa_{max}}+\rho_t\sum_{i=1}^3\sigma_{0i}{F}_{0i}
  +\rho_s(\sigma_{12}{F}_{12}+\sigma_{23}{F}_{23}+\sigma_{31}{F}_{31})]\nonumber\\
  -\sum_\mu \eta_\mu[(1-\gamma_\mu)U_\mu(x)\delta_{x+\mu,y}+(1+\gamma_\mu)U_\mu^\dagger(x-\mu)\delta_{x-\mu,y}]
 \end{eqnarray}
 where the coefficients are given by:
 \begin{eqnarray}
 &&  \eta_i={\nu\over 2u_s},~ \eta_0={\xi\over 2},~ \sigma={1\over2\kappa}-{1\over2 \kappa_{max}},\nonumber\\
 &&  \rho_t={\nu(1+\xi)\over 4u_s^2},~  \rho_s={\nu\over 2u_s^4}.
 \end{eqnarray}

 Quenched gauge field configurations are generated using the conventional
 Cabbibo-Mariani pseudo-heat bath algorithm with over-relaxation.
 Quark propagators are obtained using the so-called  Multi-mass
 Minimal Residual ($\rm M^3R$) algorithm, which can yield the
 propagators with different quark masses at one inversion~\cite{glaessner96:multimass}.
 Dirichlet boundary conditions are used in the temporal direction
 for the fermion fields. Error estimates are made using the conventional jack-knife method
 for all quantities.

 All the relevant simulation parameters are summarized
 in Table~\ref{tab:parameter}. Among these parameters, the
 spatial lattice spacing $a_s$ in physical units corresponding each $\beta$ has been
 obtained in Ref.~\cite{chuan06:tune_beta}, together with the corresponding
 parameter $u_s$; the parameter $\nu_c$ and $\nu_{ud}$
 has been obtained in Ref.\cite{chuan06:tune_v}. Finally, the
 largest hopping parameter,
 $\kappa_{max}$ is chosen such that no exceptional gauge field configurations
 are encountered. This corresponds to the lightest pion mass of about $500$-$600$MeV
 in our simulation.
 Note that our calculation is performed using three set of lattices whose physical volume are about
 the same but with different lattice spacings. The physical size $L$ in the shorter
 spatial direction is about $1.6$fm and, for the lightest pion mass $m_\pi$ in our simulation, this gives
 $m_\pi L\simeq 5$ and therefore finite volume corrections which might spoil the validity of
 L\"uscher's formulae are expected to be small.
 Three different lattice spacings allow us to extrapolate our final
 results to the continuum limit.
 In order to find the physical point for the charm quark and to
 facilitate chiral extrapolation, six nearby values are taken
 for both $\kappa^{ud}$ and $\kappa^c$ around $\kappa^{ud}_{max}$
 and $\kappa^c_{max}$, respectively.

\begin{table}
\centering \caption{Simulation parameters in this study. All
lattices have the same aspect ratio: $\xi=5$.} \label{tab:parameter}
\begin{tabular}{|c|c|c|c|}
\hline
    &$\beta=2.5$    &$\beta=2.8$    &$\beta=3.2$\\
\hline \hline
$N_{\rm conf}$  &700                &500        &200\\
\hline
$u_s^4$     &0.4236             &0.4630     &0.50679\\
\hline
$\nu_c$     &0.732              &0.79       &0.89\\
\hline
$\nu_{ud}$  &0.9305             &0.96       &1.0\\
\hline
$a_s(fm)$   &0.2037             &0.1432     &0.0946\\
\hline
$lattice$   &$8\times8 \times 12\times 40$  &$12\times 12\times 20\times 64$    &$16\times 16\times 24\times 80$\\
\hline
$\kappa^c_{max}$    &0.0577&0.0598&0.0595\\
\hline
$\kappa^{ud}_{max}$ &0.0613&0.0611&0.0606\\
\hline
\end{tabular}
\end{table}

 \subsection{Checking single-particle spectrum and dispersion relations}
 \label{subsec:dispersion}

 After inserting a complete set of states, any single-particle  correlation
 function can be written in the following form:
 \begin{eqnarray}
  C(t)=\langle O(t) O^\dagger(0)\rangle = \sum_n C_n e^{-E_n t}
 \end{eqnarray}
 We may define the effective mass function $M_{\rm eff.}(t)$ as follows:
 \begin{eqnarray}
  M_{\rm eff.}(t)=\log {C(t)\over C(t+1)},
 \end{eqnarray}
 which in the large temporal limit is dominated by a constant
 which is the mass of the lowest energy gap. Therefore, fitting
 the effective mass function to a constant in a plateau region
 yields the lowest energy gap.

 In this study, we have calculated single-particle correlation
 functions of several mesons: $D^*$, $D_1$, $\eta_c$, $J/\psi$ and
 $\pi$. $D^*$ and $D_1$ are the main objects that we want to
 study, other particles are for heavy quark mass interpolation and
 light quark mass extrapolation (chiral extrapolation).
 As we explained in the previous section, single particle
 correlation functions for $D^*$, $D_1$, $\eta_c$ and $J/\Psi$ are measured
 for both zero and non-zero three-momenta. For the pion, only zero-momentum
 correlation is measured. Since we have used anisotropic lattices with
 enhanced temporal resolutions, we obtain descent plateaus for single-particle
 correlation functions.

\begin{figure}[h]
\centering
\includegraphics[scale=0.38]{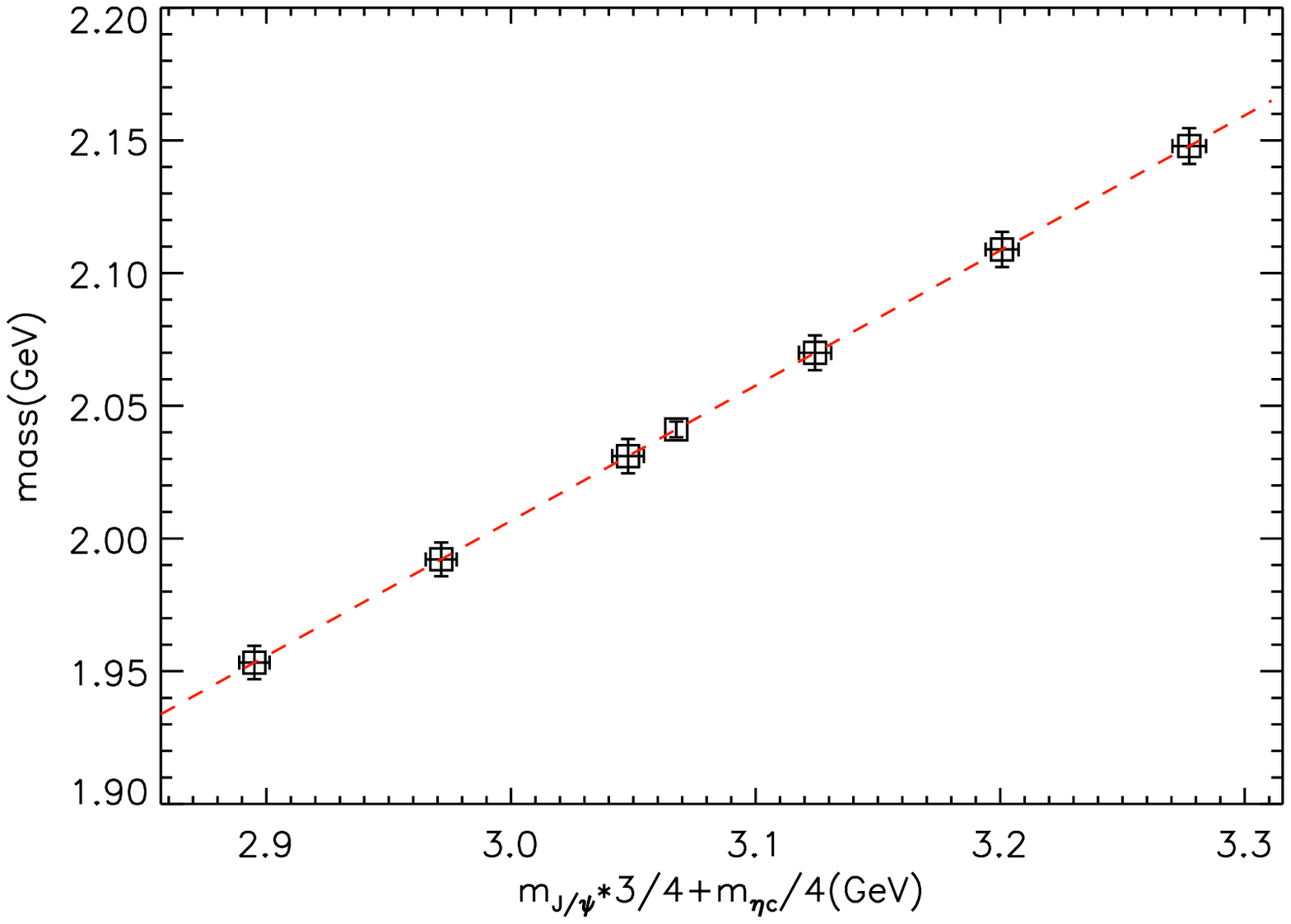}
\includegraphics[scale=0.38]{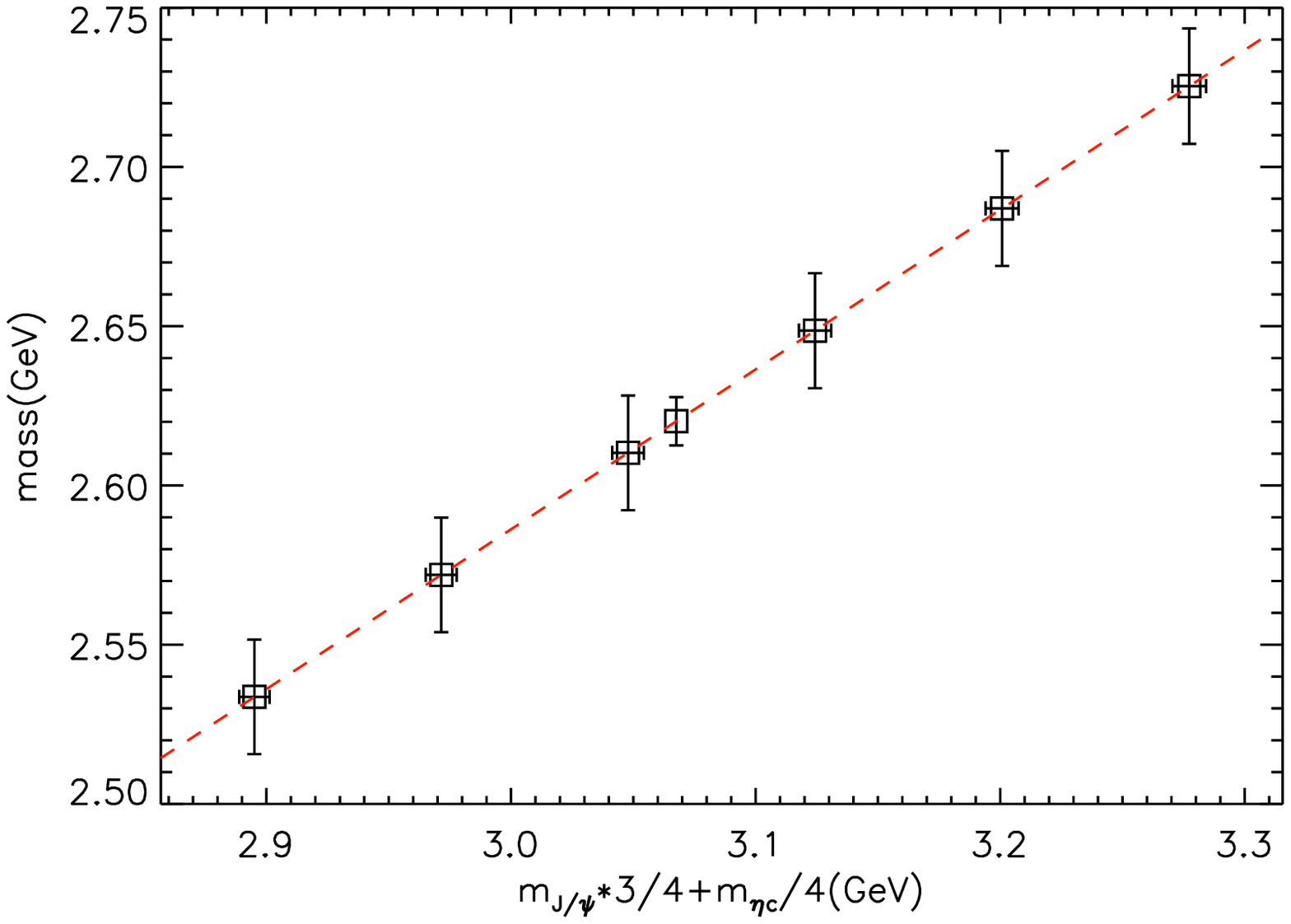}
\includegraphics[scale=0.38]{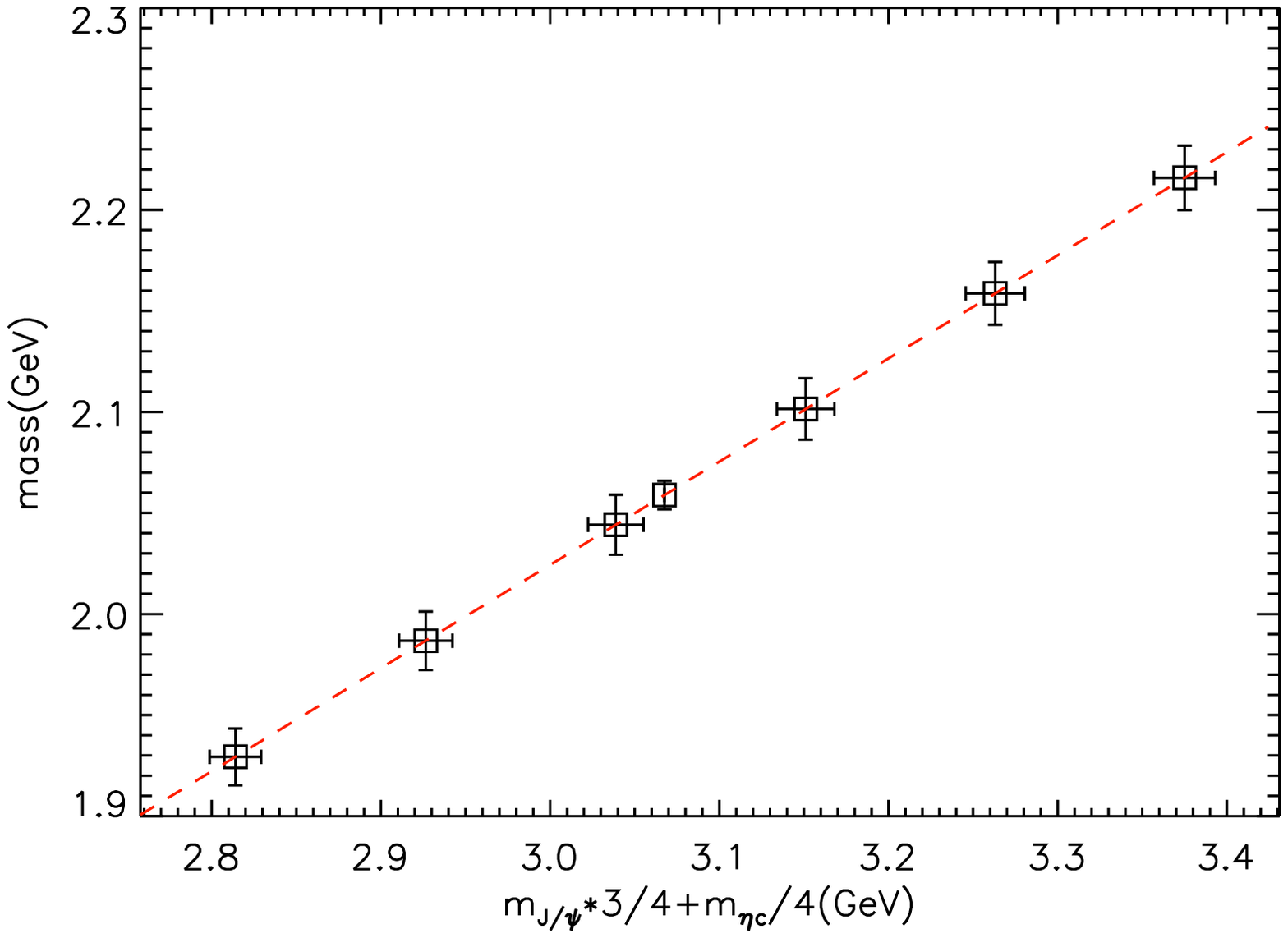}
\includegraphics[scale=0.38]{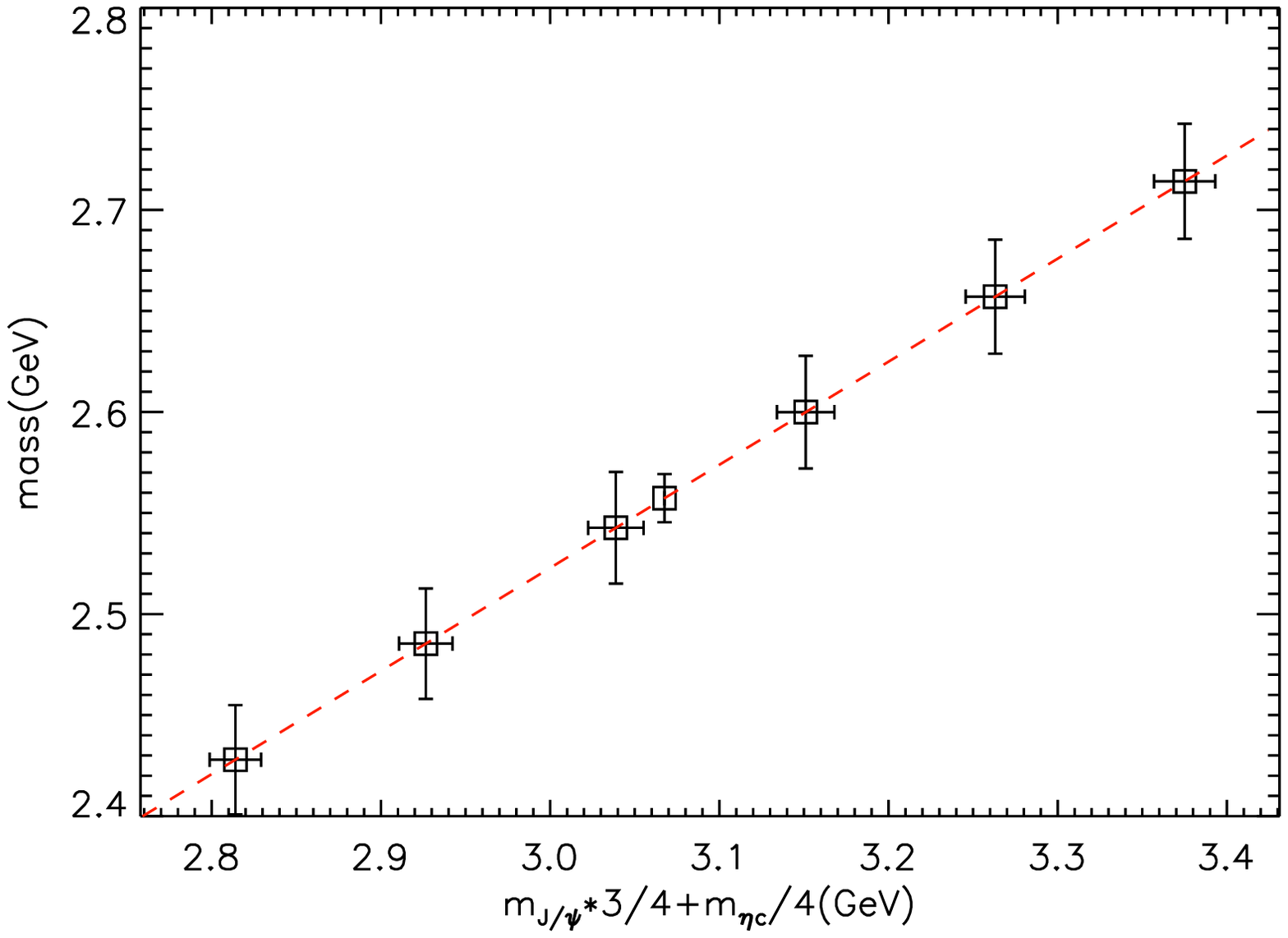}
\includegraphics[scale=0.38]{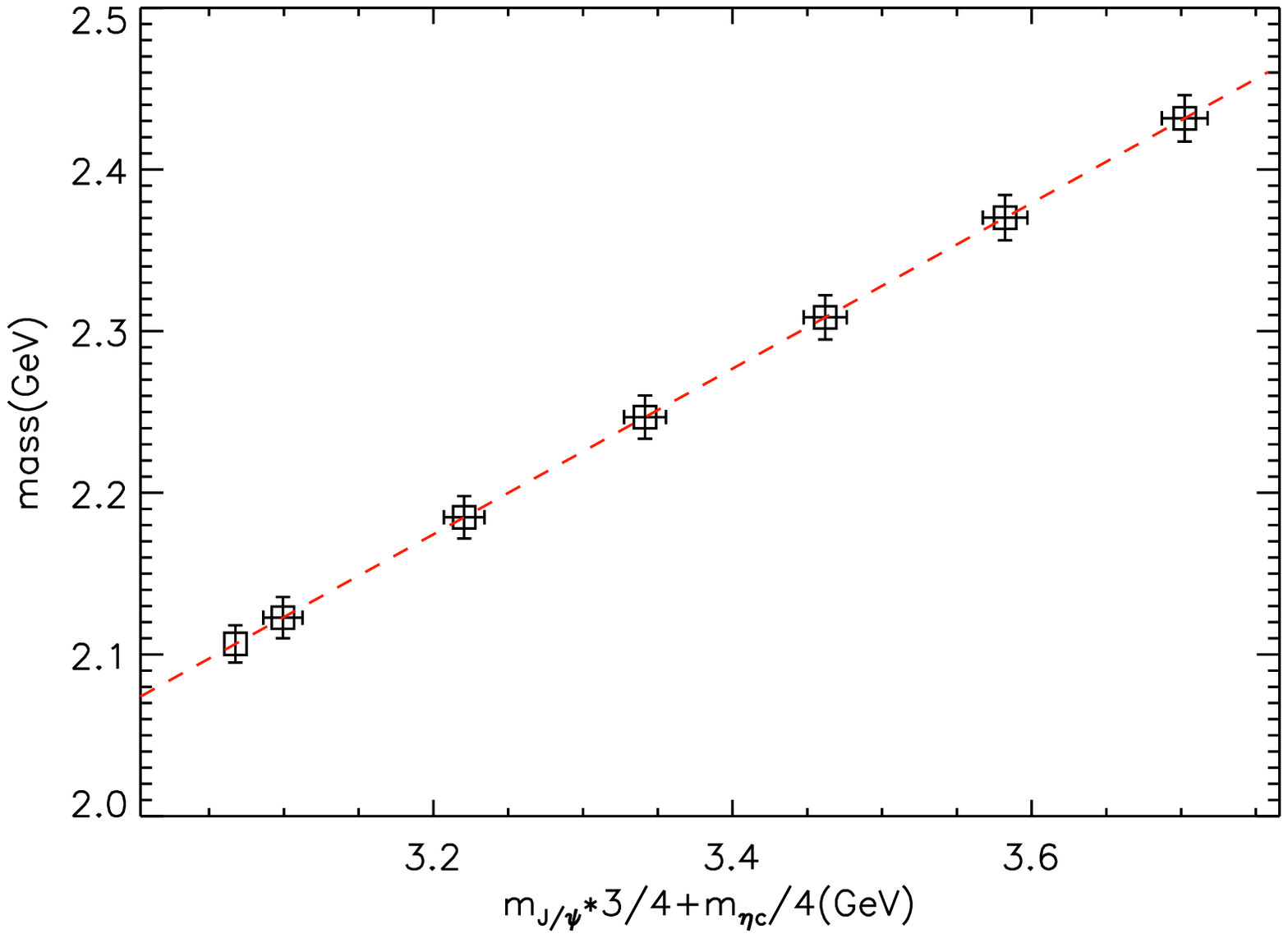}
\includegraphics[scale=0.38]{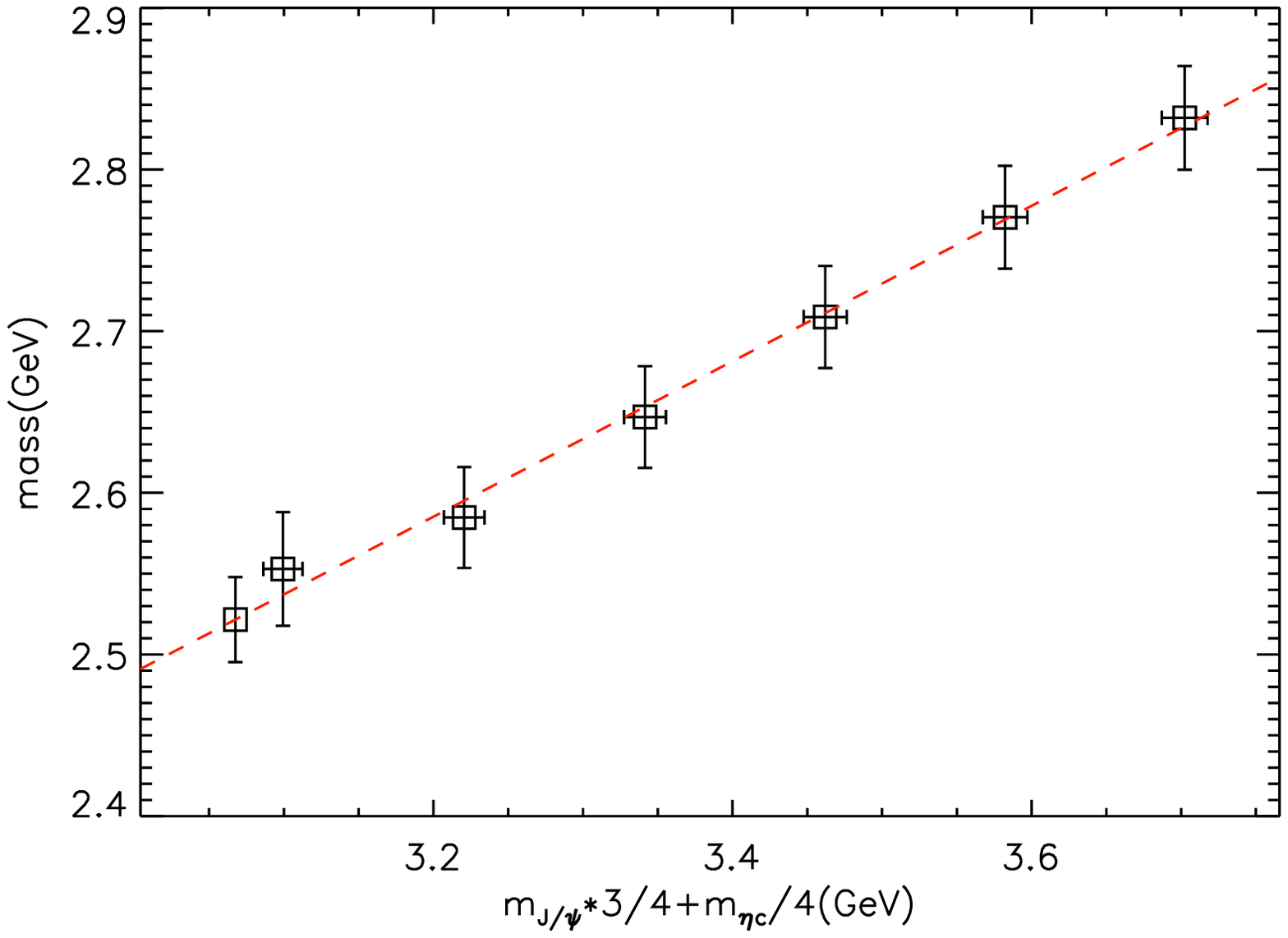}
\caption{Heavy quark mass interpolation for $m_{D^\ast}$ and
$m_{D_1}$, from top to bottom: $\beta=2.5$, $2.8$ and $3.2$.}
\label{fig:heavy}
\end{figure}

 Let us first examine the mass of the $D^*$ and $D_1$ mesons.
 After obtaining the mass values for them under various quark mass
 parameters $(\kappa^{ud},\kappa^c)$, the mass of the $\eta_c$ and $J/\Psi$ are used to fix
 the physical charm quark hopping parameter $\kappa^c$.
 For this purpose, we demand that the
 combination ${1\over 4}m_{\eta_c}+{3\over 4}m_{J/\psi}$ (spin-averaged
 charmonium mass) reproduces its physical value with the scale set by the lattice spacing.
 This procedure is shown in Fig.~\ref{fig:heavy} where the mass values of
 $D^*$ and $D_1$ are shown as functions of the charm quark mass
 parameter $\kappa^c$.
 The left panel is for $D^\ast$, and the right one is for $D_1$. In
 each figure, the 6 data points with both $x$ and $y$ error-bars are original
 data for the $D$ meson mass. The interpolated point with only $y$ error-bar is the result of
 charm quark mass interpolation. Such interpolations are performed
 for each light quark mass parameter $\kappa^{ud}$ although in Fig.~\ref{fig:heavy}
 one particular light quark mass parameter $\kappa^{ud}$ is shown.

\begin{figure}[h]
\centering
\includegraphics[scale=0.38]{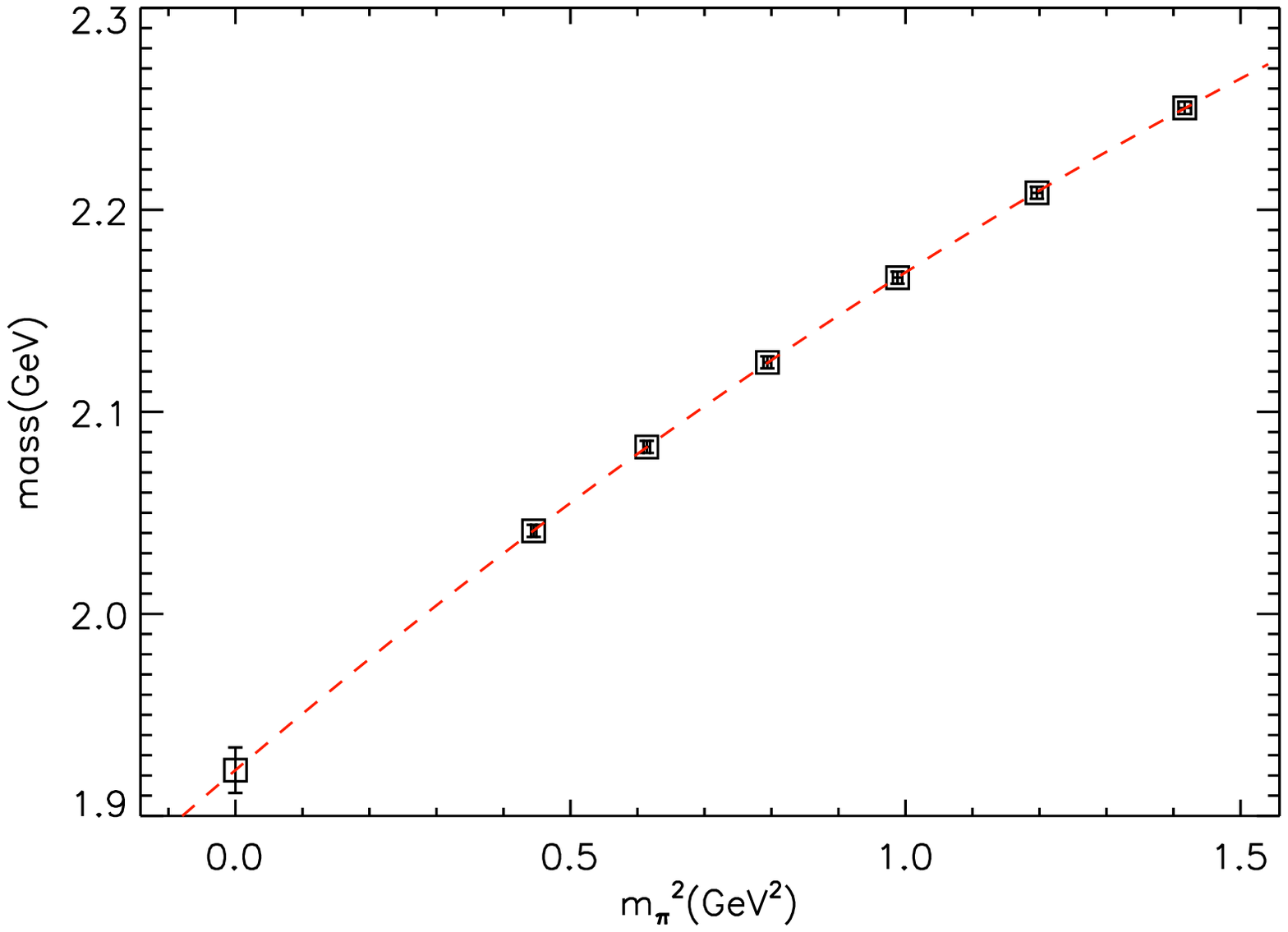}
\includegraphics[scale=0.38]{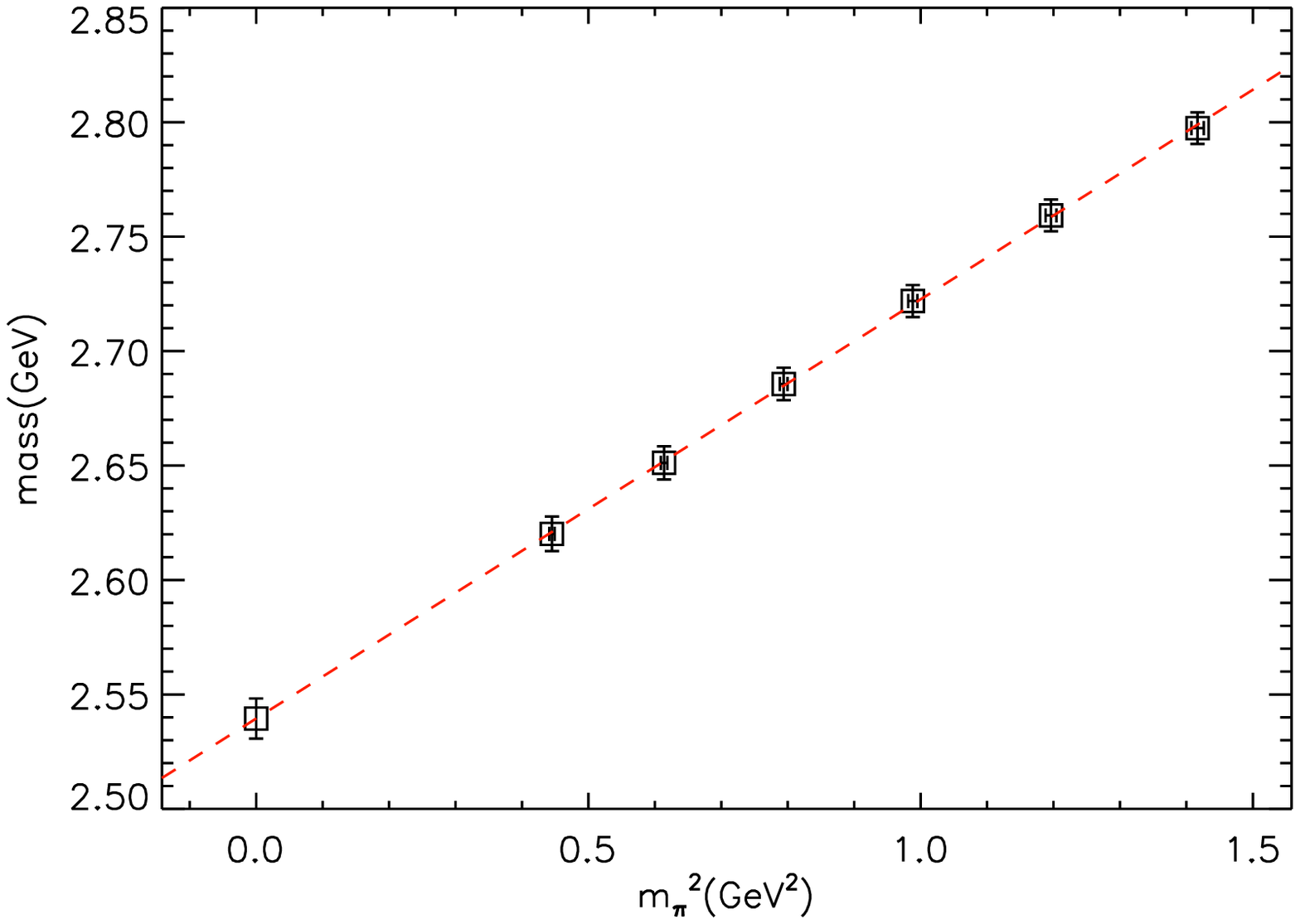}
\includegraphics[scale=0.38]{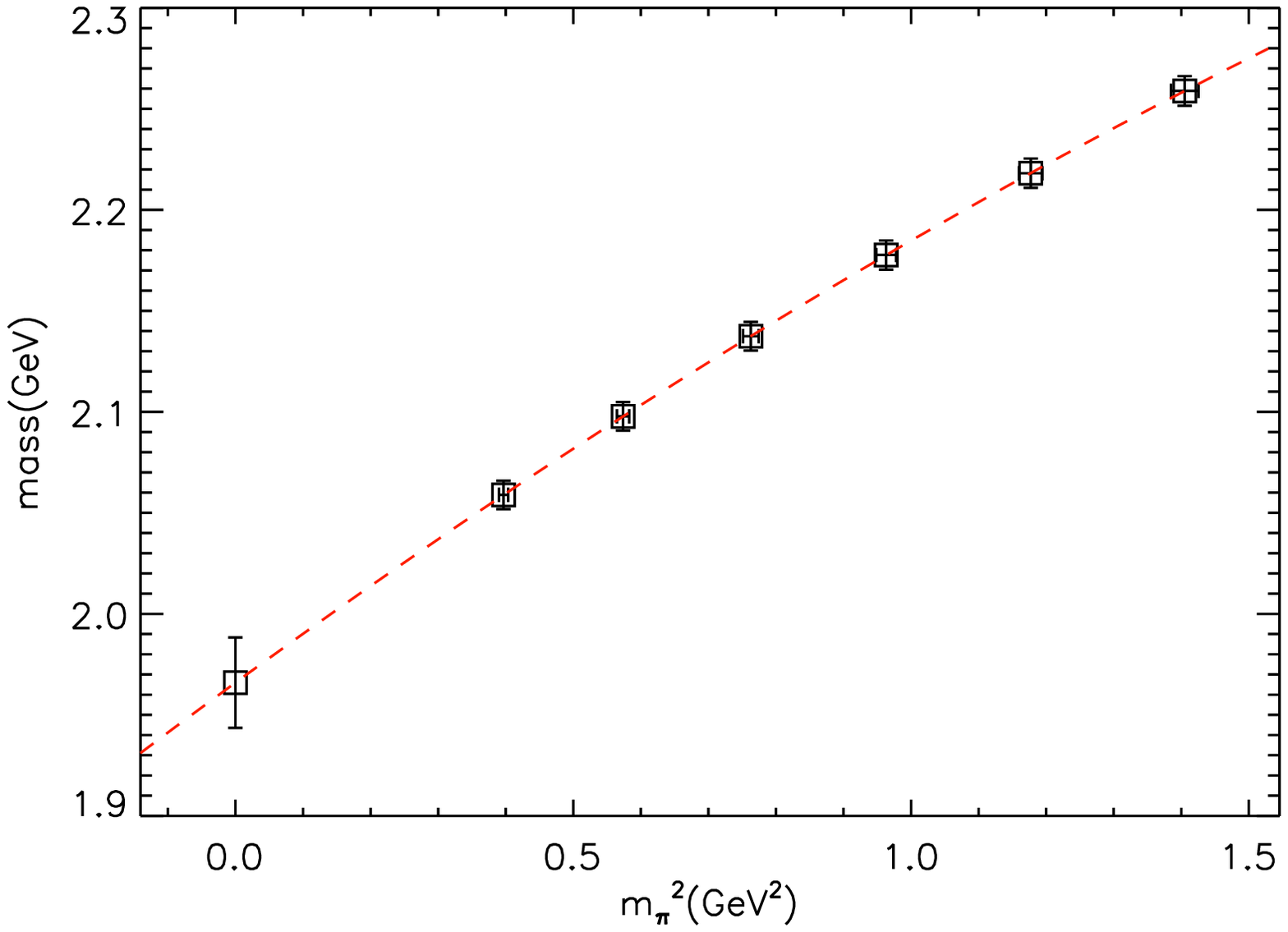}
\includegraphics[scale=0.38]{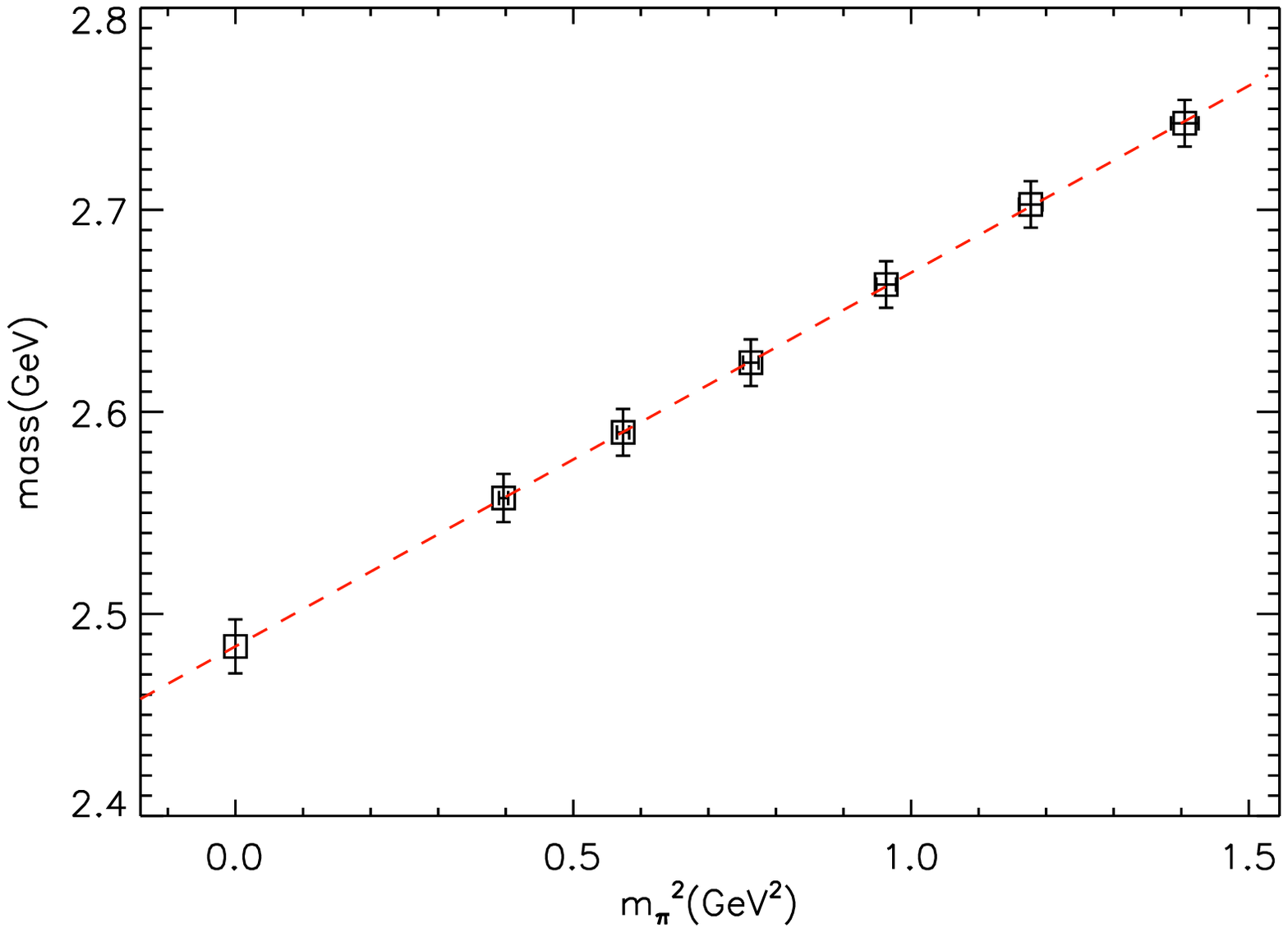}
\includegraphics[scale=0.38]{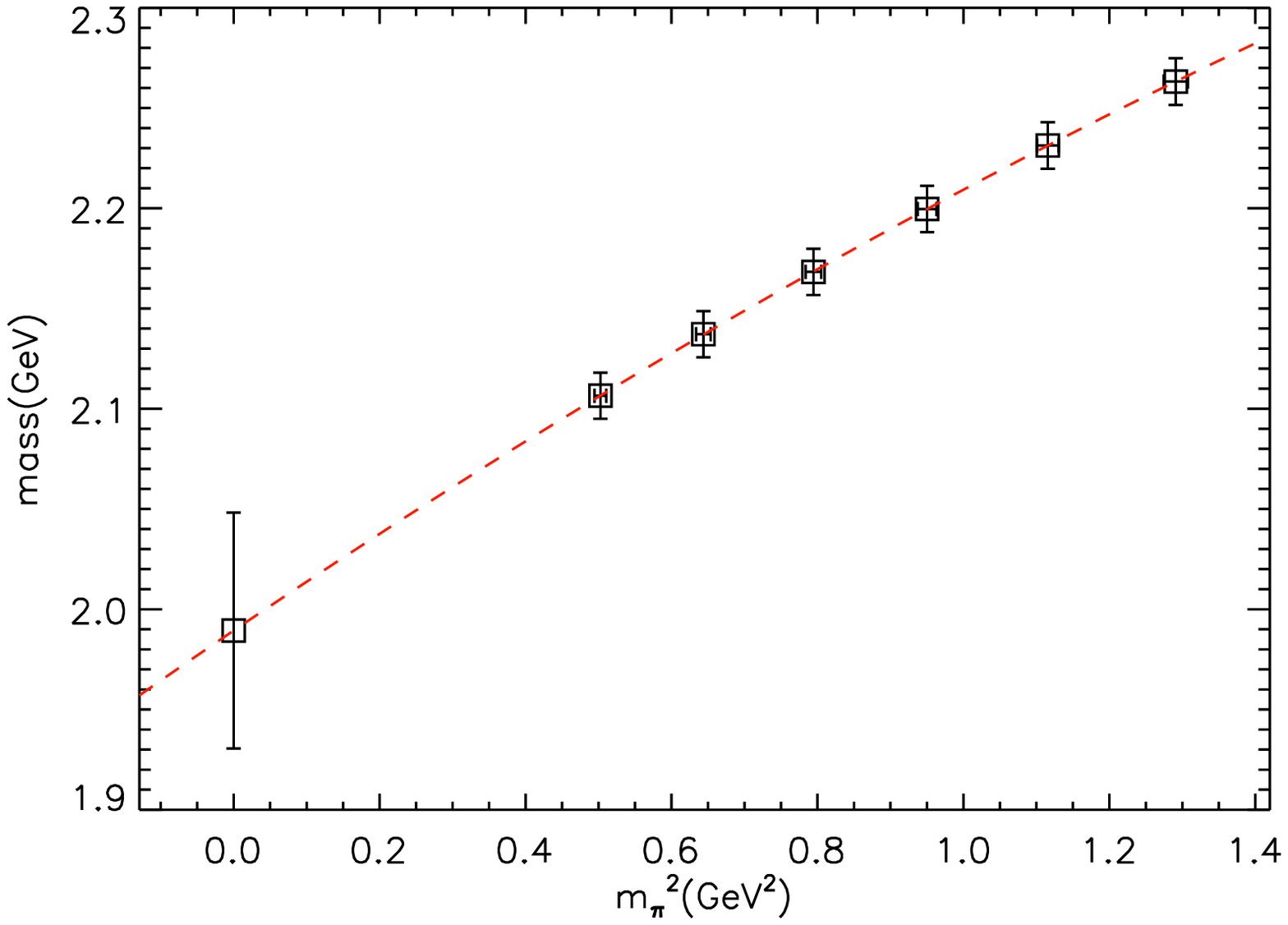}
\includegraphics[scale=0.38]{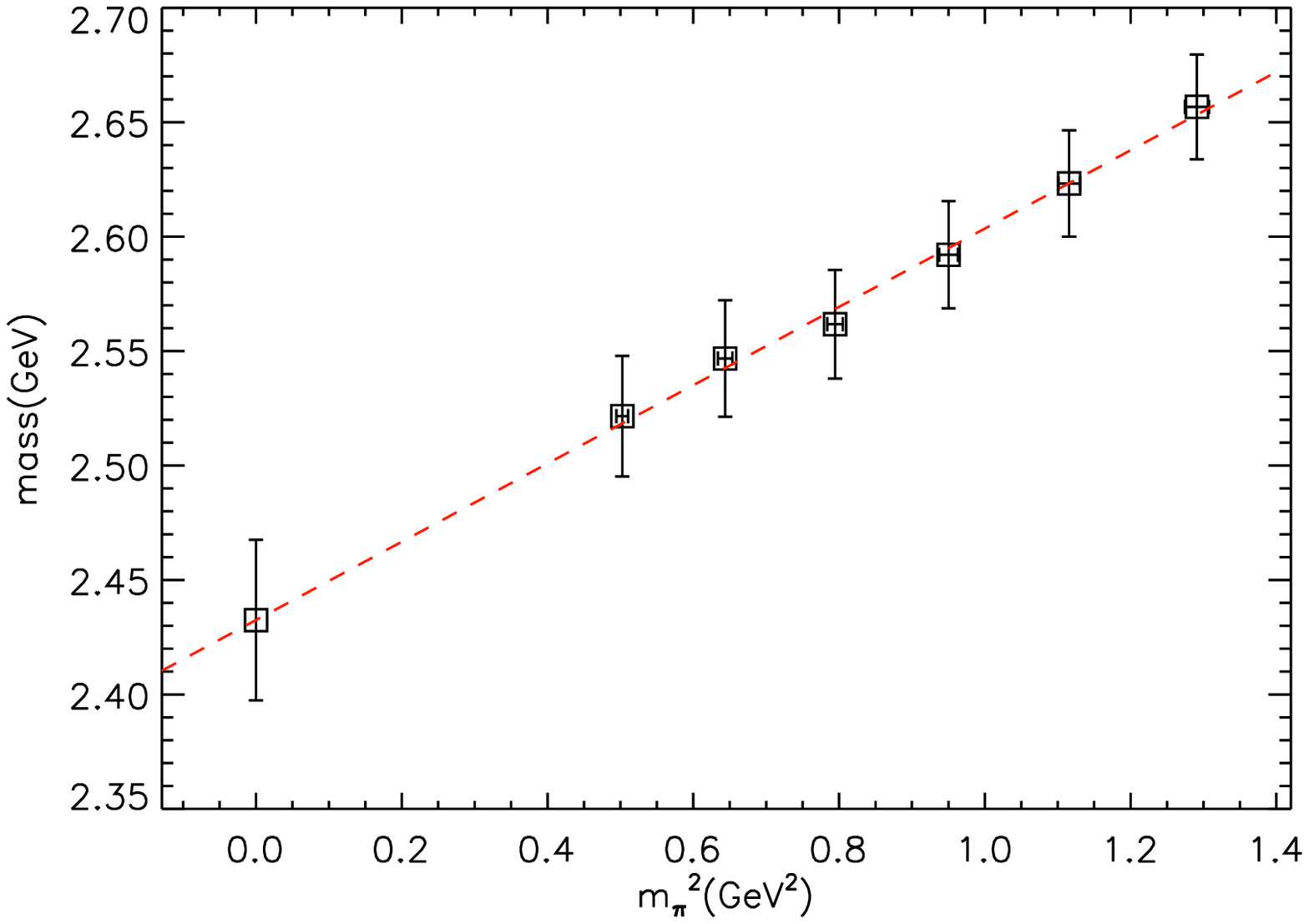}
\caption{Chiral extrapolations for $m_{D^\ast}$ and $m_{D_1}$, from
top to bottom: $\beta=2.5$, $2.8$ and $3.2$.} \label{fig:light}
\end{figure}

 After heavy quark mass interpolation for each light quark mass
 parameter $\kappa^{ud}$, the mass of the $D^*$ and $D_1$ mesons
 ($m_{D^\ast}$ and $m_{D_1}$, respectively) are extrapolated
 versus $m_\pi^2$ towards the chiral limit $m^2_\pi=0$. Since our simulation
 points are still far from the true chiral region, we adopted either linear or
 quadratic functions in $m^2_\pi$ according to the behavior of the data.
 This procedure is shown in Fig.~\ref{fig:light}.

\begin{figure}[h]
\centering
\includegraphics[scale=0.6]{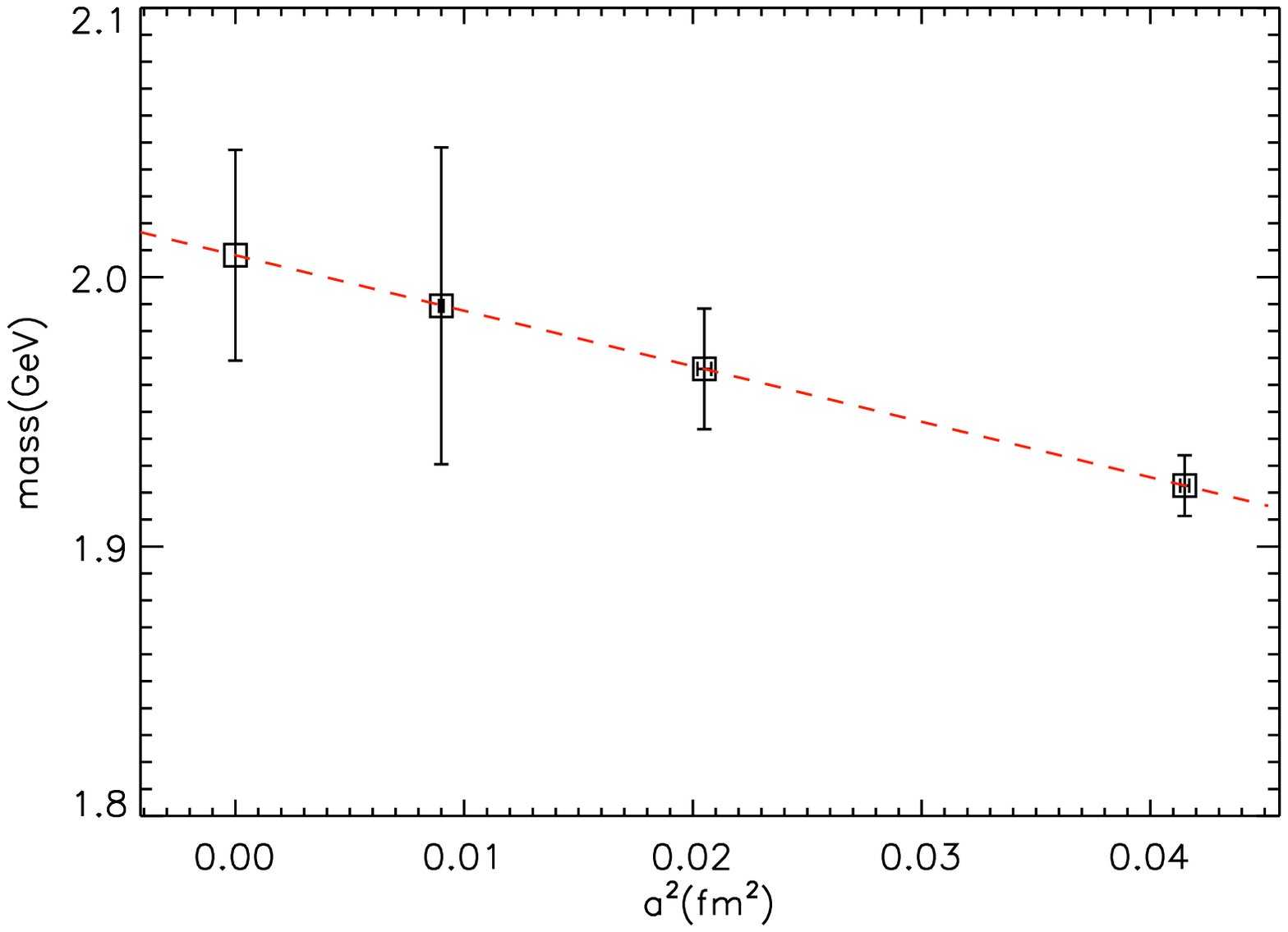}
\includegraphics[scale=0.6]{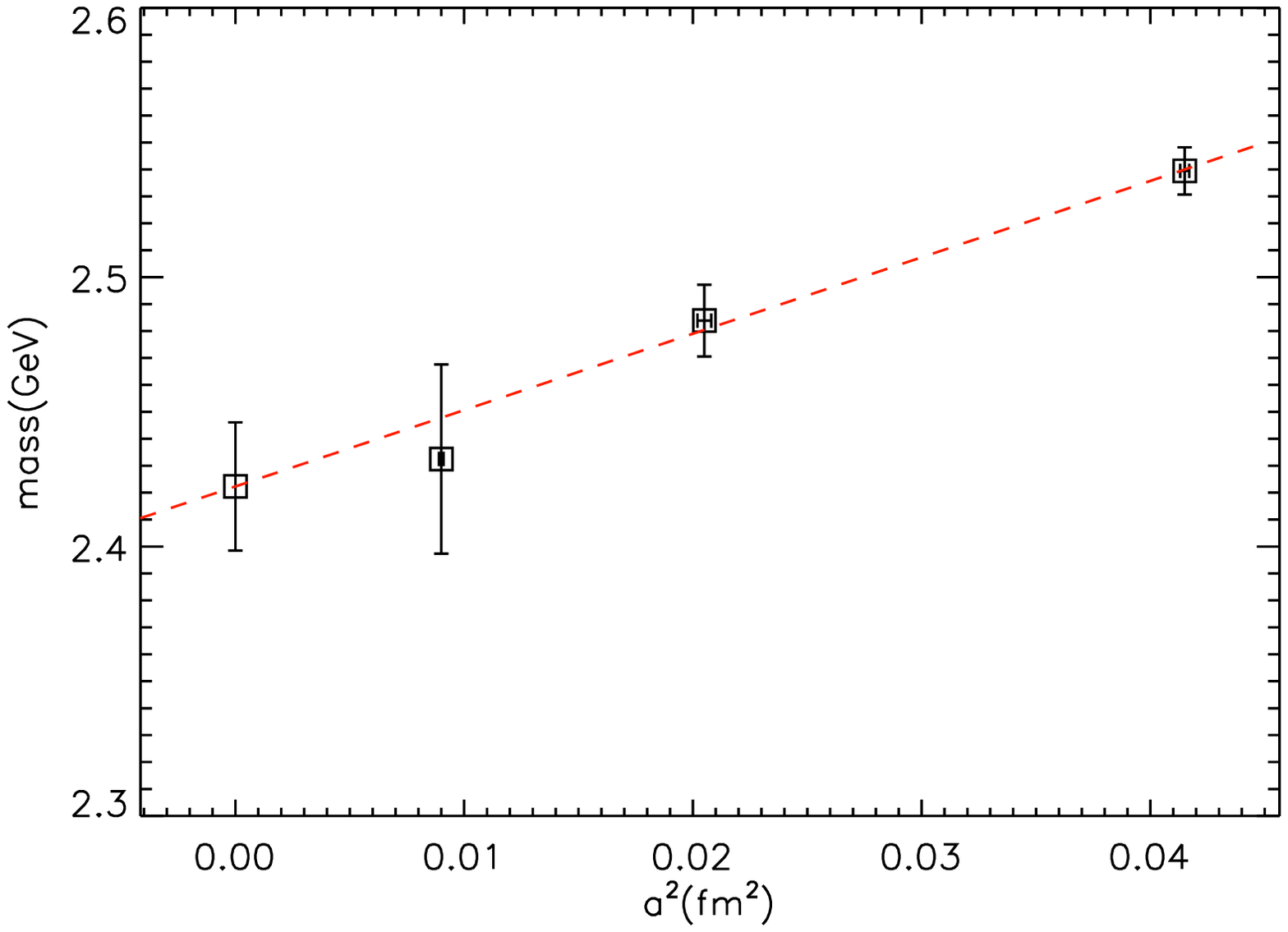}
\caption{Continuum limit extrapolations of $m_{D^\ast}$ and
$m_{D_1}$.} \label{fig:spacing}
\end{figure}

 After all these interpolations and extrapolations, we obtain the mass of
 $D^\ast$ and $D_1$ for each lattice spacing. Finally, a continuum extrapolation
 can be carried out for  $m_{D^\ast}$ and $m_{D_1}$ with
 linear function in $a^2_s$ since
 we are using an improved fermion action.
 This is illustrated in Fig.~\ref{fig:spacing}. Our final results for
 the mass of the $D$ mesons are:
 \begin{eqnarray}
  m_{D^\ast}=2.008\pm 0.039 {\rm GeV}, ~m_{D_1}=2.422\pm 0.024 {\rm GeV},
 \end{eqnarray}
 from which we can see that our results are compatible with
 experimental result within error bars.

\begin{figure}[h]
\centering
\includegraphics[scale=0.35]{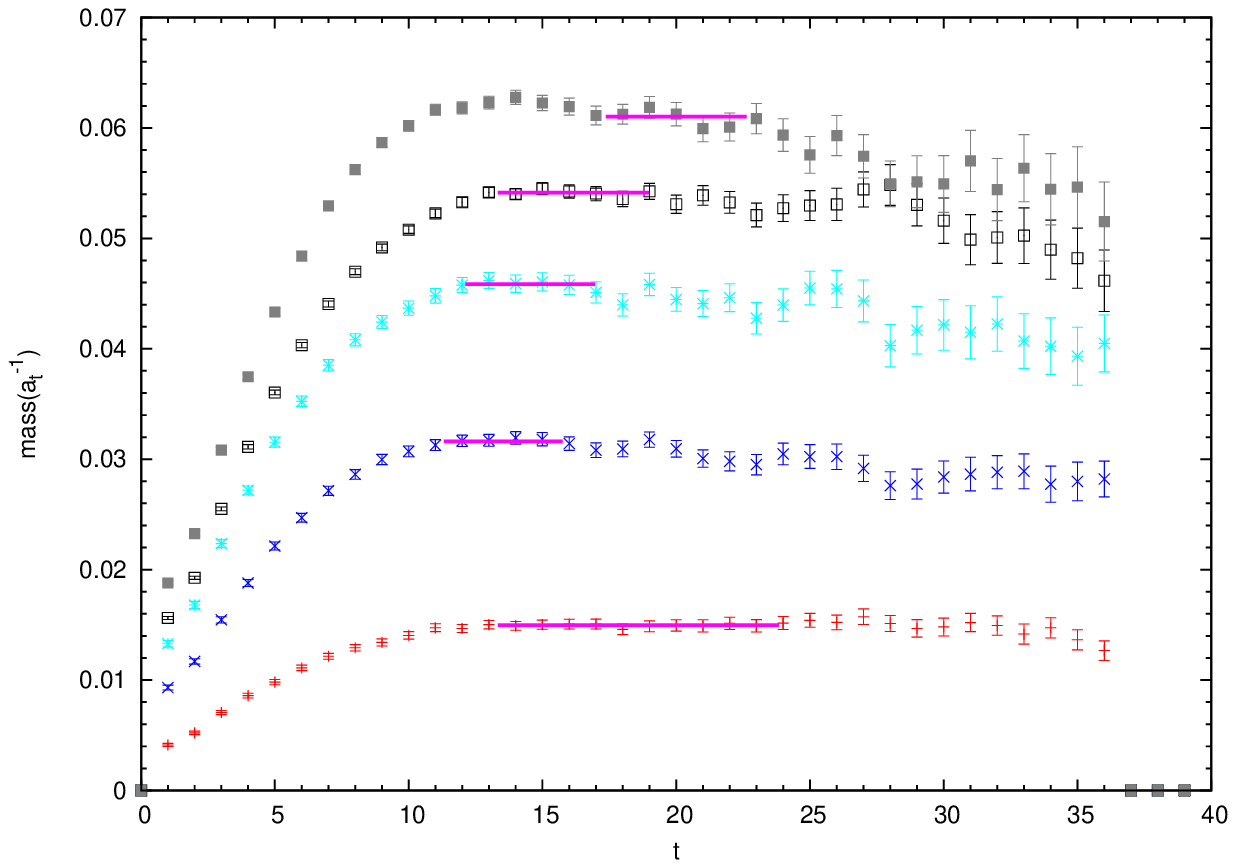}
\includegraphics[scale=0.35]{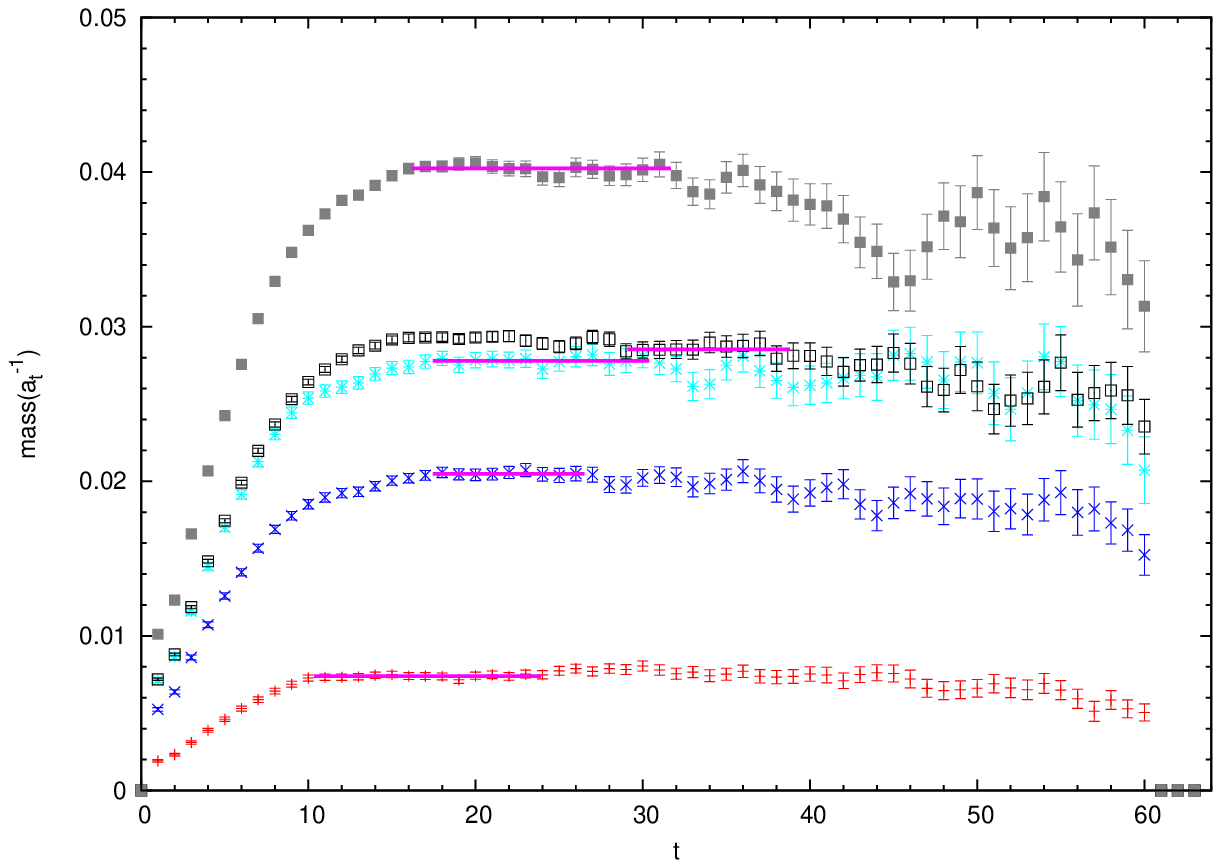}
\includegraphics[scale=0.35]{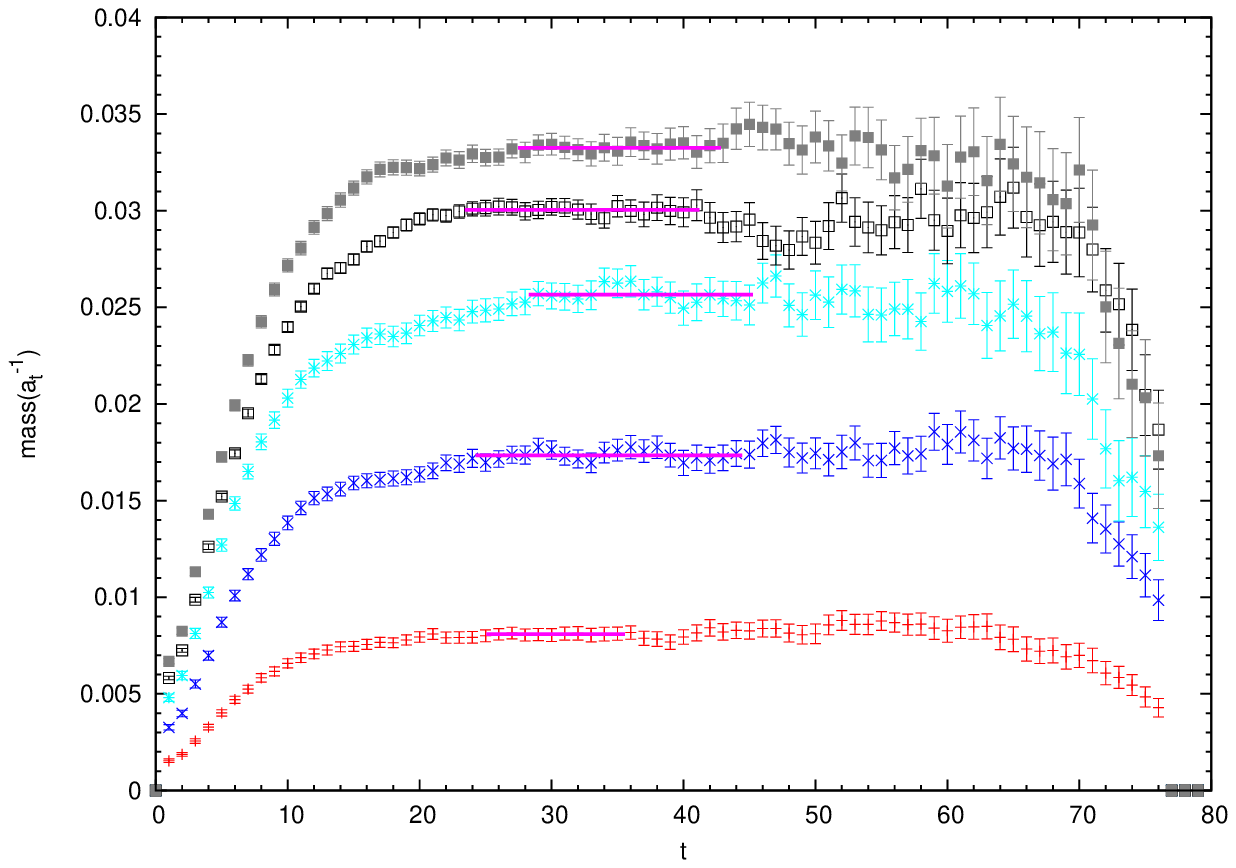}
\includegraphics[scale=0.35]{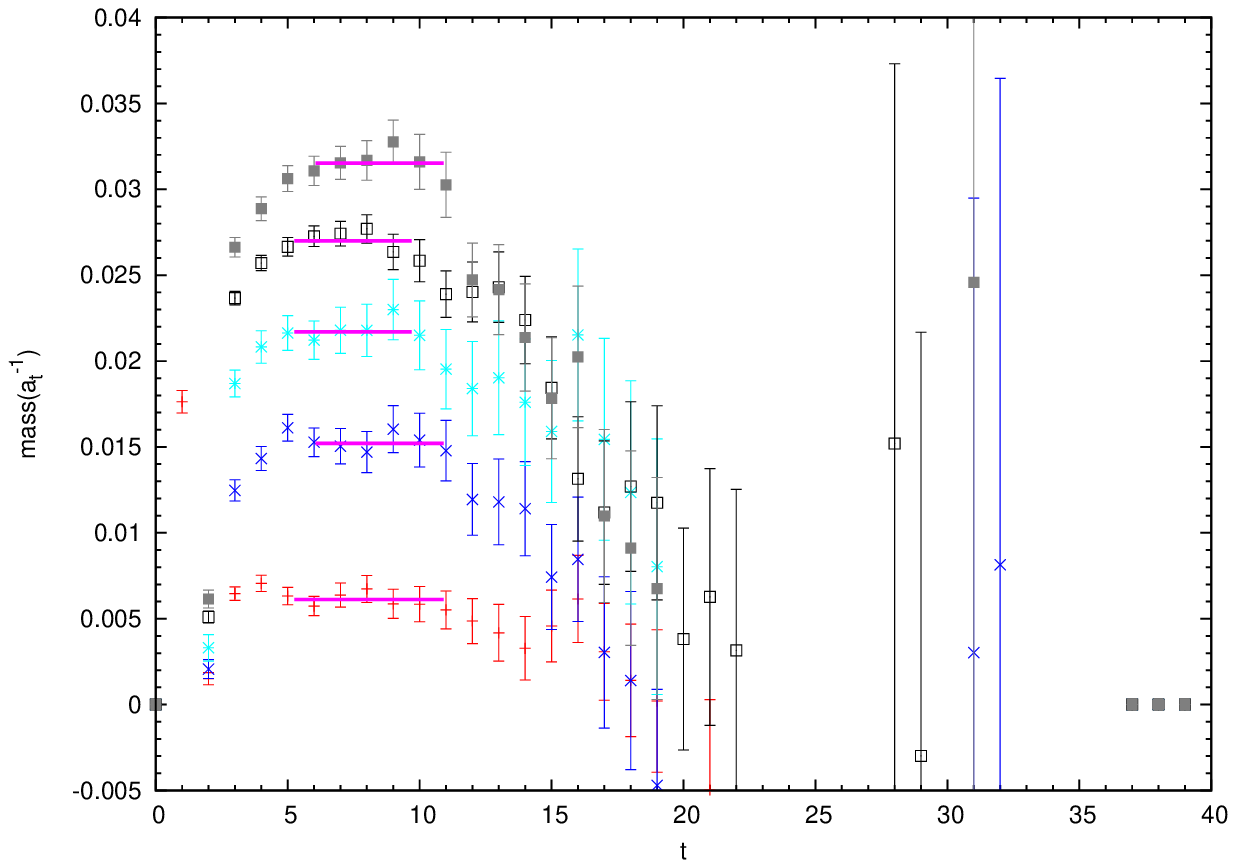}
\includegraphics[scale=0.35]{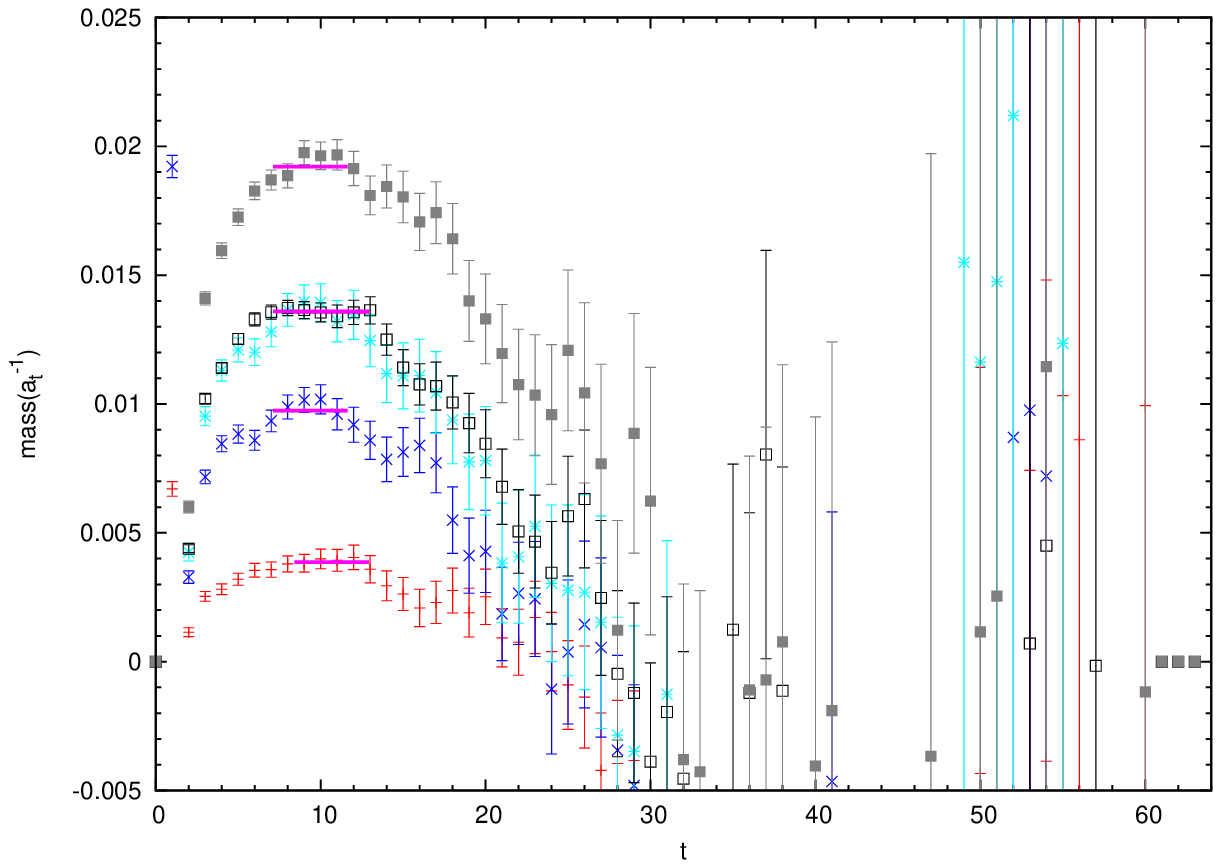}
\includegraphics[scale=0.35]{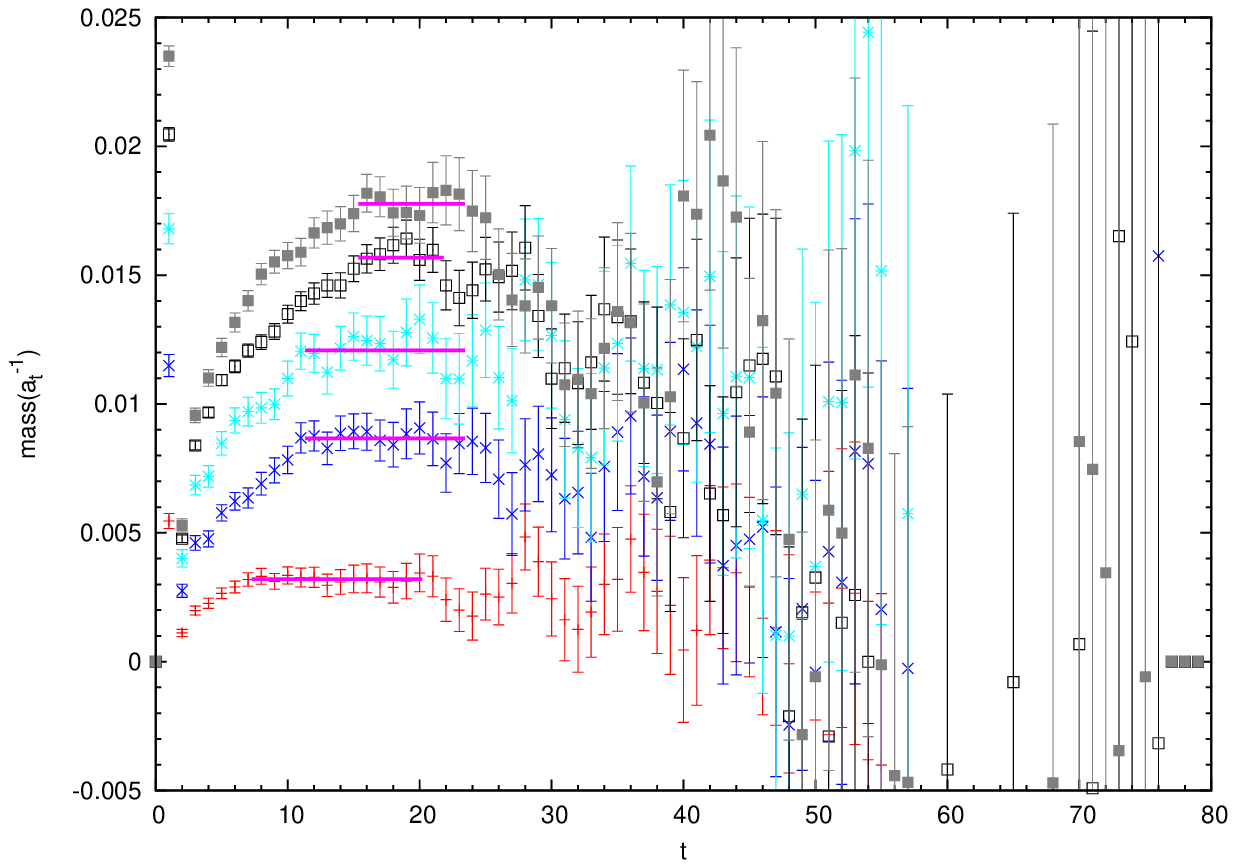}
\includegraphics[scale=0.35]{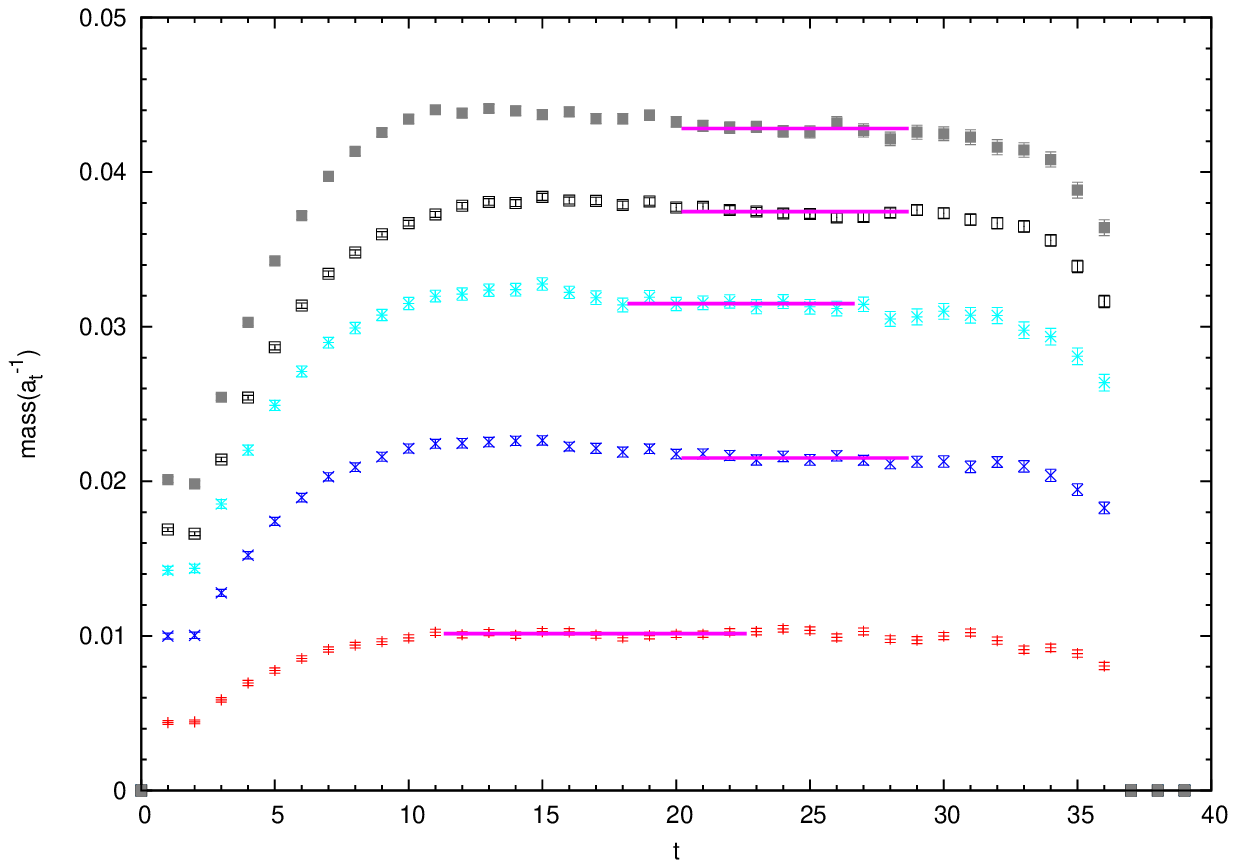}
\includegraphics[scale=0.35]{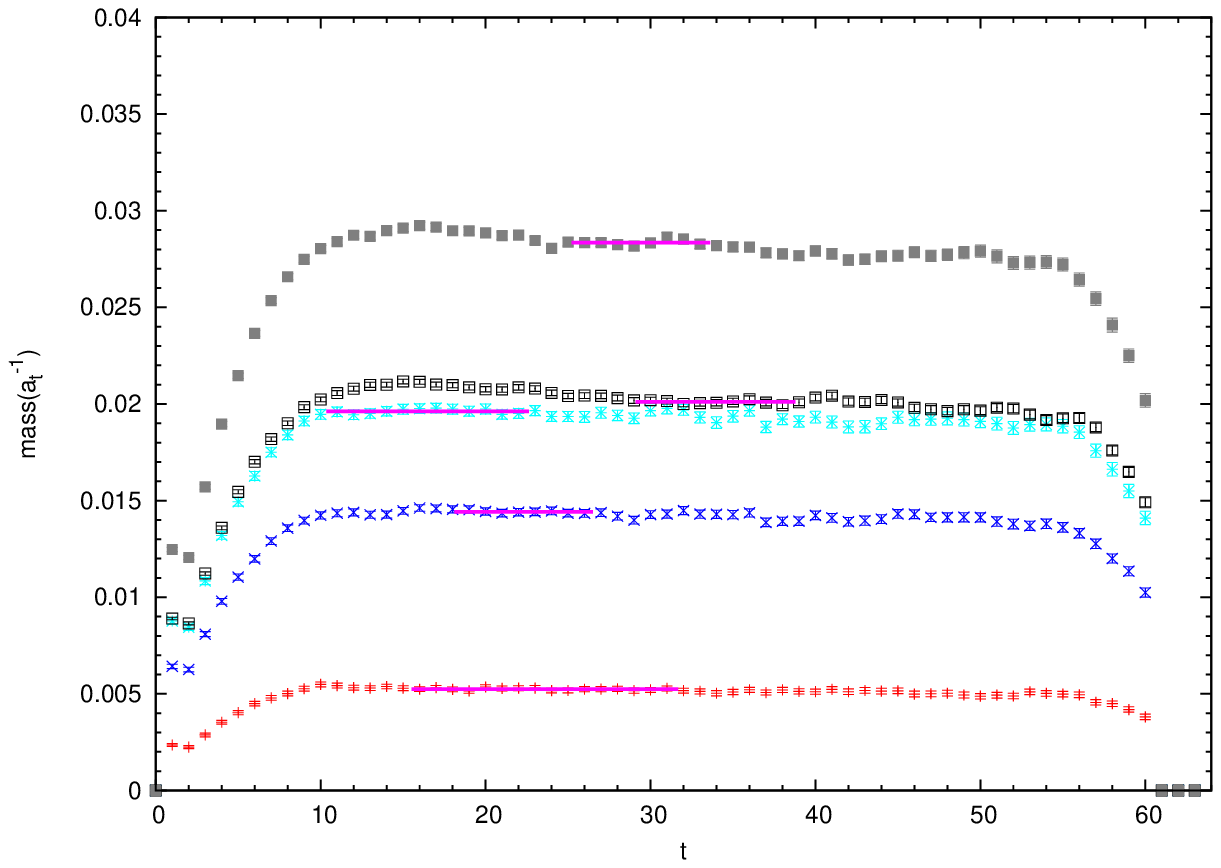}
\includegraphics[scale=0.35]{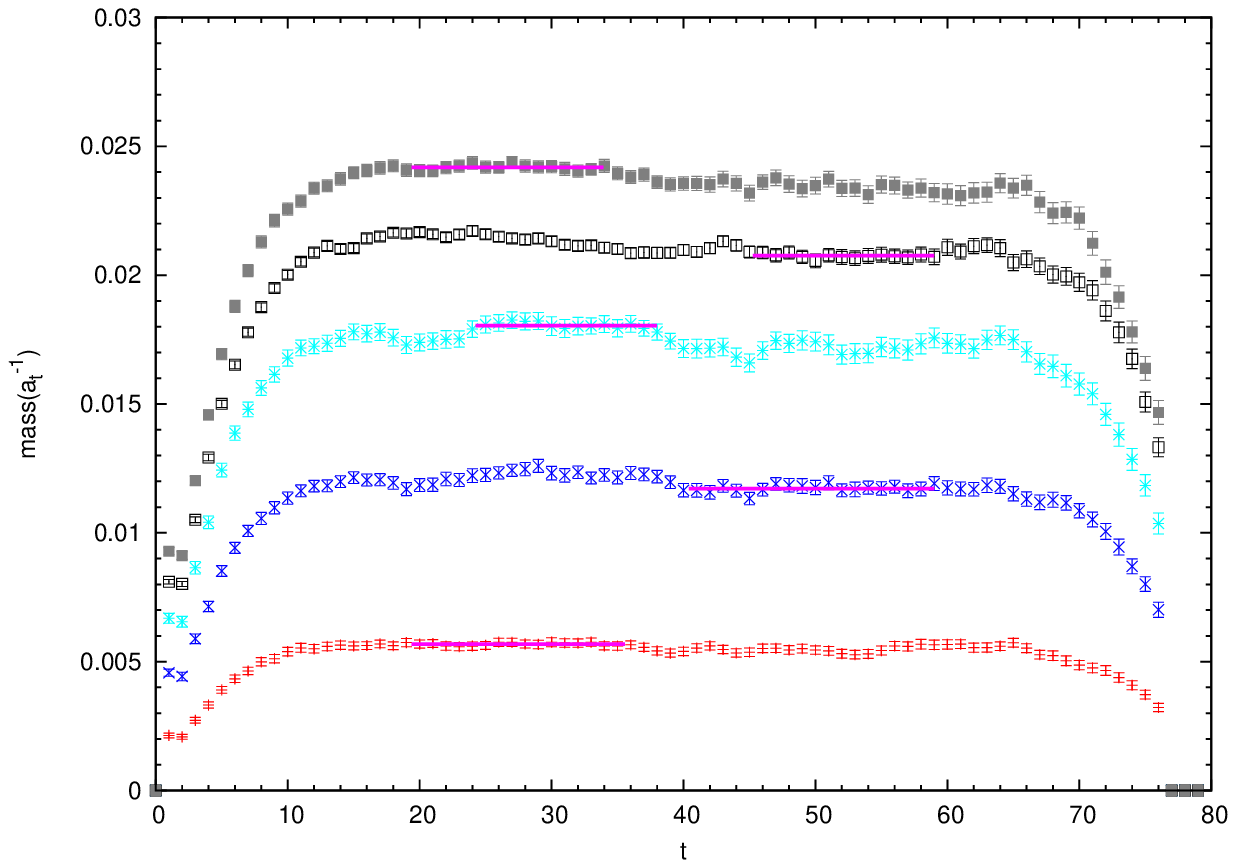}
\includegraphics[scale=0.35]{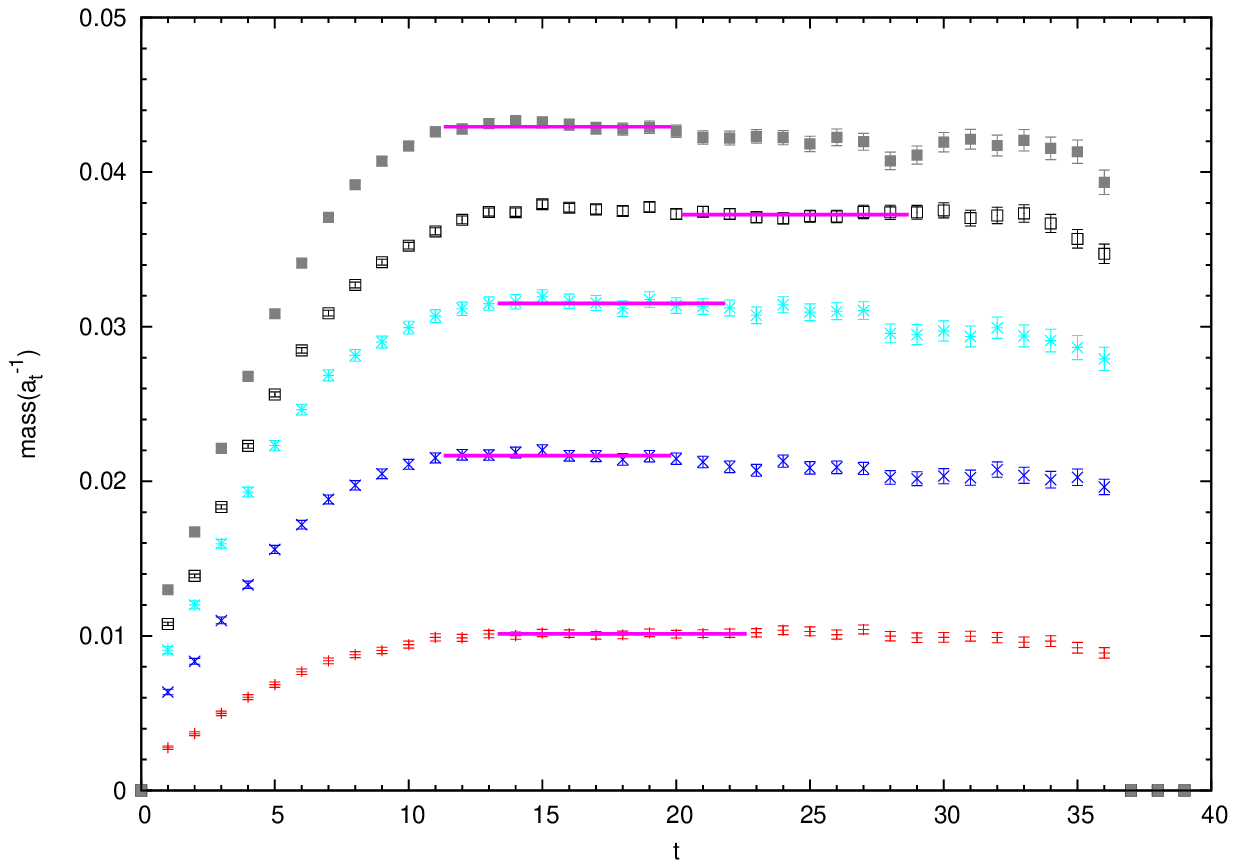}
\includegraphics[scale=0.35]{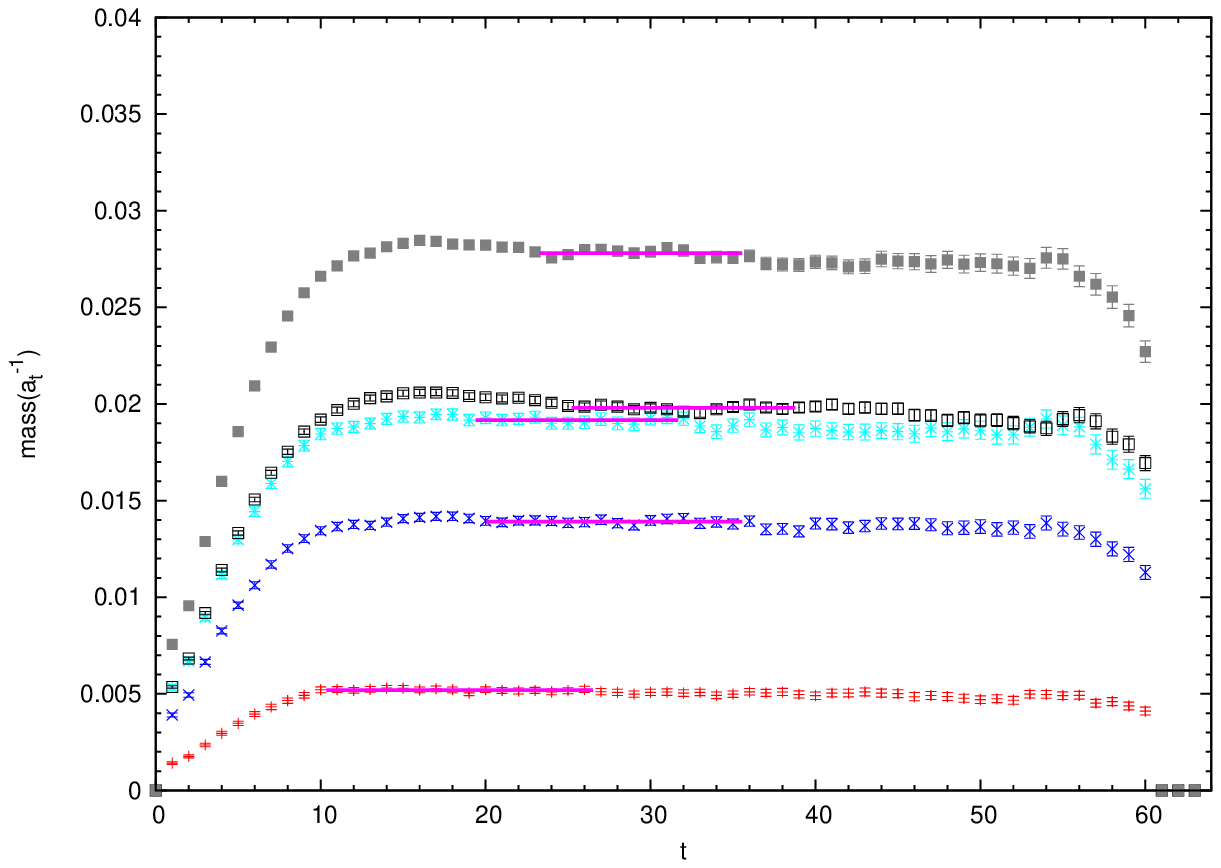}
\includegraphics[scale=0.35]{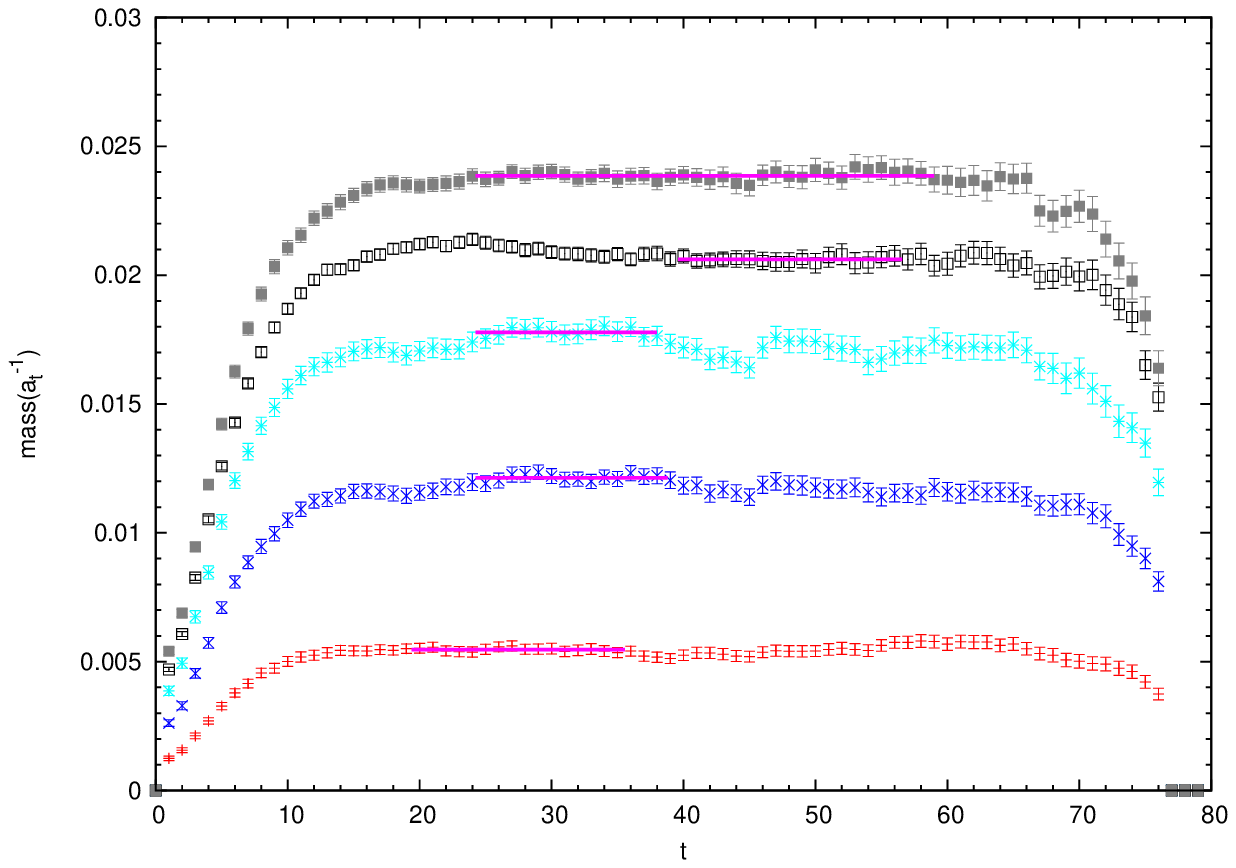}
 \caption{Effective mass plateaus for the quantity
 $\log[R(t,{\bf k})/R(t+1,{\bf k})]$ as discussed in
 subsection~\ref{subsec:dispersion},
 From top to bottom: ${D^\ast}$, ${D_1}$,
 $\eta_c$, $J/\psi$, from left to right: $\beta=2.5,2.8,3.2$}
\label{fig:ratio}
\end{figure}

 Let us now move on to the dispersion relations for
 $D^*$, $D_1$, $\eta_c$ and $J/\Psi$. The aim for
 this study is to verify that we do get single-particle asymptotic states.
 For this we need to know the energy of these particles with definite three-momentum.
 Since the correlators with non-zero three-momentum $C(t,{\bf k})$ is much noisier
 than the one with zero three-momentum $C(t,{\bf 0})$,
 it is difficult to obtain the plateau of the energy directly, particularly for the
 axial-vector meson $D_1$.
 To get around this, we form the following ratio:
 \begin{eqnarray}
  R(t,{\bf k})={C(t,{\bf k})\over C(t,{\bf 0})}\propto e^{-\delta E({\bf k})\cdot t}
 \end{eqnarray}
 where $\delta E({\bf k})=E({\bf k})-E({\bf 0})$ designates the ``kinetic energy'' of
 the particle. It turns out that, by forming this ratio, most of the
 noise is suppressed and a plateau for
 $\delta E({\bf k})$ can be extracted from:
 \begin{eqnarray}
  \delta E_{\rm eff.}(\bk,t)=\log {R(t,{\bf k})\over R(t+1,{\bf k})}.
 \end{eqnarray}
 These plateaus are illustrated in Fig.~\ref{fig:ratio}. In fact,
 for the $D^\ast$, $J/\psi$ and $\eta_c$ mesons,
 we can also get the plateau directly. There is no
 need to form the ratio. However, if we do form the ratio, the results we
 get from this ratio are fully compatible with what we get by
 direct extraction from the original correlators. For the axial-vector
 meson $D_1$, however, forming the ratio helps to suppress the noise and to develop
 the mass plateau.

\begin{figure}[h]
\centering
\includegraphics[scale=0.7]{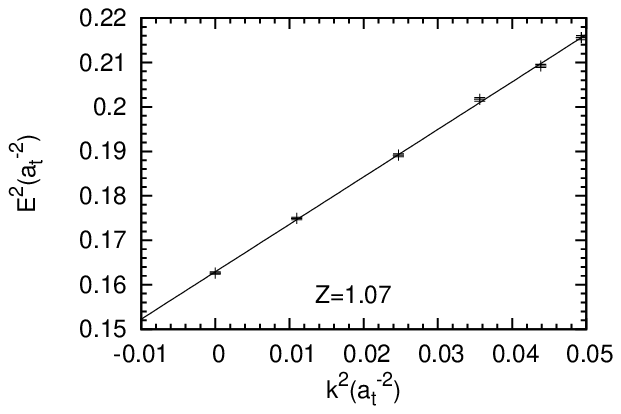}
\includegraphics[scale=0.7]{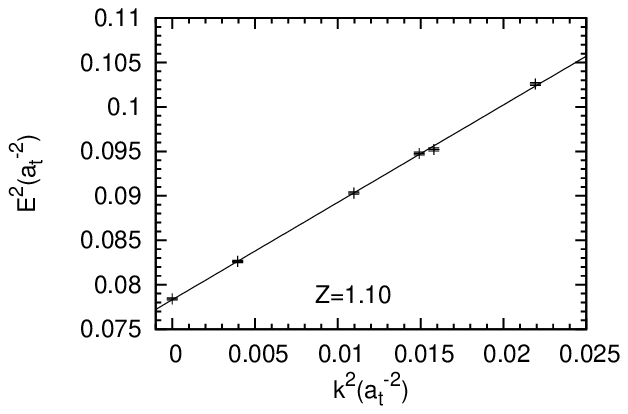}
\includegraphics[scale=0.7]{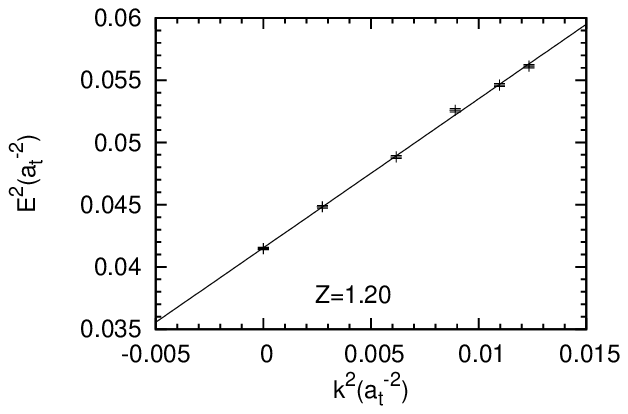}
\includegraphics[scale=0.7]{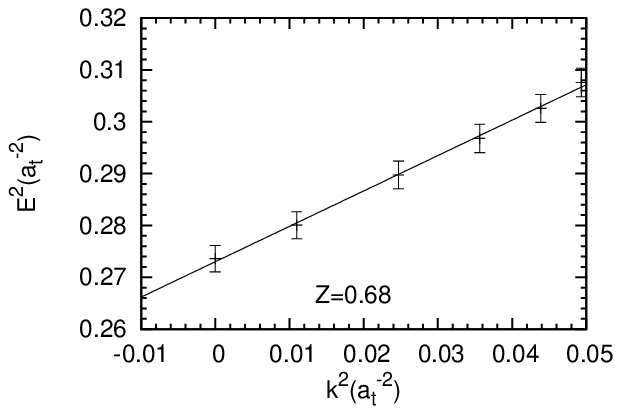}
\includegraphics[scale=0.7]{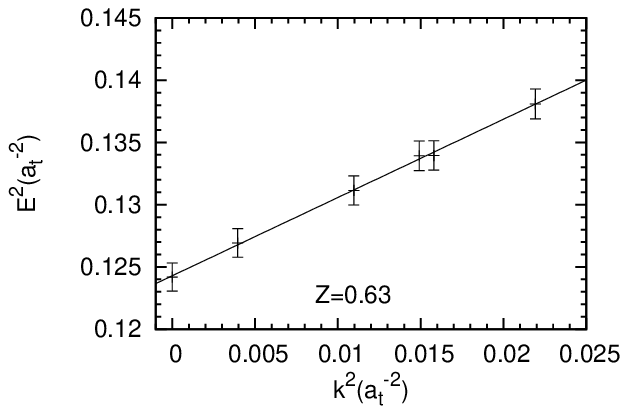}
\includegraphics[scale=0.7]{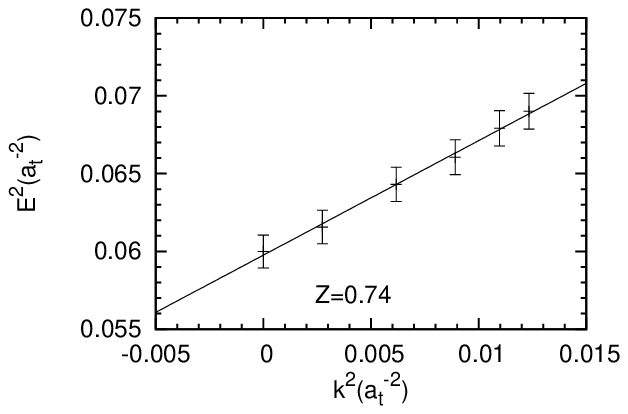}
\includegraphics[scale=0.7]{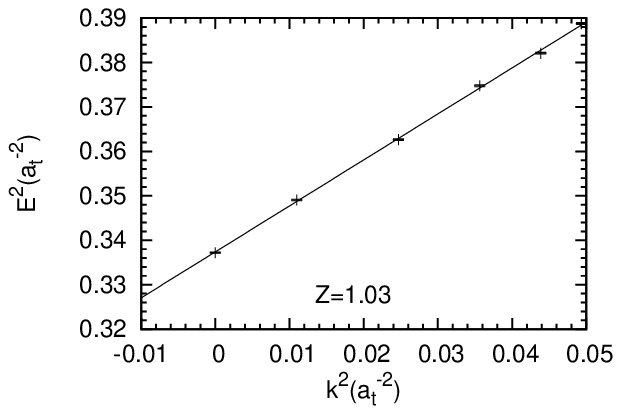}
\includegraphics[scale=0.7]{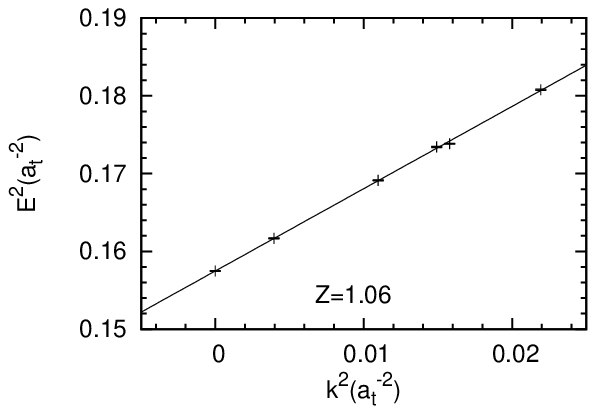}
\includegraphics[scale=0.7]{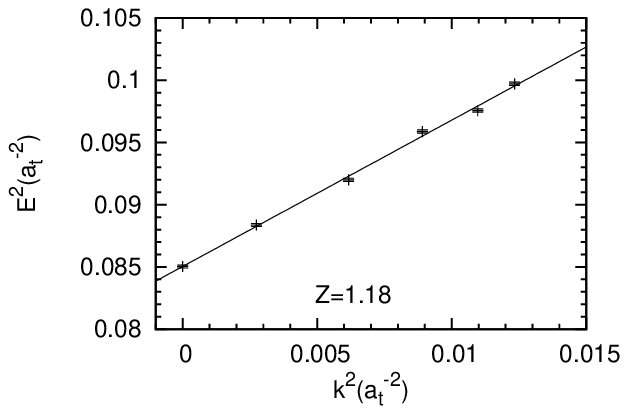}
\includegraphics[scale=0.7]{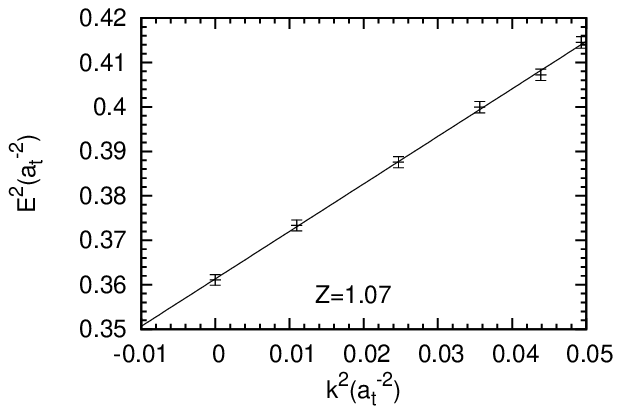}
\includegraphics[scale=0.7]{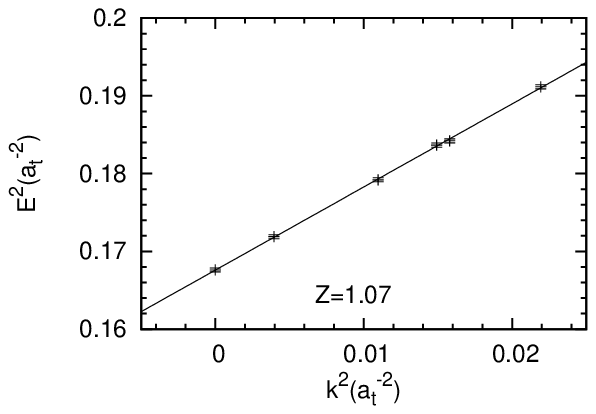}
\includegraphics[scale=0.7]{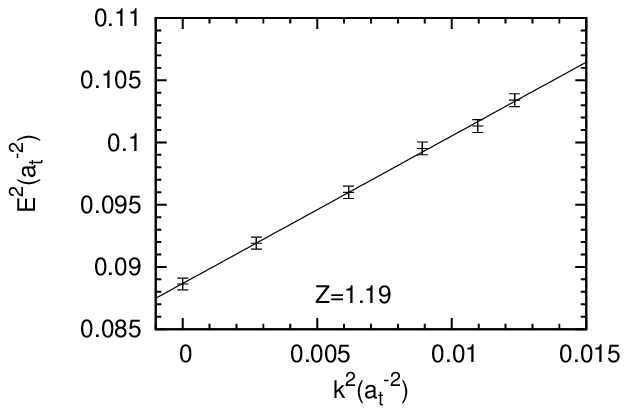}
\caption{Dispersion relations for various mesons obtained from
 single meson energies. From top to
 bottom: ${D^\ast}$, ${D_1}$, $\eta_c$, $J/\psi$;
 from left to right: $\beta=2.5,2.8,3.2$.} \label{fig:Z}
\end{figure}

 After getting the results of $\delta E({\bf k})$, the results for
 $E({\bf k})=E(0)+\delta E({\bf k})$ is also obtained from which
 one can check the dispersion relation at low-momenta:
 \be
   \label{dispersion}
  E^2({\bf k})=m^2+ Z\cdot {\bf k}^2+\cdots\;,
 \ee
 where $Z$ is a parameter to be fitted. In order to recover usual  continuum
 dispersion relation with $Z=1$, one has to tune the
 bare speed of light parameter $\nu$ in the fermion action.
 This has been done in Ref.~\cite{chuan06:tune_v}, for several values of $\beta$.
 In this study, we use the results of $\nu$ in Ref.~\cite{chuan06:tune_v}
 as input parameters. Since we have get the value of $E({\bf k})$ in
 our calculation, we can fit the our data using
 Eq.~(\ref{dispersion}), and get the value of $Z$. As an illustration, the
 results of dispersion relations are shown in Fig.~\ref{fig:Z} for
 certain input quark mass parameters. The values of $Z$ are also indicated in
 each panel. From these results, it is seen that the values of $Z$ is
 approximately equal to 1 for $\eta_c$ and $J/\Psi$,
 which suggests that our choice for the value of $\nu$ is
 approximately right.

 After all these checking, including the mass spectrum and the dispersion relations
 for the $D^*$ and $D_1$ mesons, we are confident that our finite
 box can accommodate well-established single-particle asymptotic
 states and we may now proceed to study the scattering of $D^*$ and
 $D_1$ mesons at low-momenta.

 \subsection{Results for the scattering length and effective range}

 As we argued in Sec.~\ref{sec:operators}, only one correlation
 matrix $C(t)$ have to be computed for each symmetry channel of the
 two-particle system. For each symmetry channel, we have studied
 $5$ different non-zero momentum modes. Therefore, including the
 zero-momentum mode, for each symmetry channel, the correlation
 matrix for the two-particle system is a $6\times 6$ matrix.

 To extract the two-particle energy eigenvalues, we adopt
 the usual ${\rm L\ddot{u}scher-Wolff}$ method~\cite{luscher90:finite}.
 For this purpose,  a new matrix $\Omega(t,t_0)$ is defined as:
 \begin{eqnarray}
  \Omega(t,t_0)=C(t_0)^{-{1\over2}}C(t)C(t_0)^{-{1\over 2}},
 \end{eqnarray}
 where $t_0$ is a reference time-slice. Normally one picks a
 $t_0$ such that the signal is good and stable.
 The energy eigenvalues for the two-particle system are
 then obtained by diagonalizing the matrix $\Omega(t,t_0)$.
 The $i$-th eigenvalue of the matrix has the following behavior
 in the large $(t-t_0)$ limit:
 \begin{eqnarray}
  \lambda_i(t,t_0) \propto e^{-E_i(t-t_0)}\;.
 \end{eqnarray}
 Therefore, the exact energy $E_i$ can be extracted from
 the effective mass plateau of the eigenvalue $\lambda_i$.

\begin{figure}[h]
\centering
\includegraphics[scale=0.6]{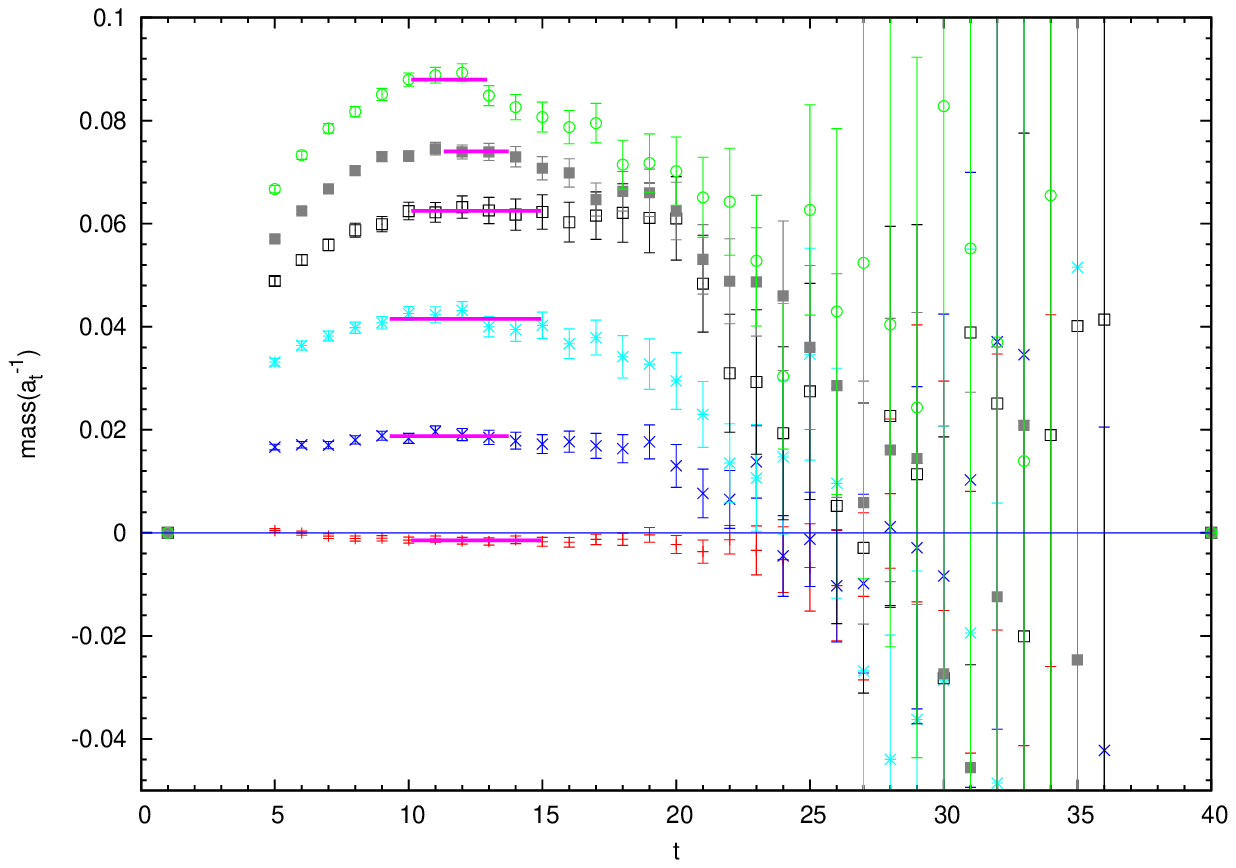}
\includegraphics[scale=0.6]{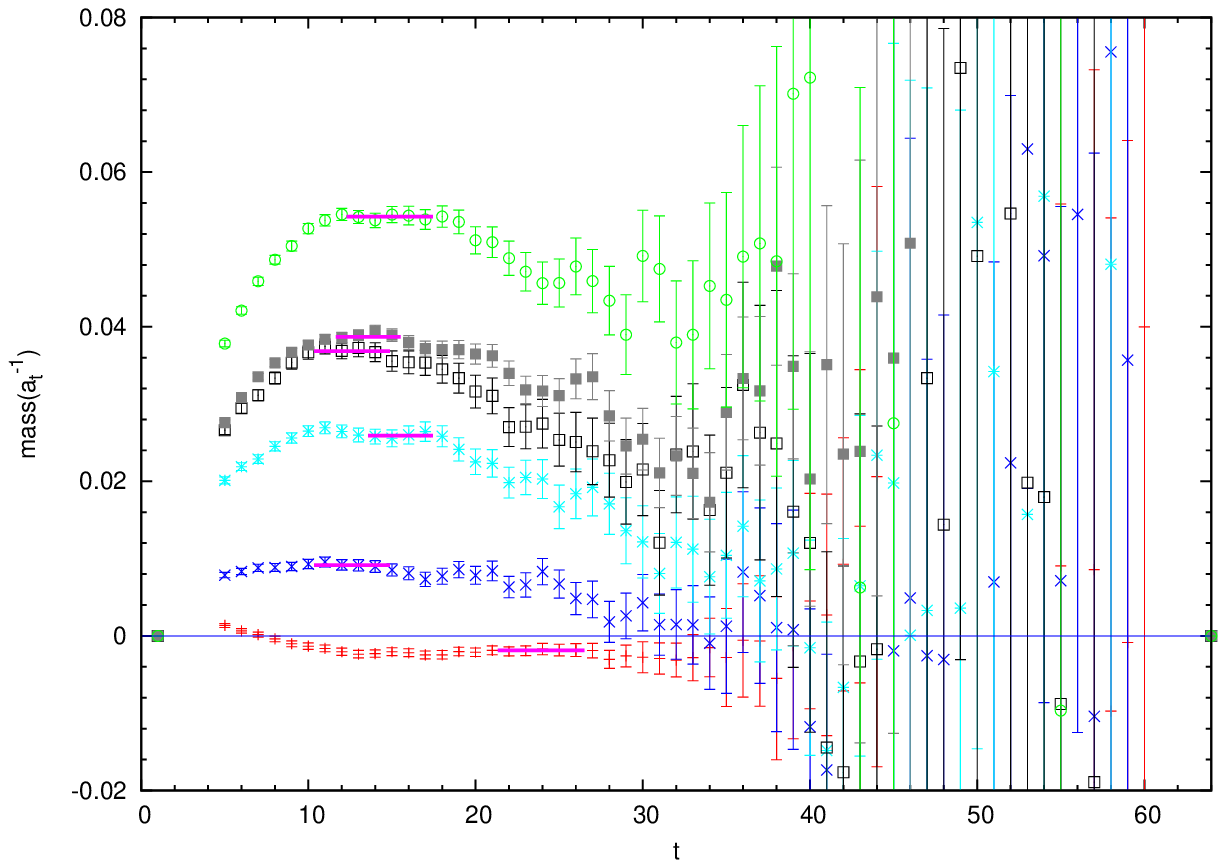}
\includegraphics[scale=0.6]{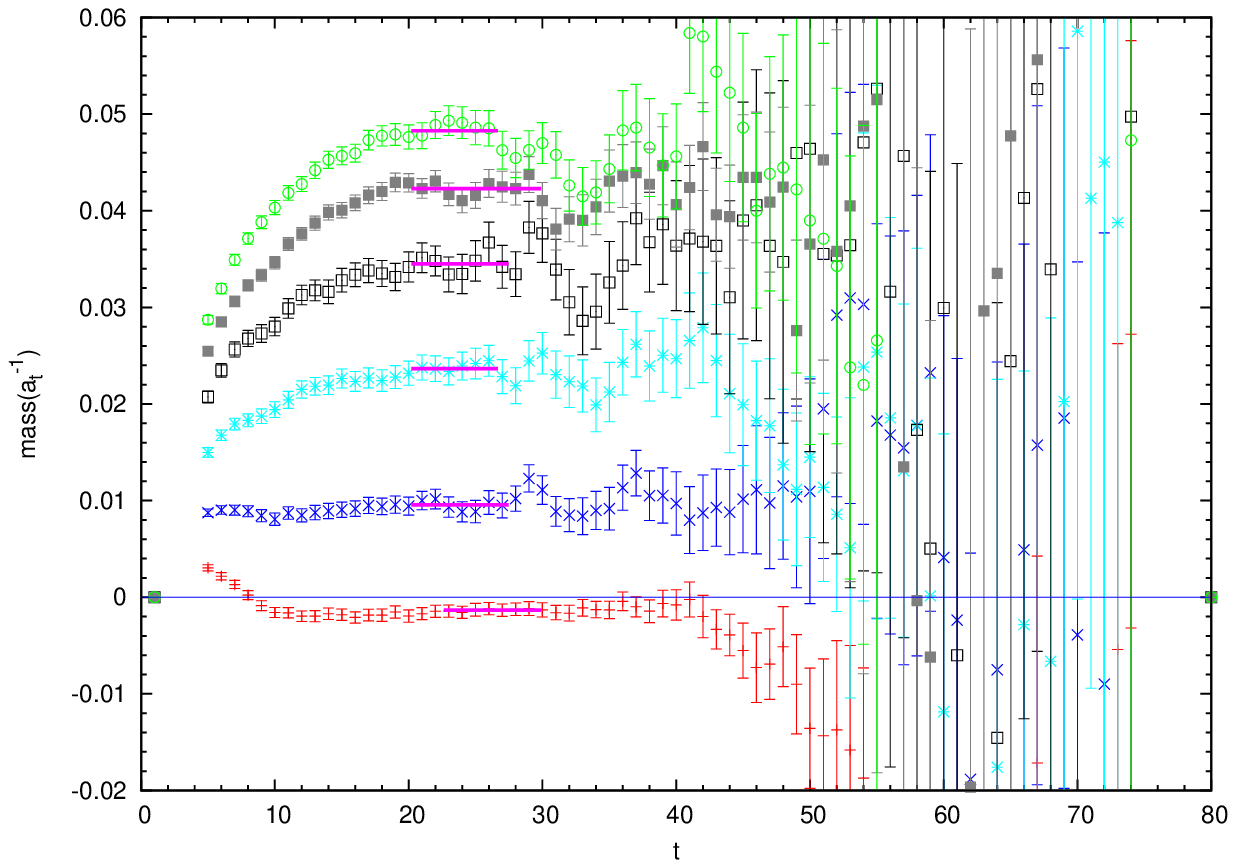}
 \caption{The effective mass plateaus of $\delta E_i$ in
 the $A^{(1)}_1$ channel as obtained from Eq.~(\ref{eq:meff_twop}).
 From top to bottom: $\beta=2.5,2.8,3.2$} \label{fig:A1_11}
\end{figure}

 The real signal for the eigenvalue in our simulation turns out
 to be so noisy that reliable plateau cannot be found
 directly. Therefore,  the following ratio was attempted:
 \begin{eqnarray}
  \calR(t,t_0)={\lambda_i(t,t_0)\over C_{D^\ast}(t)C_{D_1}(t)}\propto e^{-\delta E_i\cdot t}
 \end{eqnarray}
 where $C_{D^\ast}(t)$ and $C_{D_1}(t)$ are one-particle correlation
 function with zero momentum for the corresponding mesons.
 Therefore, $\delta E_i$ is the difference of the two-particle
 energy with the threshold of the two mesons:
 \begin{eqnarray}
  \delta E_i=E_i-m_{D^\ast}-m_{D_1}
 \end{eqnarray}
 By taking this ratio, the signal to noise ratio is greatly
 enhanced. The energy difference $\delta E_i$ can
 be extracted reliably from the following effective mass:
 \be
 \label{eq:meff_twop}
 M_{\rm eff}(t)=\ln \left({\calR(t)\over \calR(t+1)}\right)\;.
 \ee
 For the $A^{(1)}_1$ channel with correlation matrix
 given in Eq.~(\ref{eq:C_A1}), the situation is illustrated
 in Fig.~\ref{fig:A1_11} for lattices at $\beta=2.5$, $2.8$ and
 $3.2$. Six different plateaus in each
 panel correspond to different modes and this procedure is
 carried out for each pair of quark mass parameters $(\kappa^{ud},\kappa^c)$.

\begin{figure}[h]
  \centering
  \includegraphics[scale=0.6]{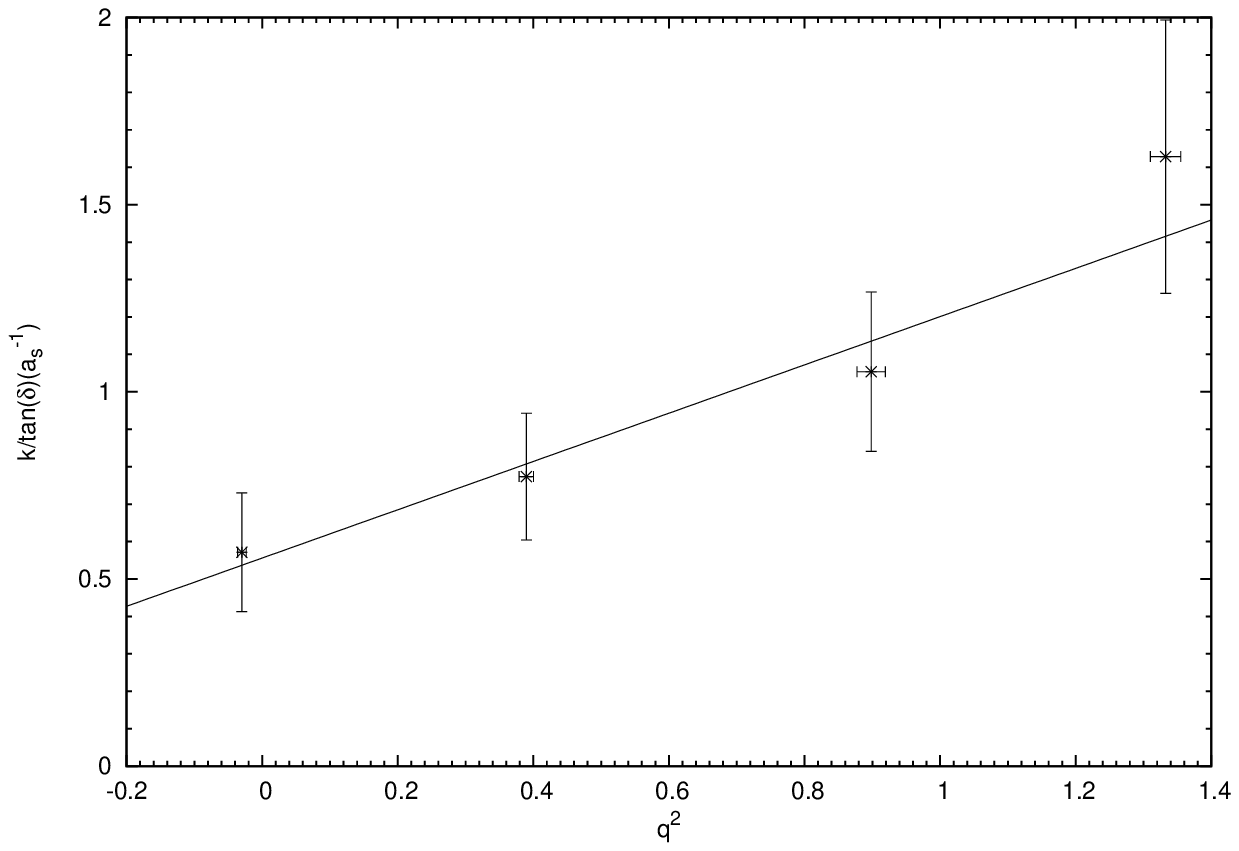}
  \includegraphics[scale=0.6]{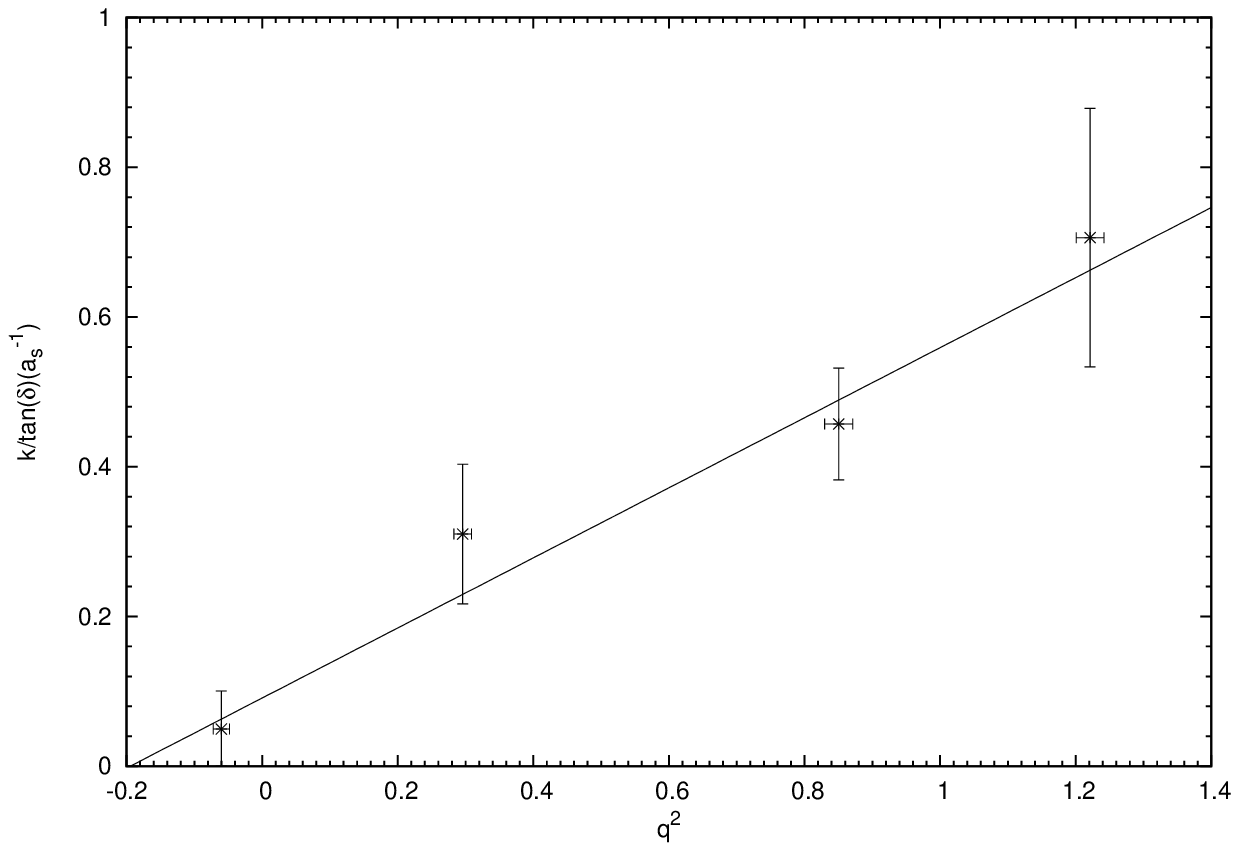}
  \includegraphics[scale=0.6]{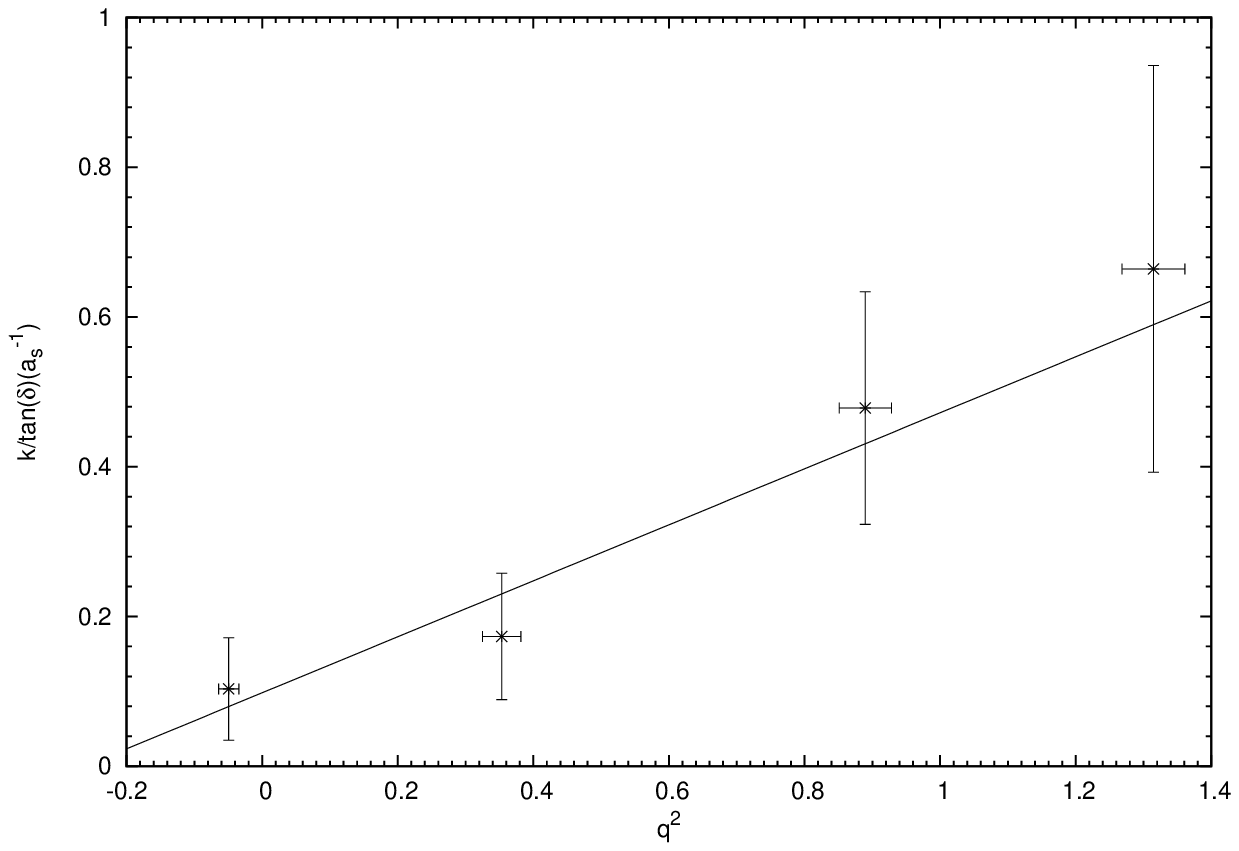}
  \caption{The quantity $k/\tan\delta(k)$ versus
  $q^2$ in the $A_1(1)$ channel. From top to bottom: $\beta=2.5$, $2.8$ and $3.2$.}
  \label{fig:k_over_tan}
\end{figure}

 With the energy difference $\delta E_i$ extracted from the simulation data,
 one utilizes the definition:
 \begin{eqnarray}
  \sqrt{m_{D^\ast}^2+\bar{\bf k}^2}+\sqrt{m_{D_1}^2+\bar{\bf k}^2}=\delta E_i + m_{D^\ast} +m_{D_1}
  \label{kbar}
 \end{eqnarray}
 to solve for $\bar{\bk}^2$ which is then plugged into
 the modified L\"uscher's formula (i.e.
 Eq.~(\ref{eq:phase_rectangular}).
 Close to the scattering threshold, the quantity $k/\tan\delta(k)$
 has the following expansion:
 \begin{eqnarray}
  {k\over \tan\delta (k)}={1\over a_0}+{1\over2} r_0 k^2 +\cdots\;,
  \label{kovertan}
 \end{eqnarray}
 where $a_0$ is the scattering length and $r_0$ is the effective range.
 The l.h.s of Eq.(\ref{kovertan}) can also be calculated using L\"uscher's formula.
 Therefore, we can fit our data with Eq.~(\ref{kovertan}), from which the values of
 $a_0$ and $r_0$ are obtained. Since Eq.~(\ref{kovertan}) is only
 valid when $k$ is small, we use the data for the lowest $4$ modes in the fitting.
 For a particular choice of $(\kappa^{ud}, \kappa^c)$,
 this fitting procedure is shown in Fig.~\ref{fig:k_over_tan}
 for three values of $\beta$ in our simulation.

\begin{figure}[h]
\centering
\includegraphics[scale=0.38]{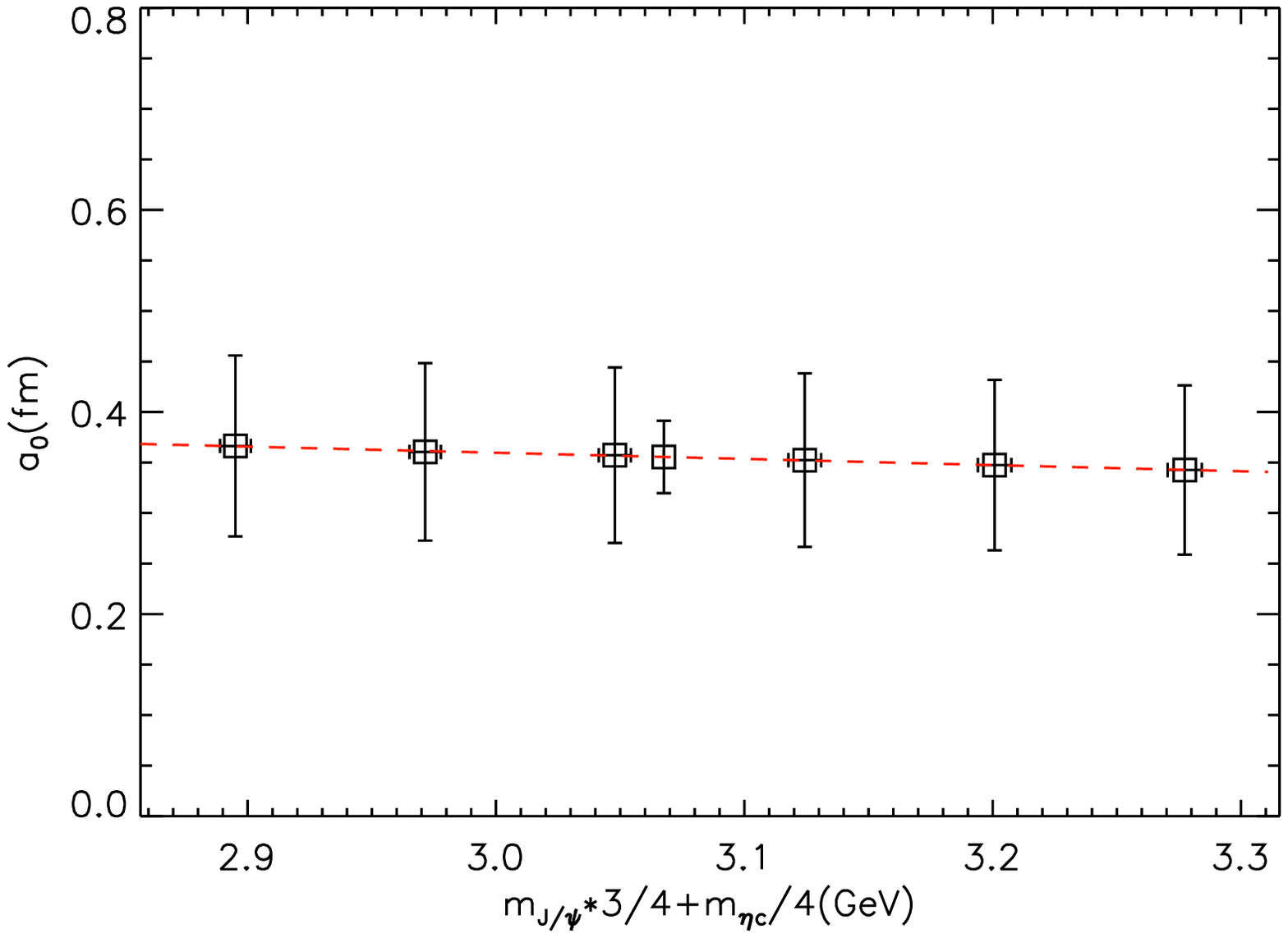}
\includegraphics[scale=0.38]{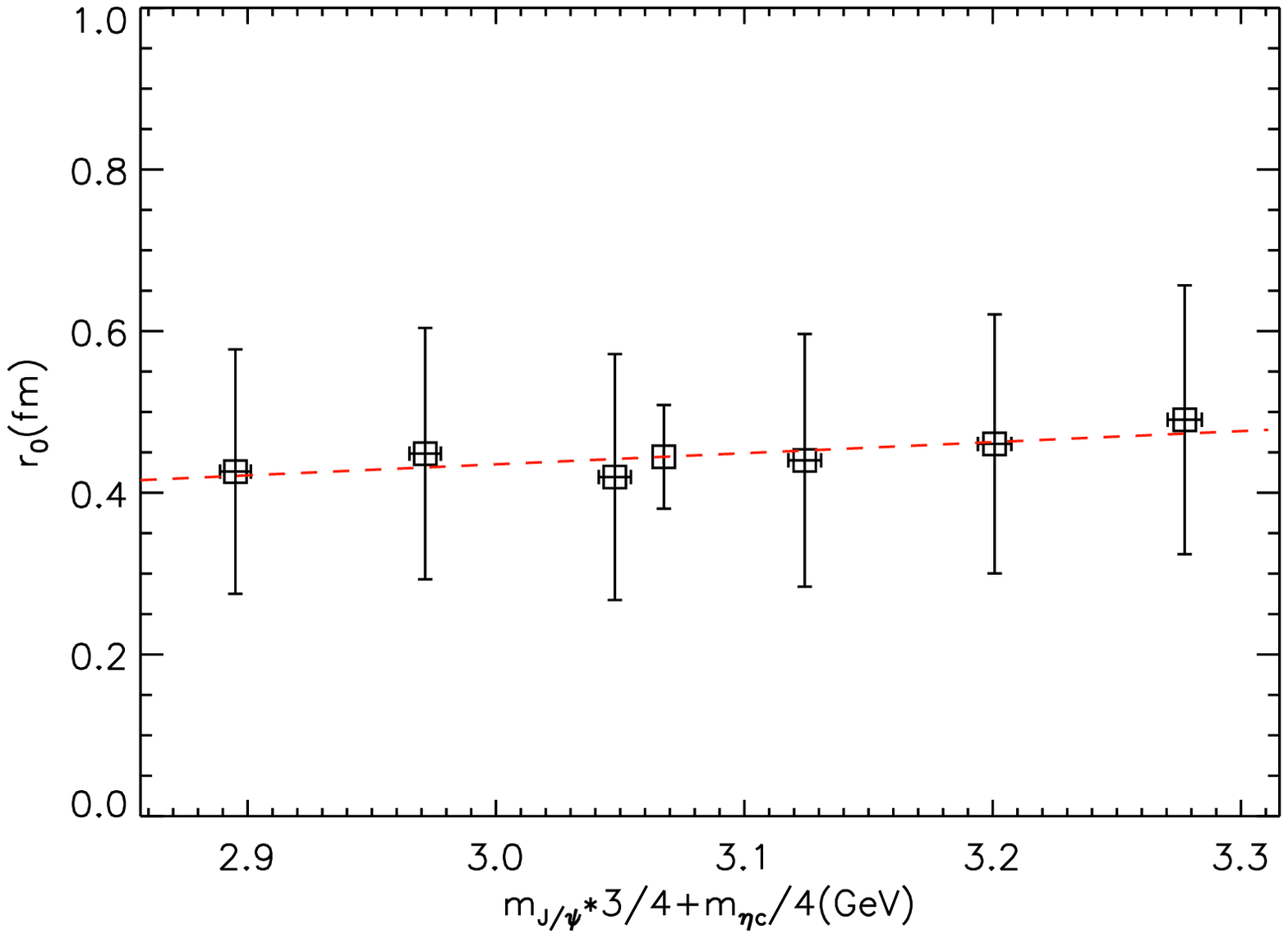}
\includegraphics[scale=0.38]{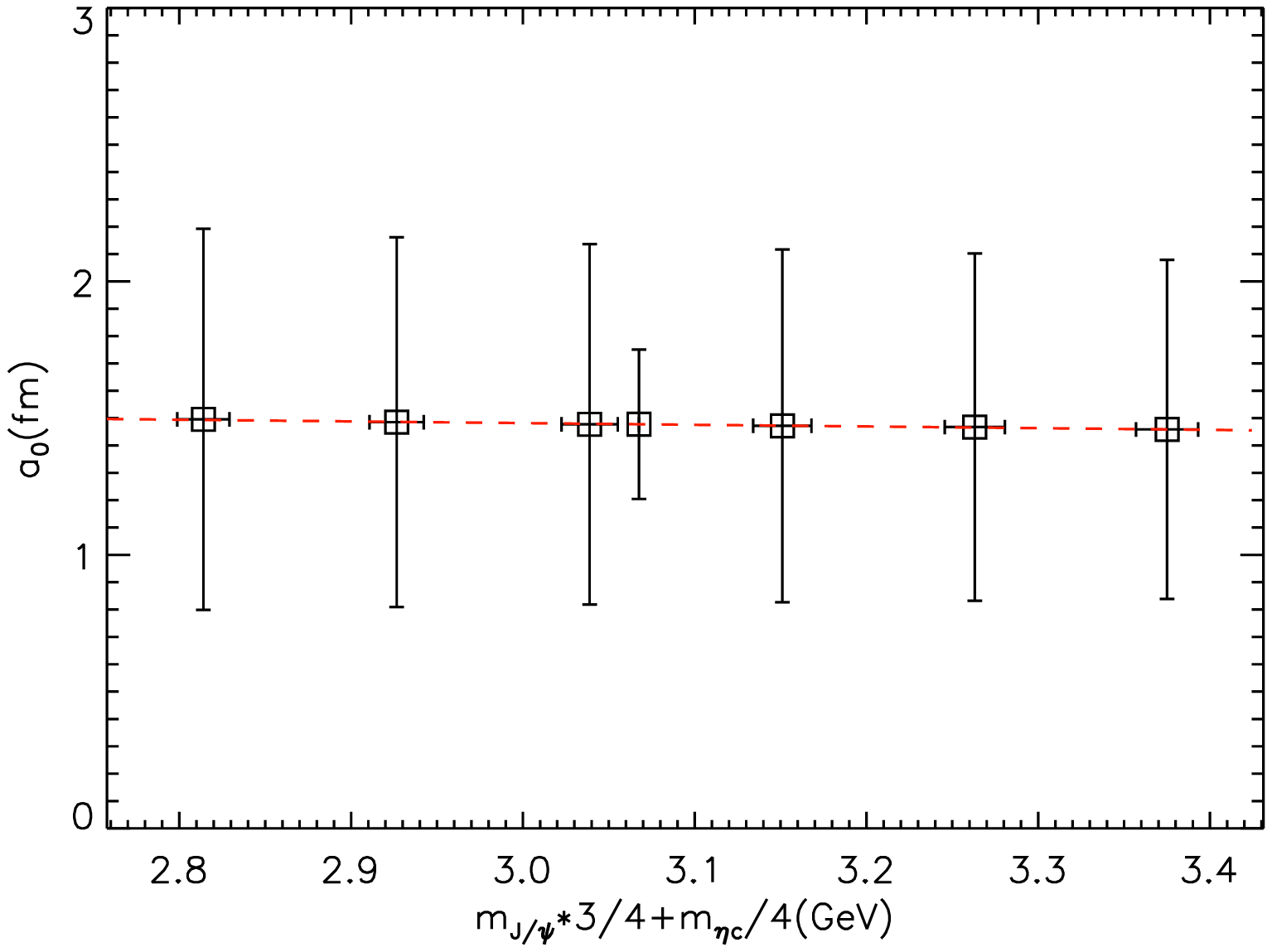}
\includegraphics[scale=0.38]{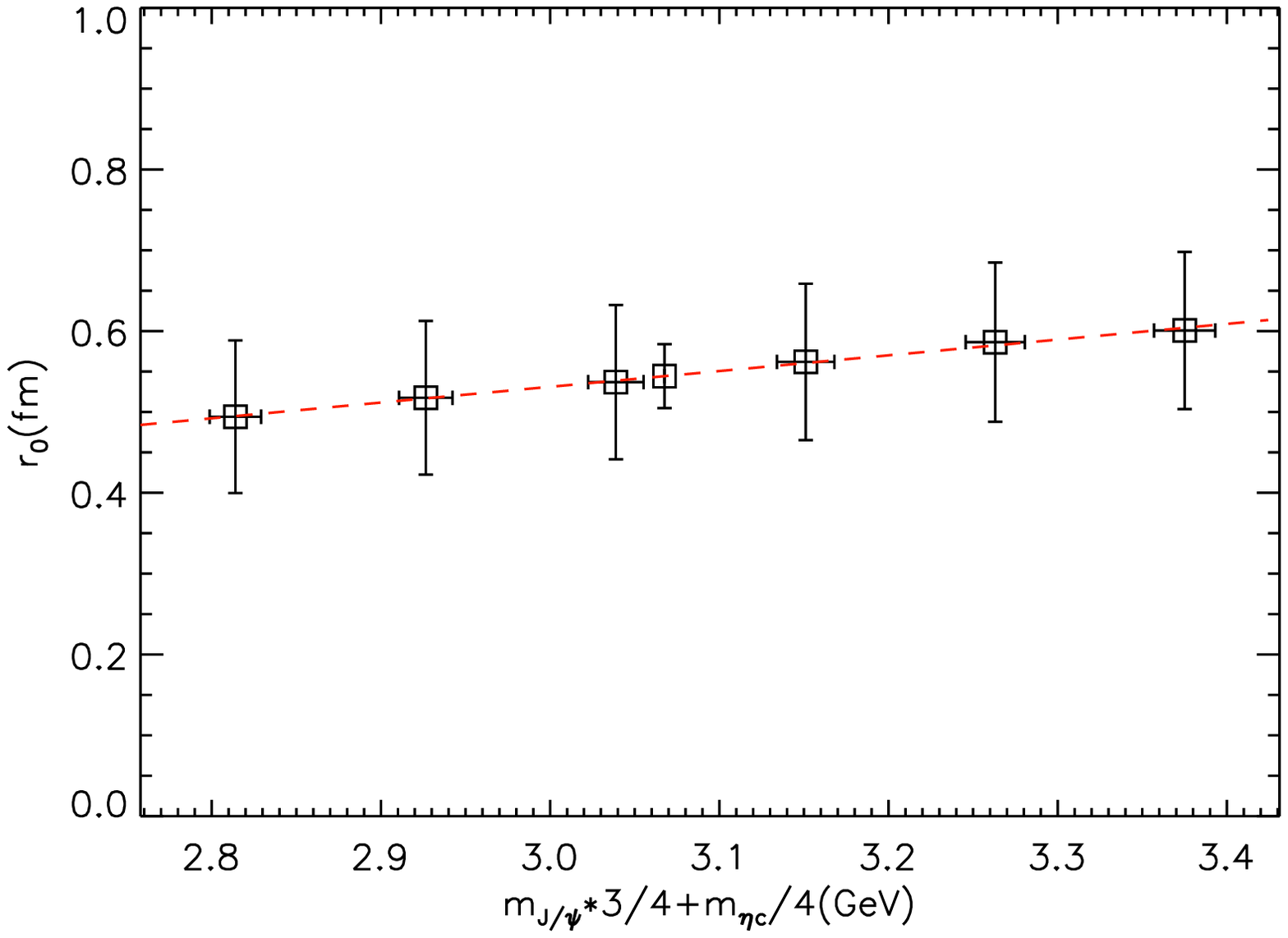}
\includegraphics[scale=0.38]{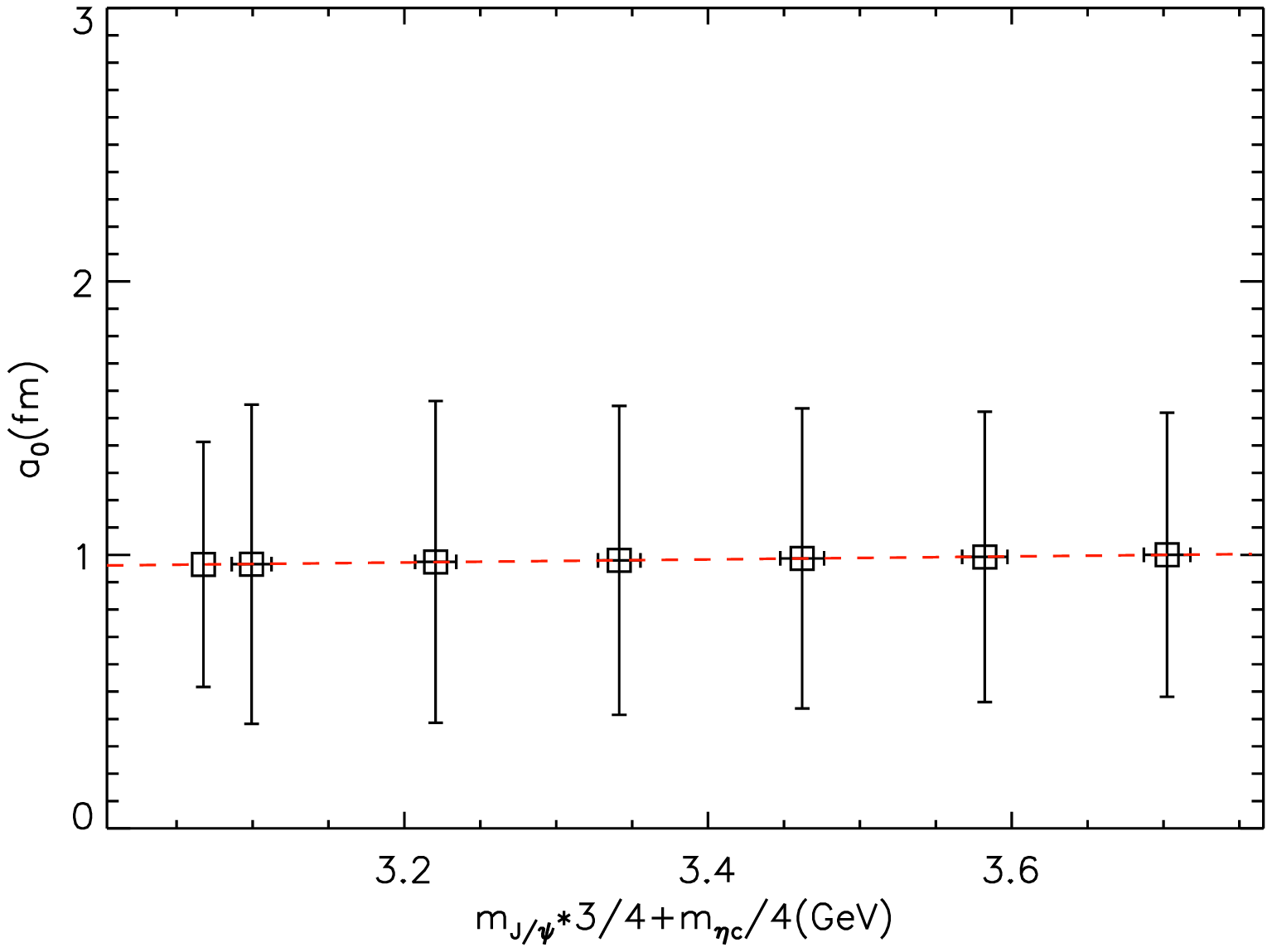}
\includegraphics[scale=0.38]{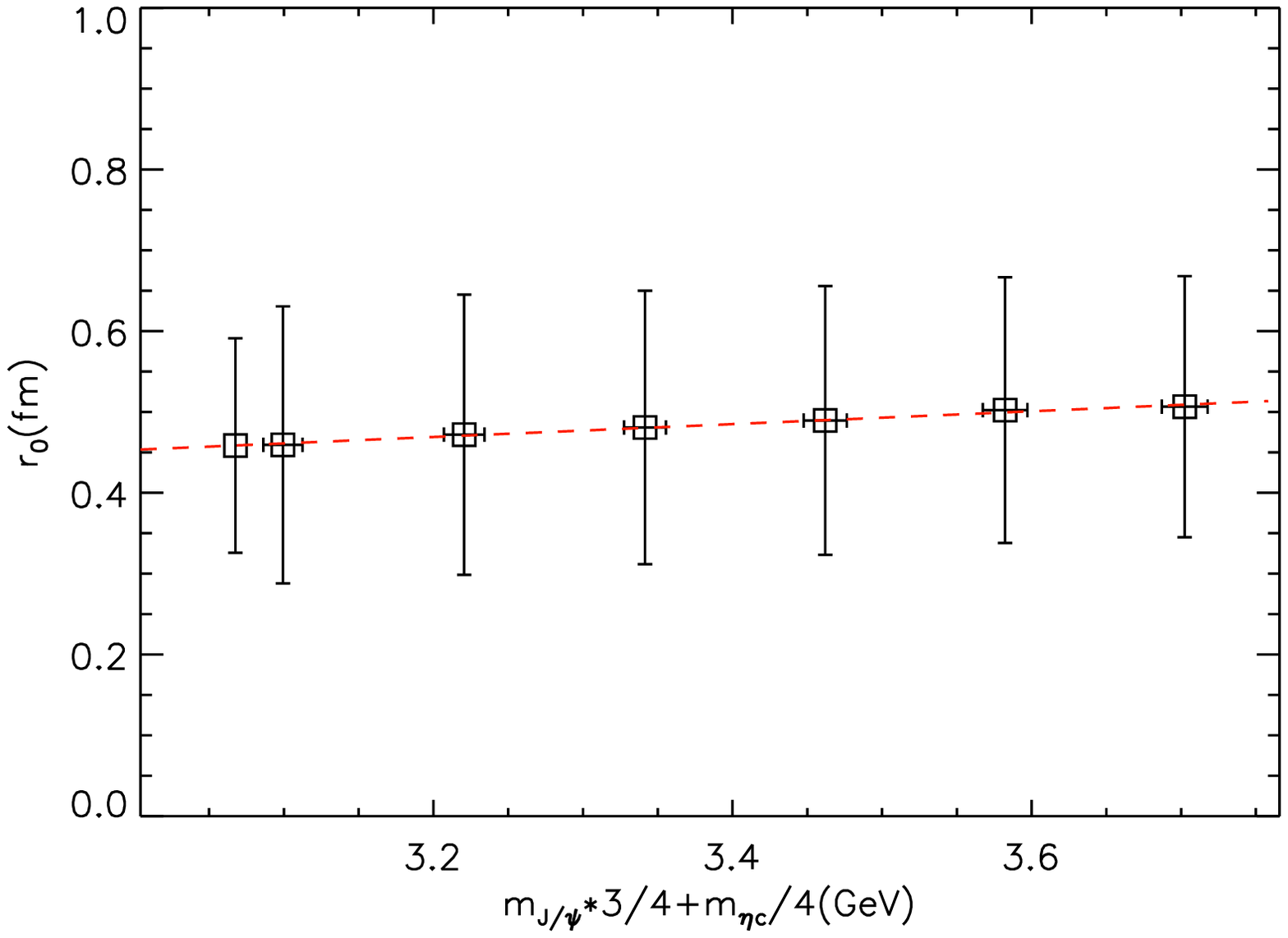}
 \caption{Interpolations for $a_0$ and $r_0$ when the heavy quark mass is brought
 to its physical value. From
 top to bottom: $\beta=2.5$, $2.8$ and $3.2$.} \label{fig:a0r0_heavy}
\end{figure}

\begin{figure}[h]
\centering
\includegraphics[scale=0.38]{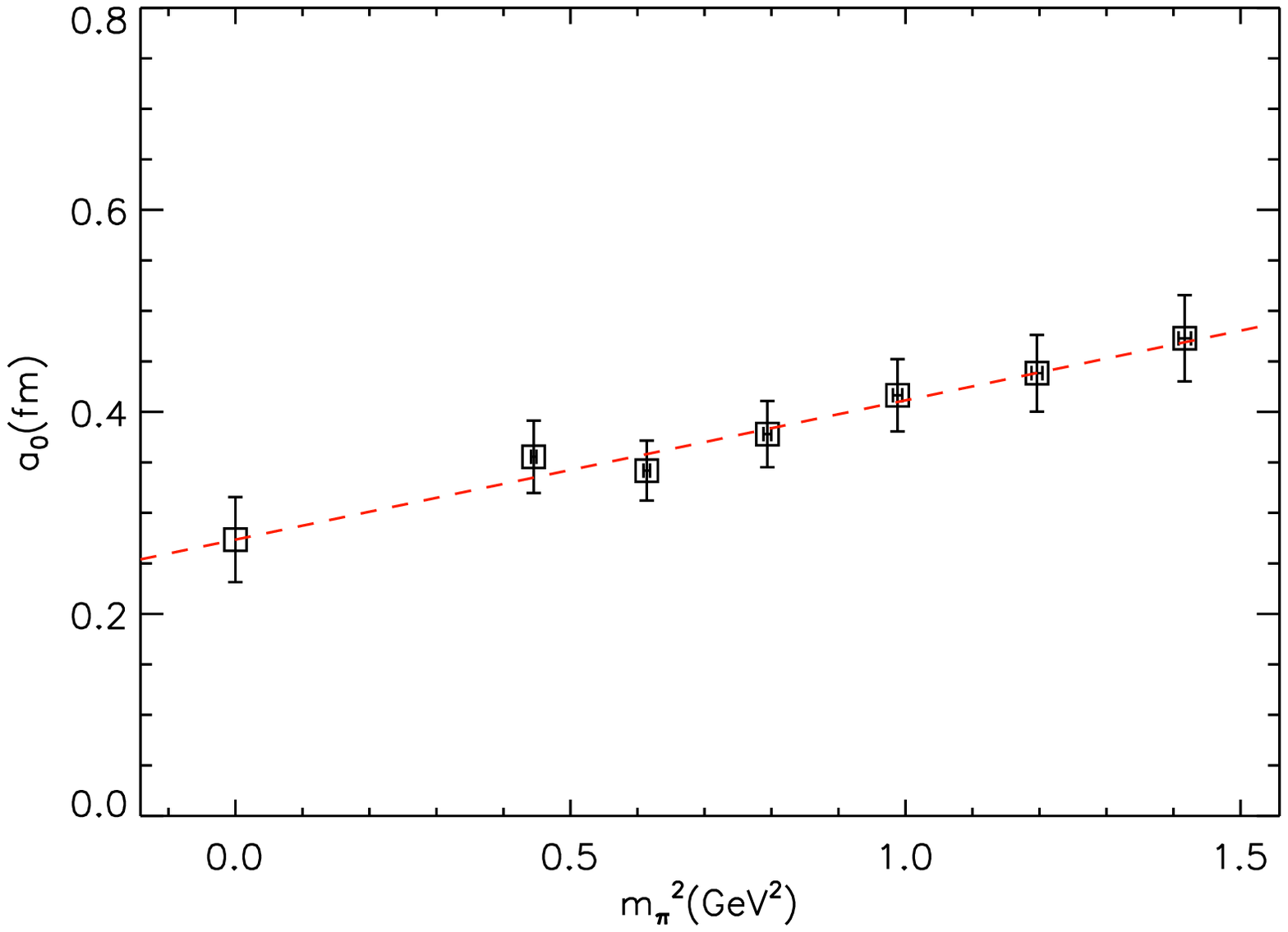}
\includegraphics[scale=0.38]{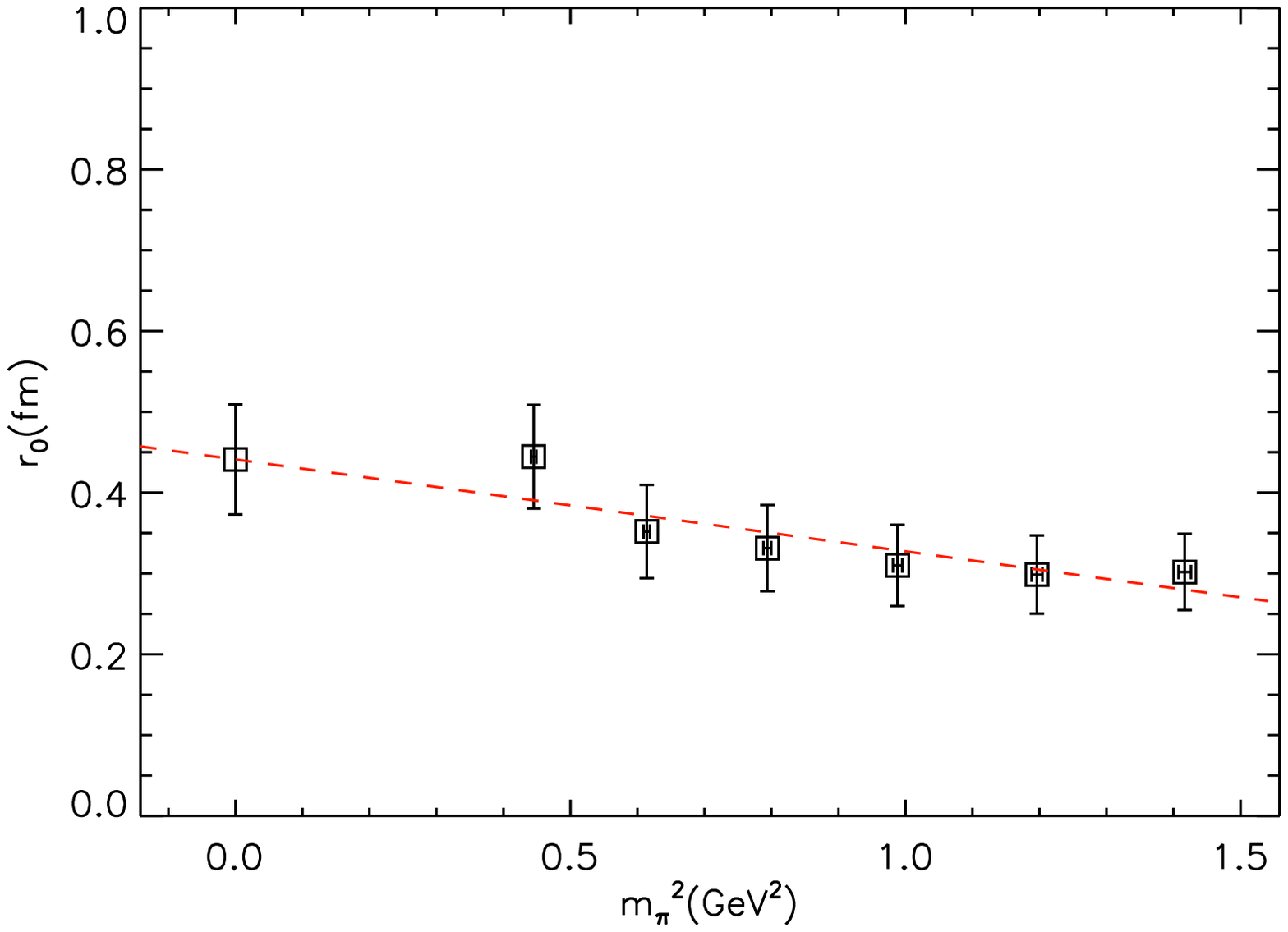}
\includegraphics[scale=0.38]{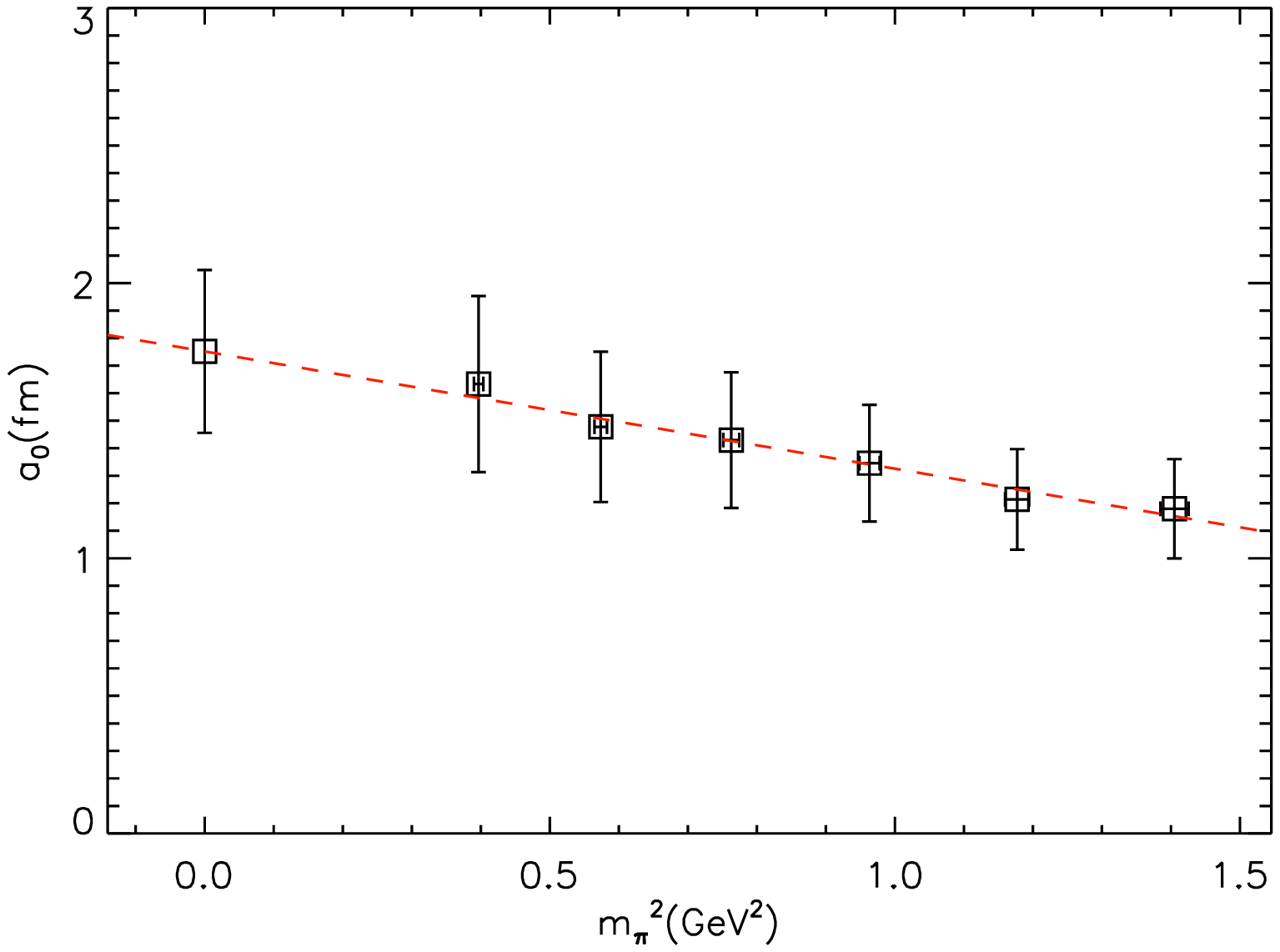}
\includegraphics[scale=0.38]{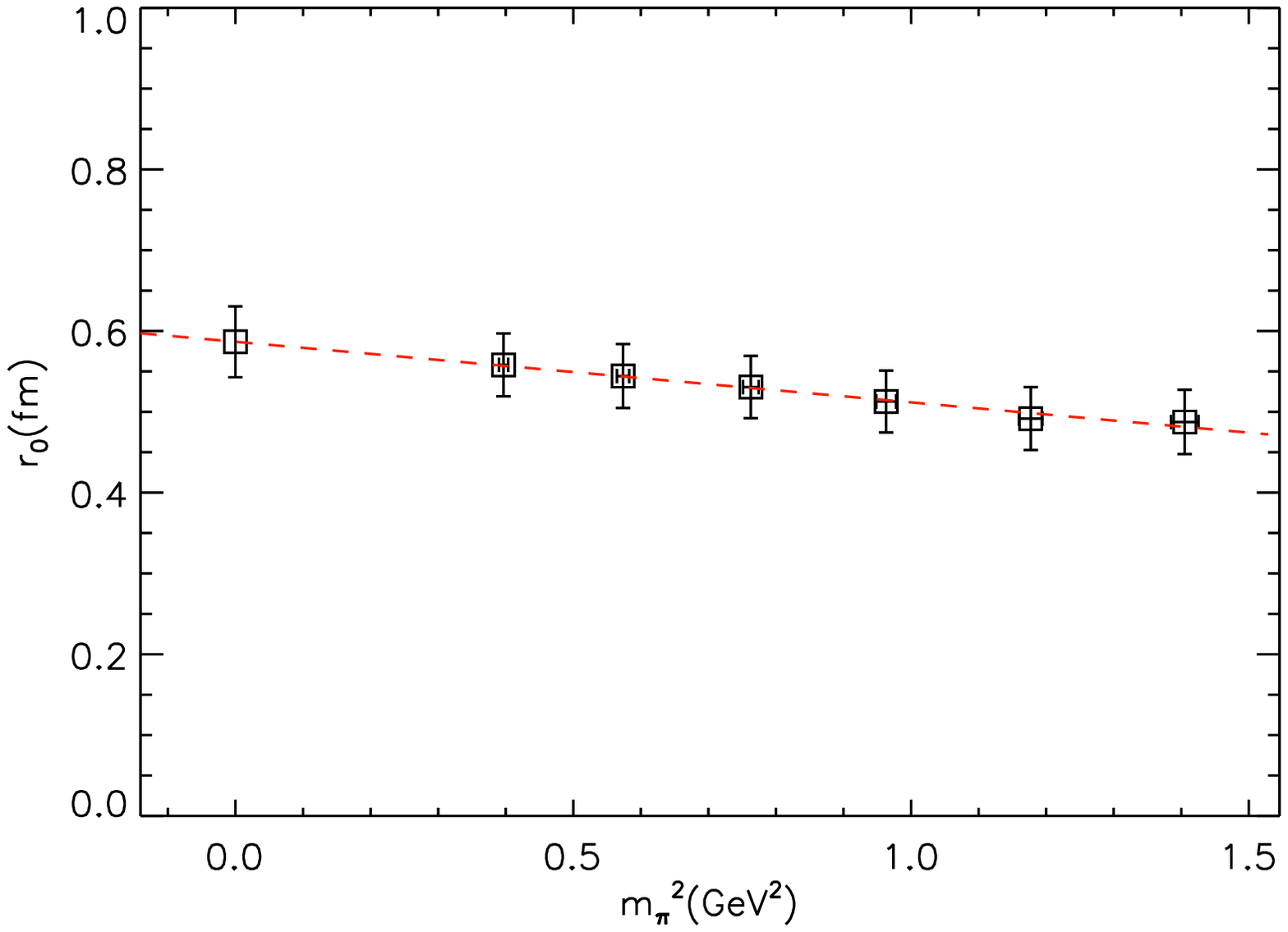}
\includegraphics[scale=0.38]{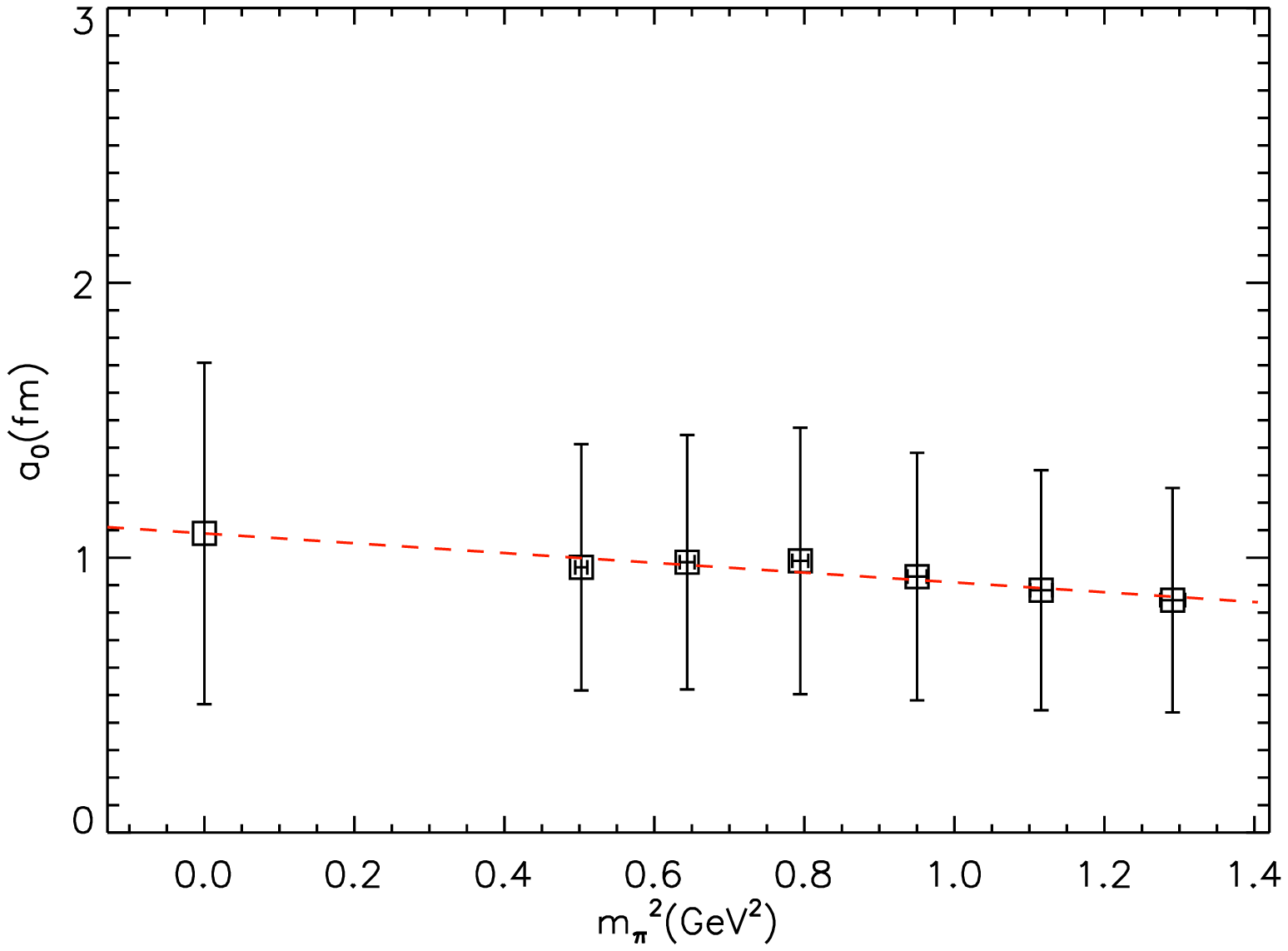}
\includegraphics[scale=0.38]{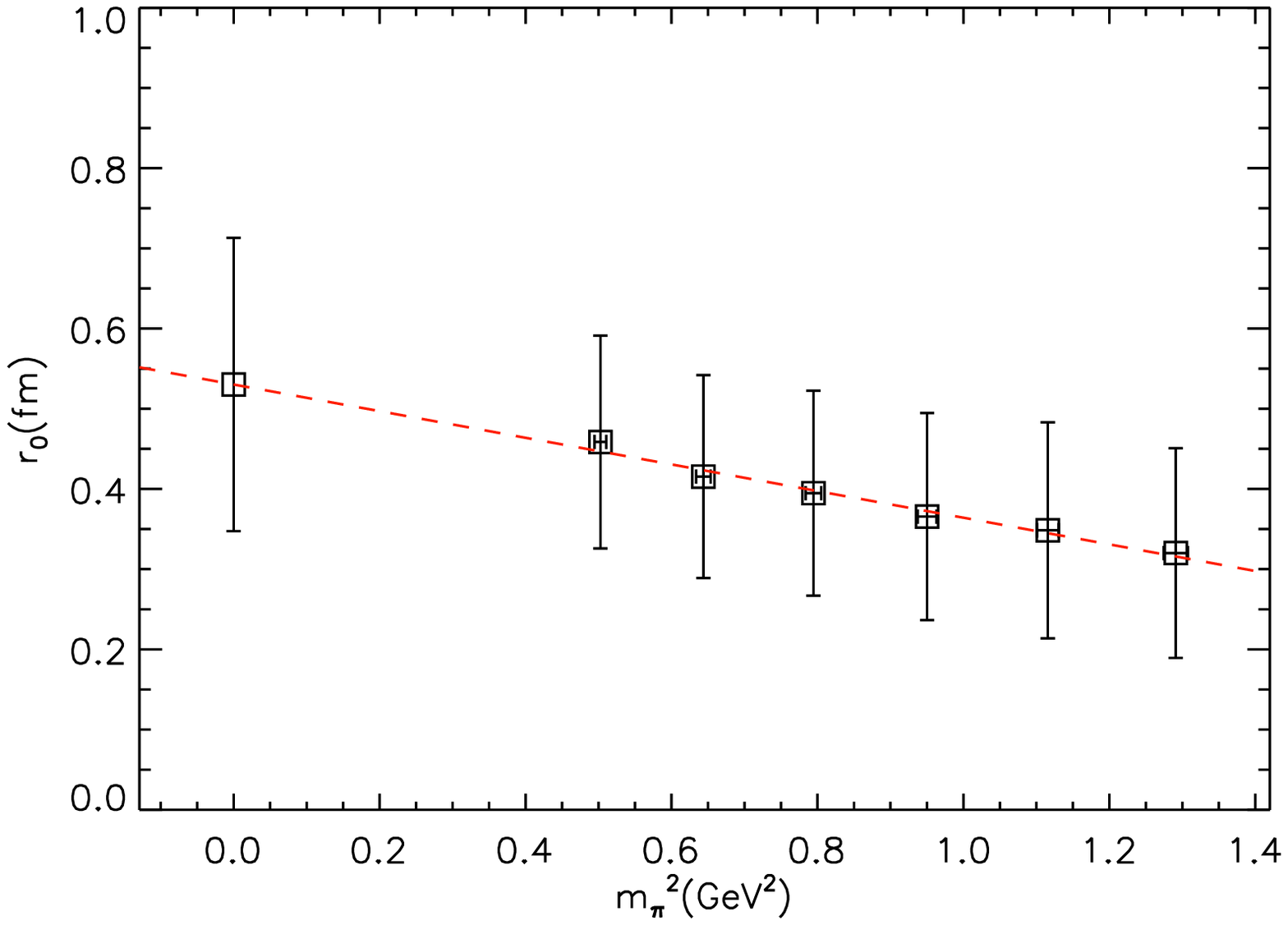}
\caption{Chiral extrapolations for $a_0$ and $r_0$ at various
$\beta$ values. From top to bottom: $\beta=2.5$, $2.8$ and $3.2$.}
\label{fig:a0r0_light}
\end{figure}

\begin{figure}
\centering
\includegraphics[scale=0.6]{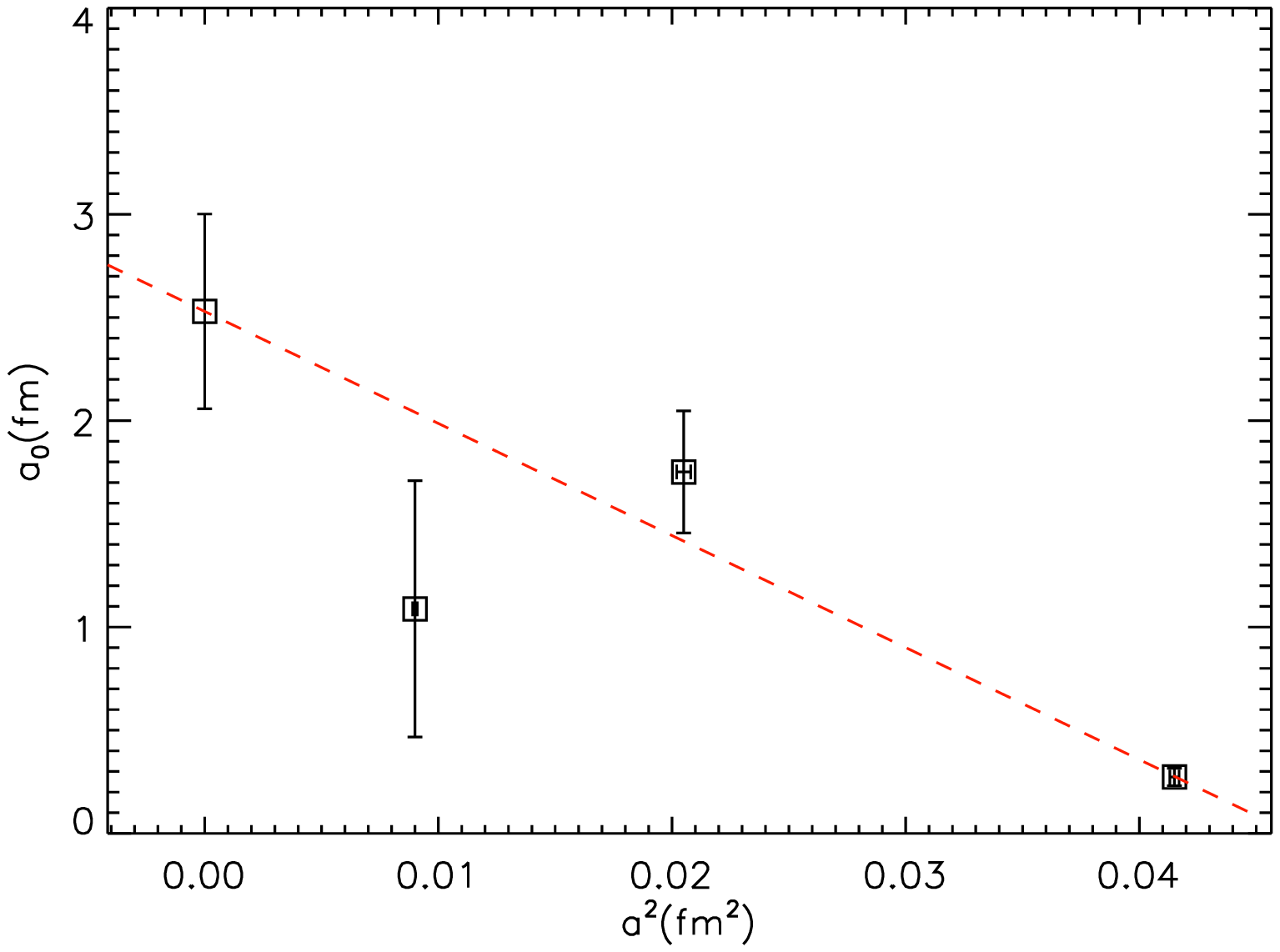}
\includegraphics[scale=0.6]{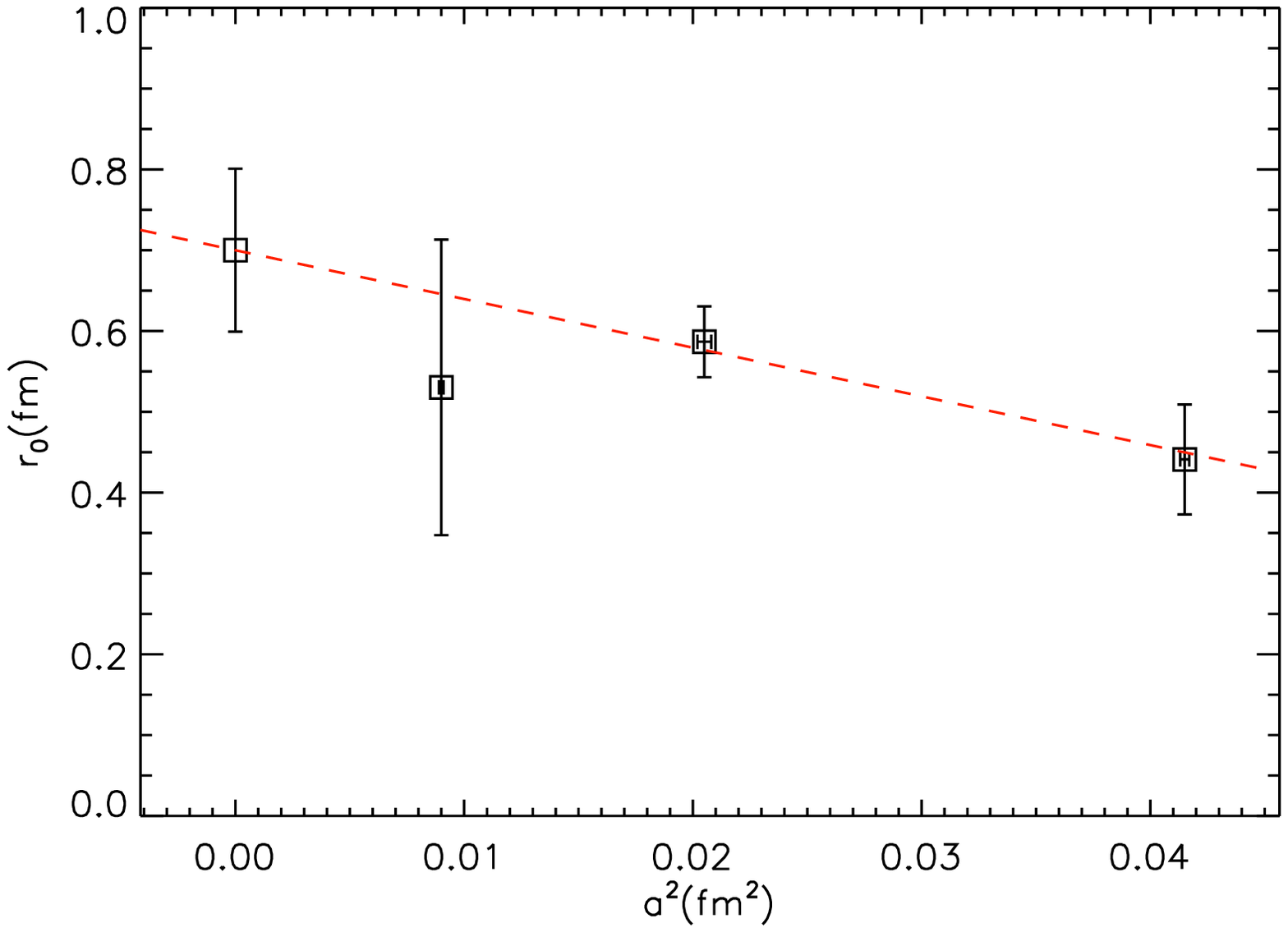}
\caption{Continuum extrapolation for $a_0$ and $r_0$.}
\label{fig:a0r0_spacing}
\end{figure}

 After getting the value of $a_0$ and $r_0$ for each pair of quark
 mass parameter $(\kappa^{ud},\kappa^c)$, the results  are
 interpolated versus $\kappa^c$ to the physical charm quark mass
 which is determined by the experimental value of
 ${1\over 4} m_{\eta_c}+{3\over4}m_{J/\psi}$.
 This is shown in Fig.~\ref{fig:a0r0_heavy}.
 The interpolated data are then taken for the chiral extrapolation.
 In this step, the results for $a_0$ and $r_0$ are extrapolated versus $m^2_\pi$
 towards the chiral limit as shown in Fig.~\ref{fig:a0r0_light}.
 Finally, continuum limit is taken by a linear extrapolation in $a^2_s$
 for the results of $a_0$ and $r_0$ obtained after chiral extrapolation.
 The final results for the scattering length $a_0$ and the effective range $r_0$
 in this channel is shown in figure \ref{fig:a0r0_spacing}.
 After these extrapolations, we obtain the scattering length $a_0$
 and the effective range $r_0$ for the $A_1$ channel:
 \begin{eqnarray}
  a_0=2.53\pm 0.47{\rm fm}, ~r_0=0.70\pm 0.10{\rm fm}
  \;.
 \end{eqnarray}
 This result is for $A^{(1)}_1$ channel which, in the notion of continuum
 quantum numbers, corresponds to  $J^P=0^-$.
 The signal in other channels is much noisier than that
 of $A^{(1)}_1$ channel and it seems that more statistics and/or
 better interpolation operators are
 needed for a reliable extraction of the scattering parameters.
 Results for the scattering length $a_0$ and the effective
  range $r_0$ at various light quark mass parameters for three
  values of $\beta$ are also listed in Table~\ref{tab:test} for
  reference. The results after the chiral extrapolations
  and the final results in the continuum limit are also shown
  in the table.

 \subsection{Possibility of a shallow bound state}

\begin{table}
  \centering
  \begin{tabular}{|c|c|c|}
    \hline
    $\beta$ & $q^2$ & $\cot \sigma(q^2)$\\
    \hline
    2.5 & -0.026(0.003)& 5.23(0.65) \\
    \hline
    2.8 & -0.064(0.005) & 0.16(0.18)\\
    \hline
    3.2 & -0.053(0.016) & 0.92(0.93)\\
    \hline
  \end{tabular}
  \caption{Results for the lowest $q^2$ and the corresponding
  values for $\cot\sigma(q)$ as given by Eq.~(\ref{eq:sigma_rectangular}) for
  different values of $\beta$ in the simulation. Corresponding errors for the quantities
  are also given in the parenthesis.}
  \label{tab:cot_sigma}
\end{table}
 To explore the possibility of a bound state, we recall that
 for a bound state to exist, $q^2$ has to be negative and
 in fact $q^2\rightarrow -\infty$ as $L\rightarrow\infty$.
 This results in the condition: $\cot\sigma(q)=-1$ as
 discussed in the subsection~\ref{subsec:formulae}, Eq.~(\ref{eq:sigma_rectangular}).
 On the other hand, a scattering state will have:
 $q^2\simeq (1/L)$ as $L\rightarrow\infty$.
 Results for the lowest (negative) $q^2$  and the
 corresponding values of $\cot\sigma(q)$ as computed from Eq.~(\ref{eq:sigma_rectangular})
 are listed in Table~\ref{tab:cot_sigma}. It is seen that our results
 for $\cot\sigma(q)$ for the lowest (negative) $q^2$
 are all {\em positive}. The absolute values for the lowest $q^2$
 are also not large. Our results obtained so far seems to be more consistent with
 a scattering state than a bound state.

 One could investigate this possibility from another point of view,
 namely by the values of scattering length
 and effective range. The value of effective range $r_0$ obtained
 is much less than the size of our box so that using
 L\"uscher's formalism is justified. Since we are studying the
 scattering near the threshold, it is appropriate to study
 the problem using non-relativistic quantum mechanics.
 Within non-relativistic quantum mechanics, it is
 known that,
 \footnote{See, for example, ``Quantum Mechanics (non-relativistic
 theory)", 3rd ed., L.D.~Landau and E.M.~Lifshitz, Pergamon Press,
 \S 133.
 Note also that our definition on the scattering length differs from
 theirs by a sign.}
 if a shallow bound state emerges in $s$-wave potential scattering
 at low-energies, the scattering length of the system will
 diverge. In fact, if the potential acquires an infinitely shallow
 bound state, the scattering length should approach
 {\em negative infinity}~\cite{Sasaki:2007kr}. Our lattice results for the
 scattering lengths indicate that it is quite large but
 positive. This usually happens when the potential is
 on the verge of developing a shallow bound state.
 Note that this argument is generally valid for a wide variety of potentials.

 If we further approximate the potential by a square-well potential,
 we could even estimate the depth $V_0$ and the range of the potential $R$ from
 our lattice results on $a_0$ and $r_0$. We find that,
 $R=r_0=0.70(10)$fm and $V_0=73(21)$MeV. These values for a
 square-well potential also gives no bound states. If we fix
 $r_0R=0.7$fm, the first bound state will occur at about
 $V_0\simeq 92$MeV.

\section{Conclusions}
 \label{sec:conclude}
 In this paper, we present our quenched anisotropic lattice study
 for the scattering of $D^*$ and $D_1$ mesons near the threshold.
 The calculation is based on a finite-size technique due to
 L\"uscher which enables us to extract the scattering phases
 from the exact two-particle energies measured in Monte Carlo
 simulations. Our study focuses on the $s$-wave scattering in the
 channel $J^P=0^-$ and the scattering
 threshold parameters, i.e. scattering length $a_0$ and effective range
 $r_0$ are obtained. After the chiral and continuum extrapolations, we
 obtain: $a_0=2.53(47)$fm and $r_0=0.70(10)$fm, indicating that
 the interaction between a $D^*$ and a $D_1$ meson is attractive
 in this channel. As for the other channels, although we have also
 computed the correlation matrices, but the signal is too noisy
 to obtain definite results. Better operators and more statistics
 are probably needed in further studies.

 Based on our results for the scattering phases near the threshold,
 we have also discussed the possibility of a shallow bound state
 in this channel. We investigate the quantity $\cot\sigma$ which should
 approach $(-1)$ for a bound state. Our results for this quantity are all
 positive. Our results for scattering length are also positive. Based on these
 indications, it seems that, although the interaction between the two
 charmed mesons is attractive,
 it is unlikely that they form a genuine bound state right below the
 threshold. The lowest two-particle state is likely to be a scattering state.
 This result might shed some
 light on the nature of the recently discovered
 $Z^+(4430)$ state by Belle.
 However, we should emphasize that, our lattice calculation is done
 in a particular channel only and it is within the quenched
 approximation. Obviously, to further clarify the nature
 of the structure $Z^+(4430)$, lattice studies in other symmetry channels
 and preferably with dynamical fermions are much welcomed.

 \section*{Acknowledgments}

 The author would like to thank
 Prof. H.~Q.~Zheng, Prof. S.~H.~Zhu and
 Prof. S.~L.~Zhu from Peking University for valuable discussions.
 This work is supported in part by NSFC under grant No.10835002, No.10675005 and No.10721063.

 \newpage

\begin{table}
  \centering
  \begin{tabular}{|c|c|c|c|c|c|c|c|}
    \hline
    $\beta$ & $\kappa^{ud}$ & $a_0(fm)$ & $r_0(fm)$ & $a_0(fm)$ & $r_0(fm)$ & $a_0(fm)$ & $r_0(fm)$\\
    \hline
    \hline
     &0&0.36(04) &0.44(06)& & & & \\
    \cline{2-4}
     &1&0.34(03) &0.35(06)& & & & \\
    \cline{2-4}
     &2&0.38(03)&0.33(05)& & & & \\
    \cline{2-4}
    \raisebox{1.5ex}[0pt]{2.5}&3&0.42(04)  &0.31(05)&\raisebox{1.5ex}[0pt]{0.27(04)}&\raisebox{1.5ex}[0pt]{0.44(07)} & & \\
    \cline{2-4}
     &4&0.44(04)  &0.30(05)& & & & \\
    \cline{2-4}
     &5&0.47(04)  &0.30(05)& & & & \\
    \cline{1-6}
     &0&1.63(32) &0.56(04)& & & & \\
    \cline{2-4}
     &1&1.48(27) &0.54(04)& & & & \\
    \cline{2-4}
     &2&1.43(25) &0.53(04)& & & & \\
    \cline{2-4}
    \raisebox{1.5ex}[0pt]{2.8}&3&1.35(21)  &0.51(04)&\raisebox{1.5ex}[0pt]{1.75(30)}&\raisebox{1.5ex}[0pt]{0.59(04)} &
          2.53(47)&0.70(10) \\
    \cline{2-4}
     &4&1.21(18)  &0.49(04)& & & & \\
    \cline{2-4}
     &5&1.18(18)  &0.49(04)& & & & \\
    \cline{1-6}
     &0&0.96(45) &0.46(13)& & & & \\
    \cline{2-4}
     &1&0.98(46) &0.42(13)& & & & \\
    \cline{2-4}
     &2&0.99(48) &0.39(13)& & & & \\
    \cline{2-4}
    \raisebox{1.5ex}[0pt]{3.2}&3&0.93(45)  &0.37(13)&\raisebox{1.5ex}[0pt]{1.09(62)}&\raisebox{1.5ex}[0pt]{0.53(18)} & & \\
    \cline{2-4}
     &4&0.88(44)  &0.35(13)& & & & \\
    \cline{2-4}
     &5&0.84(41)  &0.32(13)& & & & \\
    \cline{1-6}
    \hline
  \end{tabular}
  \caption{Results for the scattering length $a_0$ and the effective
  range $r_0$ at various light quark mass parameters for three
  values of $\beta$. The results after the chiral extrapolations
  and the final results in the continuum limit are also shown.}
  \label{tab:test}
\end{table}

 \end{document}